\documentclass{aa}  

\usepackage[colorlinks=true,     linkcolor=blue, citecolor=blue, filecolor=blue, urlcolor=blue]{hyperref} 

\usepackage{color}
\usepackage{longtable}
\usepackage{multicol}
\usepackage{gensymb}
\usepackage{subfig}
\usepackage{graphicx}

\usepackage{txfonts}

\begin{document}

\title{Star formation at different stages of ram-pressure stripping as observed through far-ultraviolet imaging of 13 GASP galaxies}

   \titlerunning{Ultraviolet imaging of 13 GASP galaxies}

   \authorrunning{Koshy George\inst{1}\fnmsep\thanks{koshyastro@gmail.com}}

\author{K. George\inst{1}\fnmsep\thanks{koshyastro@gmail.com}, B. M. Poggianti\inst{2},  B. Vulcani\inst{2}, M. Gullieuszik\inst{2}, J. Postma\inst{3}, J. Fritz\inst{4}, P. C{\^o}t{\'e}\inst{5}, Y. L. Jaffe\inst{6}, A. Moretti\inst{2}, A. Ignesti\inst{2}, G. Peluso\inst{2}, N. Tomi\'{c}i\'{c}\inst{7}, A. Subramaniam\inst{8}, S. K. Ghosh\inst{9}, S.N. Tandon\inst{10}}

\institute{University Observatory, LMU Faculty of Physics, Scheinerstrasse 1, 81679 Munich, Germany
\and
INAF-Astronomical Observatory of Padova,  vicolo dell'Osservatorio 5 35122 Padova, Italy
\and
University of Calgary, Calgary, Alberta, Canada
\and
Instituto de Radioastronomia y Astrofisica, UNAM, Campus Morelia, A.P. 3-72, C.P. 58089, Mexico
\and
National Research Council of Canada, Herzberg Astronomy and Astrophysics Research Centre, Victoria, Canada
\and
Departamento de F\'isica, Universidad T\'ecnica Federico Santa Mar\'ia, Avenida Espa\~na 1680, Valpara\'iso, Chile
\and
Department of Physics, Faculty of Science, University of Zagreb, Bijenicka 32, 10 000 Zagreb, Croatia
\and
Indian Institute of Astrophysics, Koramangala II Block, Bangalore, India
\and
Tata Institute of Fundamental Research, Mumbai, India
\and
Inter-University Center for Astronomy and Astrophysics, Pune, India
}

\abstract{Galaxies undergoing ram-pressure stripping develop gaseous tails that can extend several kiloparsecs outside the galaxy disc. Under favourable conditions, star formation can occur in the stripped tail, but there are cases where it does not happen, and this can be attributed to several factors, including the properties of the intracluster medium and different stages of stripping, although a clear consensus has not yet been reached. We used FUV and H$\alpha$ imaging from the GASP survey to investigate how different stages of stripping affect star formation properties in the tail and disc of 13 galaxies undergoing stripping. FUV imaging provides a direct opportunity to study star formation, unlike H$\alpha$, which is an indirect tracer and can have other mechanisms responsible for the emission. The 13 galaxies have different stripping strengths, as identified from the MUSE integral field spectroscopy. The star-forming knots in the disc and tails show a good correspondence between the measured FUV and H$\alpha$ flux. This is especially true for strong and extreme cases of stripping, which have developed extended ionised gaseous tails featuring clumpy structures. The mechanism behind the H$\alpha$ emission on the tails of these regions, which correlates well with FUV emission, is photoionisation caused by young massive stars. Optical emission line ratio maps enable us to understand the emission mechanism, which can be attributed to star formation, LINER activity, or a combination of both phenomena and AGN. The star-forming regions in the emission line maps correspond well to the areas with significant FUV flux in these galaxies. FUV imaging reveals that six galaxies exhibit minimal star formation in their tails. In two cases, star formation is limited to the central regions, and their discs are truncated. In galaxies with truncated discs, star formation is confined to a smaller region on the disc, as indicated by the FUV flux, compared to H$\alpha$. Galaxies with strong stripping, marked by significant FUV and H$\alpha$ emission along their tails, are undergoing recent star formation and are likely recent infalls. In contrast, galaxies with truncated discs confine star formation to the centre, likely because they have completed a cluster crossing that depleted most of their outer gaseous disc. Galaxies with little FUV flux along their tails show unresolved H$\alpha$ emission, particularly in the extended H$\alpha$ tail where no FUV emission is present. The unresolved H$\alpha$ emission along the tail may be the result of processes other than star formation.}


\maketitle

\section{Introduction}

Galaxy clusters, the most massive collapsed structures in the Universe, are located at the intersection point of the filaments through which galaxies infall over time. The infalling galaxies are usually in radial orbits and move at high velocities within the cluster before becoming virialised after a few cluster crossings. A high-speed encounter between an infalling gas-rich spiral galaxy and the intracluster medium (ICM) of galaxy clusters, with temperatures of $\sim$ 10$^7$-10$^8$ K and electron densities of $\sim$ 10$^{-4}$-10$^{-2}$ cm$^{-3}$, can result in the hydrodynamic process of ram-pressure stripping \citep{Gunn_1972}. This process, which partially or completely strips the cold disc gas ($\sim$ 10$^2$ K), depletes the fuel for further star formation and is considered an efficient process responsible for quenching star formation in dense environments \citep{Boselli_2022}. This scenario is supported by the observed reduction in gas content of spiral galaxies in clusters \citep{Davies_1973,Giovanelli_1985,Solanes_2001,Jaffe_2015,Lee_2022} as well as by the lack of star-forming spiral galaxies at the cores of rich galaxy clusters in the local Universe \citep{Dressler_1980,Kauffmann_2004,Vulcani_2010,Paccagnella_2016}.\\

Galaxies undergoing ram-pressure stripping create extended gas tails that are a few kiloparsec in length found at a distance from the galaxy disc \citep{Jachym_2019,Serra_2024}. Observations have revealed that such gaseous tails often contain star-forming knots that can create spectacular features in the stripping direction \citep{Poggianti_2019a,Giunchi_2023a,Giunchi_2023b}. In the extreme case, the combination of star formation in stripped tails together with the disc give the galaxy the appearance of a jellyfish when observed in broadband optical and ultraviolet (UV). The baryon content within the stripped tails undergoes different phase transitions that start with the stripping of neutral hydrogen (H\textsc{i}) from the galaxy disc, which is then cooled to form molecular hydrogen (H$_2$) within the tail and can then condense to form star-forming knots that ionise the gas. Ram-pressure stripping can therefore be detected in H\textsc{i}, H$_2$, radio continuum, broad-band optical, H$\alpha$, UV, and X-ray observations \citep{Sun_2006,Owen_2006,Cortese_2007,Chung_2007,Sivanandam_2010,Owers_2012,Fumagalli_2014, Ebeling_2014, Rawle_2014, Poggianti_2016, Bellhouse_2017, Gullieuszik_2017,Moretti_2018a,George_2018,Boselli_2018,Poggianti_2019a,Poggianti_2019b,Moretti_2020,Moretti_2023,Roberts_2021,Roberts_2022,George_2023,Serra_2023,George_2024}. Another scenario has also been proposed for the stripping process, in which diffuse molecular gas is stripped from the disc, condenses into giant molecular clouds, and eventually leads to star formation \citep{Cramer_2020}. Ram-pressure stripping is an efficient process that depletes the fuel for star formation in galaxies in dense environments, and it has been observed in clusters beyond the local Universe \citep{Ebeling_2014,Boselli_2019,Durret_2022,Vulcani_2024,Xu_2025}. \\

Star-forming regions in galaxies are normally found in the disc where cold gas undergoes gravitational collapse. It is puzzling that star formation occurs in situ within the stripped tail, given that it happens in a hostile environment outside the galaxies where the cold gas mixes with the surrounding hot ICM, which can ionise the gas through thermal conduction. Star formation produces H$\alpha$ emission, which is caused by gas ionised by OB stars with lifetimes less than 10 Myr. The H$\alpha$ flux is an indirect indicator of star formation, and it can also be influenced by other processes that ionise gas in the stripped tails. Unlike other indicators, far-ultraviolet (FUV) emission directly traces ongoing or recent star formation, which originates from stellar photospheres with lifetimes less than 200 Myr \citep{Kennicutt_2012}.
There are cases of galaxies observed to have strong H$\alpha$ tails but with little or no flux in the UV, which implies that H$\alpha$ emission in the stripped tails can also occur due to processes other than star formation. \citep{Boselli_2016, Boissier_2012,Yagi_2007,Yagi_2010,Yagi_2017,Poggianti_2019a,Laudari_2022}. A comparison of the flux from different regions in various wavelengths can provide clues to the, star formation nature and the time scales of star formation in stripped galaxy tails, as H$\alpha$ and UV emission trace different stellar population ages. Improving the sample of ram-pressure stripped galaxies with H$\alpha$ tails that have UV observations directly tracing ongoing or recent star formation is therefore useful. Until recently, UV observations of galaxies undergoing ram-pressure stripping were limited to a few clusters in the local Universe \citep{Chung_2009, Smith_2010,Hester_2010,Fumagalli_2011, Boissier_2012, Kenney_2014,Boselli_2018,Cramer_2019,Boselli_2021,Waldron_2023}, with observations of selected galaxies in clusters beyond the local Universe only being made recently \citep{George_2018,George_2023,Gullieuszik_2023}.\\

Observations of many galaxies undergoing ram-pressure stripping in a single cluster are rare, whereas finding ram-pressure stripped galaxies from several clusters is relatively easy. Therefore, UV imaging of ram-pressure stripped galaxies in different clusters can be used to conduct a statistical analysis of star formation in galaxies and any cluster-specific variations in the star formation properties of galaxies undergoing ram-pressure stripping. As part of the GAs Stripping Phenomena in galaxies with MUSE (GASP) survey \citep{Poggianti_2017}, we have observed ram-pressure stripped galaxies in clusters with a redshift (z) $\sim$ 0.04–0.07 using the far-ultraviolet (FUV) band of the Ultraviolet Imaging Telescope (UVIT) on board Astrosat. This paper presents FUV imaging observations of 13 GASP galaxies in ten galaxy clusters and includes the identification of UV knots from the stripped tail and the galaxy disc and a comparison with the MUSE H$\alpha$ flux in the UV-detected knots. The  aims of the study are to identify areas of ongoing star formation in the stripped tail and disc of these galaxies and determine whether H$\alpha$ emitting regions along the stripped tail are also detected in FUV imaging, as well as to compare the emitting regions on the disc. The primary goal of the GASP survey is to examine the gas removal process in galaxies, as observed using the MUSE integral-field spectrograph on VLT\footnote{Very Large Telescope} \citep{Poggianti_2017}. The motivation behind observing selected GASP galaxies in UV is to gain a better understanding of ongoing star formation in stripped gas and the progression of star formation in galaxies, both in the tail and disc. The UV also helps study any lack of star formation in the outer region of the galaxy disc, which can be due to disc truncation as a result of ram-pressure stripping \citep{Koopmann_1998,Koopmann_2004}, and at the centre due to energetic feedback from active galactic nucleus (AGN; see \citealt{George_2019,George_2023}).\\

Previous studies using UV observations of galaxies from the GASP survey have focused on five strong and extreme cases of ram-pressure stripping. These galaxies, with three face-on and two edge-on orientations, provided a different view of star formation along the stripped tail \citep{George_2018,Poggianti_2019a,George_2023,Tomicic_2024}. The UV sample of ram-pressure stripped galaxies has increased manyfold with the identification of numerous candidates from UV imaging observations of five galaxy cluster fields \citep{George_2024}.\\

We discuss the observations in section 2, provide the results in section 3, and summarise the study in section 4. Throughout this paper we adopt a Chabrier initial mass function, a concordance $\Lambda$ CDM cosmology with $H_{0} = 70\,\mathrm{km\,s^{-1}\,Mpc^{-1}}$, $\Omega_{\rm{M}} = 0.3$, $\Omega_{\Lambda} = 0.7$. 

\section{Observations, data, and analysis}

\subsection{13 GASP galaxies}

The GASP galaxies are classified according to the strength of their optical stripping signatures, with a JClass as described in \citet{Poggianti_2016}. JClass 5 and 4 exhibit the strongest evidence and are the most secure candidates for ram-pressure stripping, featuring the classical jellyfish morphology with tentacles of stripped material, while JClass 3 are probable cases of stripping, and JClass 2 and 1 are the weakest, tentative candidates. The FUV imaging data presented here for 13 GASP galaxies, which have a range of JClass (JClass=5-to-1), exhibit varying degrees of ram-pressure stripping features. These galaxies were selected from the sample of 64 GASP galaxies based on different JClass, along with the viewing constraints imposed by the satellite. A more refined classification of galaxies undergoing ram-pressure stripping in the GASP survey is being implemented by \citet{Poggianti_2025}, based on JType from the H$\alpha$ imaging data of the galaxies. According to this new scheme, galaxies with weak signs of stripping, characterised by only weak signs of extraplanar gas, are classified as JTypes 0.3 and 0.5. Galaxies exhibiting significant gaseous tails or unilateral debris are categorised as strong stripping with JType 1. Galaxies with gaseous tails that are at least as long as their stellar disc diameter are classified as extreme stripping JType 2 and display a spectacular jellyfish-like morphology. Galaxies with a truncated H$\alpha$ disc, hosting gas only in the central regions, are assigned JType = 3, which likely indicates an advanced stage of stripping. JType = -9 denotes unknown cases. Table 1 lists the JType of 13 galaxies undergoing ram-pressure stripping with different intensities as is clear from their assigned JTypes. Out of 13 galaxies, six have a JType of 2, five have a JType of 1, two have a JType of 3. Originally classified as an extreme case of ram-pressure stripping with a JClass of 5, JO190 was included in our FUV imaging observations. This galaxy has been assigned an unknown status with a JType of -9 according to the latest revised classification scheme. This suggests that the features observed in this galaxy are unlikely to have resulted from stripping. We present the galaxy FUV imaging for completeness, but it is not part of our analysis.\\

The UVIT on board the multi-wavelength astronomy satellite Astrosat \citep{Agrawal_2006}, was used to perform FUV imaging observations of 14 galaxies. The UVIT consists of twin telescopes: a FUV (130-180nm) telescope and a near-ultraviolet (NUV) (200-300nm), optical (VIS) (320-550nm) telescope, which operates with a dichroic beam splitter. The telescopes, 38cm in diameter, generate circular images over a 28$\arcsec$ diameter field simultaneously in all three channels \citep{Kumar_2012}. There are options for a set of narrow and broadband filters, out of which we used the F154W filter for FUV imaging observations\footnote{UVIT NUV channel stopped working since March 2018 \citep{Ghosh_2021}}. Table 1 provides details on the UVIT observations of the 14 galaxies \footnote{PI: Koshy George, Proposal ID: A10${\_}$058}. The UVIT FUV imaging observations have an angular resolution of $\sim$ 1\farcs4 \citep{Annapurni_2016,Tandon_2017a}. FUV imaging data is corrected for distortion \citep{Girish_2017}, flat field, and satellite drift using the software CCDLAB \citep{Postma_2017}. The images from multiple orbits are coadded to create the master image. The astrometric calibration is performed using the {\tt astrometry.net} package, where solutions are obtained using the USNO-B catalogue \citep{Lang_2010}. The photometric calibration is done using the zero point values generated for photometric calibration stars, as described in \citet{Tandon_2017b} and updated in \citet{Tandon_2020}. The UV magnitudes are in AB system. 

The galaxy cluster fields hosting the galaxies were also observed at optical B and V filters as part of the WINGS and OmegaWINGS surveys \citep{Fasano_2006,Gullieuszik_2015,Moretti_2014,Moretti_2017}. The galaxies were observed using MUSE, a spatially resolved integral field unit spectrograph on the VLT, under the GASP programme \citep{Poggianti_2017}. We use the MUSE $\mathrm{H}{\beta}$ (4861.33 {\AA}), [OIII] (5006.84 {\AA}), [NII] (6583.45 {\AA}), $\mathrm{H}{\alpha}$ (6562.82 {\AA}) and [SII] (6716.44 {\AA}, 6730.81 {\AA}) emission line flux maps of these galaxies in this study. The MUSE data of the galaxies from the GASP survey reach a surface brightness detection limit of V $\sim$ 27 mag arcsec$^{2}$ and logH$\alpha$ $\sim$ 17.6 erg s cm arcsec$^{2}$ at the 3$\sigma$ confidence level.

In this study, we utilise the continuum-removed H$\alpha$ emission line flux map of galaxies, referred to as the H$\alpha$ image for convenience, which was obtained from MUSE observations, to compare with UV observations. The 1$\arcmin$ $\times$ 1$\arcmin$ field of view covers the galaxies, except for three that require two MUSE pointings to cover the disc and the tail. We note that H$\alpha$ imaging from VLT/MUSE has a plate scale of 0.2 arcsec pixel$^{-1}$, with an angular resolution of $\sim$ 1$\arcsec$ \citep{Poggianti_2017}, whereas the UVIT plate scale is 0.4 arcsec pixel$^{-1}$, with an angular resolution of $\sim$ 1.4$\arcsec$ \citep{Tandon_2017b}. \\

\begin{table*}
\caption{\label{t7} Details of galaxies ordered by JType.}
\centering
\label{galaxy details}
\tabcolsep=0.18cm
\begin{tabular}{|l|c|c|c|l|c|c|c|c|c|c|c|c|} 
\hline
GASP ID & RA$_{J2000}$ & Dec$_{J2000}$  & Int:time & cluster & logM$_\star$/M$_\odot$ &  z    & JType & $V$/$\sigma_{cl}$ & BCG$_{sep}$  & BCG$_{sep}$ \\
 & (h:m:s) & ($^\circ$:$\arcmin$:$\arcsec$)  & (sec)  &  &   &                      &        &        &  (Mpc)       &  ($r_{200}$)  \\
\hline
JO69    &     21:57:19.291  &  $-$07:46:44.16     & 17698 & Abell 2399   & 9.89    & 0.0550 &1  &    1.259  & 0.18 & 0.119  \\ 
JO70    &     21:56:04.100  &  $-$07:19:38.21    & 16171 & Abell 2399   &  10.47   & 0.0577 &1  &    0.073 & 2.37  & 1.527   \\  
JO93    &     23:23:11.724  &  $+$14:54:06.70     & 14110 & Abell 2593   &   10.53  & 0.0370 &1  &   2.696  & 1.11 & 0.916  \\ 
JO134   &     12:54:38.247  &  $-$30:09:26.30     & 10251 & Abell 3530   & 9.03    &0.0166 &1  &     16.392  & 1.08 & 0.771 \\ 
JO200   &     00:42:05.019  &  $-$09:32:04.07     & 11136 & Abell 85   &  10.82  & 0.0527 &1  &     0.992 & 0.93 &   0.460  \\ 
JW56    &     13:27:03.026  &  $-$27:12:58.04    & 19029  & Abell 1736 &  9.02   & 0.0387  &2  &   2.326 &  0.29 &  0.154 \\ 
JO85    &     23:24:31.412  &  $+$16:52:05.93    & 18613  & Abell 2589 &  10.65  & 0.0354 &2  &   1.687  & 0.49  & 0.178  \\ 
JO147   &     13:26:49.731  &  $-$31:23:44.79    & 18699 & Abell 3558   &  11.02   & 0.0506 &2  &       0.726  & 0.89  & 0.455 \\ 
JO149   &     13:28:10.548  &  $-$31:09:50.43    & 19009 & Abell 3558   &  8.87   & 0.0438 &2  &       1.404 & 1.14  & 0.586 \\  
JO160   &     13:29:28.584  &  $-$31:39:25.46    & 16252  & Abell 3558 & 10.07   & 0.0483 &2  &   0.005 & 1.23  & 0.632  \\ 
JO171   &     20:10:14.753  &  $-$56:38:29.49    & 18647  & Abell 3667 &  10.62  & 0.0521 &2  &    0.869  & 1.37 & 0.619  \\ 
JO36    &     01:12:59.405  &  $+$15:35:29.59    & 17463  & Abell 160   &  10.81   & 0.0407 &3  &     0.945 & 0.31 &   0.192 \\ 
JO23    &     01:08:08.070  &  $-$15:30:42.73     & 10947 & Abell 151   &  9.67   & 0.0551 &3  & -- & 0.75  & 0.450 \\  
JO190   &     22:26:53.670  &  $-$30:53:10.66    & 14384  & Abell 3880   & 9.34    & 0.0132 &-9  & 25.158  & 1.53 & 1.279 \\ 
\hline
\end{tabular}
\tablefoot{\label{t7} Key to columns: (1) GASP ID taken from \citet{Poggianti_2016}; (2) and (3) equatorial coordinates of the galaxy centre; (4) UVIT FUV F154W filter integration time; (5) host galaxy cluster name; (6) logarithm of the galaxy stellar mass (in solar masses from \citet{Vulcani_2018}); (7) galaxy redshift;  (8) Jellyfish type from Poggianti et al 2025; (9) $V$/$\sigma_{cl}$ of galaxy; (10) \& (11) separation between galaxy and BCG in Mpc and $r_{200}$.}
\end{table*}

We used the emission line diagnostic ratio diagrams \citep{Baldwin_1981} (BPT) created from MUSE data to distinguish gas ionisation processes operating in different regions of the galaxy. We adopt the classification scheme of \citet{Kewley_2001,Kauffmann_2003,Sharp_2010}, which uses $\mathrm{H}{\alpha}$, [SII], [OIII] and [NII] emission line flux maps to create line diagnostic diagrams and distinguish regions dominated by star formation, composite (AGN+SF), LINER, and AGN. The LINER-like emission can be attributed to the ionisation resulting from a series of mechanisms, including a low-luminosity AGN \citep{Ferland_1983,Halpern_1983}, post-AGB stars \citep{Binette_1994}, and an ICM-ISM interaction that heats the ISM through shock and conduction \citep{Bartolini_2022}, turbulence, or diffuse ionised gas \citep{Tomicic_2021a,Tomicic_2021b},
(Radovich et al. in prep).

Far-ultraviolet and H$\alpha$ emission from young star-forming regions contain contributions from stars of different spectral types and hence of different ages on the stellar main sequence. The time scales of star formation in star-forming regions of a galaxy can be understood from the presence or absence of H$\alpha$ and FUV flux. Regions of star formation with ages $>$ 10 Myr will have FUV detection with low or negligible H$\alpha$ (flux from O,B,A spectral type stars on the stellar main sequence), where as regions with star formation of ages less than 10 Myr will have detection in both FUV and H$\alpha$ (flux from O,B spectral type stars). We checked for the correspondence of FUV and H$\alpha$ emitting regions along the stripped tails and disc of the 13 galaxies. We used isophotes made from the MUSE continuum maps and used in other GASP papers to isolate the galaxy's main body and its tails. These are discussed in detail in section 3.1 of \citet{Gullieuszik_2020}. The isophotes are first overlaid as a contour on the FUV and H$\alpha$ map of the galaxies. The region covered within the contour can be considered as the galaxy’s main body and the region outside can be considered as part of the tail. We quantify the FUV flux along the tails and disc, then compare it to the detected H$\alpha$ in galaxies undergoing ram-pressure stripping. The H$\alpha$ and FUV images are at different spatial resolution, which is changed to a common resolution before performing a quantitative analysis. FUV image is astrometrically resampled to the H$\alpha$ image plate scale (0.2$\arcsec$ pixel$^{-1}$) using SWARP from the ASTROMATIC software package \citep{Bertin_2002}. The higher resolution H$\alpha$ image (1$\arcsec$) is smoothed with a Gaussian 2D kernel\footnote{$\sigma$=$\sqrt{\sigma_{FUV}^{2}-\sigma_{H\alpha}^{2}}$} of $\sigma$=2.07 pixels (0.415$\arcsec$) to match with the low resolution FUV image (1.4$\arcsec$). 

The FUV emitting regions are identified and their correspondence with H$\alpha$ emitting regions on the stripped tails and disc of the galaxies is made by creating segments. Segmentation maps are created from the H$\alpha$ and FUV imaging of the 13 galaxies using ProFound \citep{Robotham_2018}. ProFound is a source finding package that generates a dilated segmentation map of pixels belonging to an individual source and performs photometry to measure the associated flux from the region. This technique differs from the traditional approach, which uses circular or elliptical aperture-based photometry. The isophote of the source, which has irregular shapes called segments\footnote{Segment can be considered as a knot. We note that a segment can also contain unresolved emission and may not be a true knot.}, was iteratively dilated to measure the background-subtracted total flux. The iterative dilation process ensures that segments do not grow into each other, thus preventing blending. For the segments thus identified, we obtained the central RA, Dec, total flux and the total number of pixels within which the flux is measured. We created a segmentation map of each galaxy by running ProFound on the H$\alpha$ image, identifying various segments on the disc and stripped tails of the galaxy. Segments with a geometric centre within the contour defining the galaxy's main body are considered part of the disc, while the rest are assigned to the tail. We note there can be projection effects that make some of the tail segments to appear as part of the disc. The segmentation map created from the H$\alpha$ image was then used to extract the positional information of segments, within which we subsequently extracted the FUV fluxes. We used the H$\alpha$ image to identify segments, as we know from the spectroscopic analysis this emission is certainly related to the analysed galaxies, and also because the signal-to-noise of H$\alpha$ images is larger, making our detections easier. The segmentation maps created for the 13 galaxies are shown in Fig \ref{fig:segmap} with H$\alpha$ contour overlaid in black colour. We used the ratio of FUV and H$\alpha$ flux as a colour scheme to the segmentation map. The segments belonging to 13 galaxies are used in further analysis.  

\section{Results}

\subsection{FUV and H$\alpha$ flux ratios along the stripped tails and discs}

\begin{figure*}
\centering
\begin{multicols}{3}
    \includegraphics[width=6.0cm]{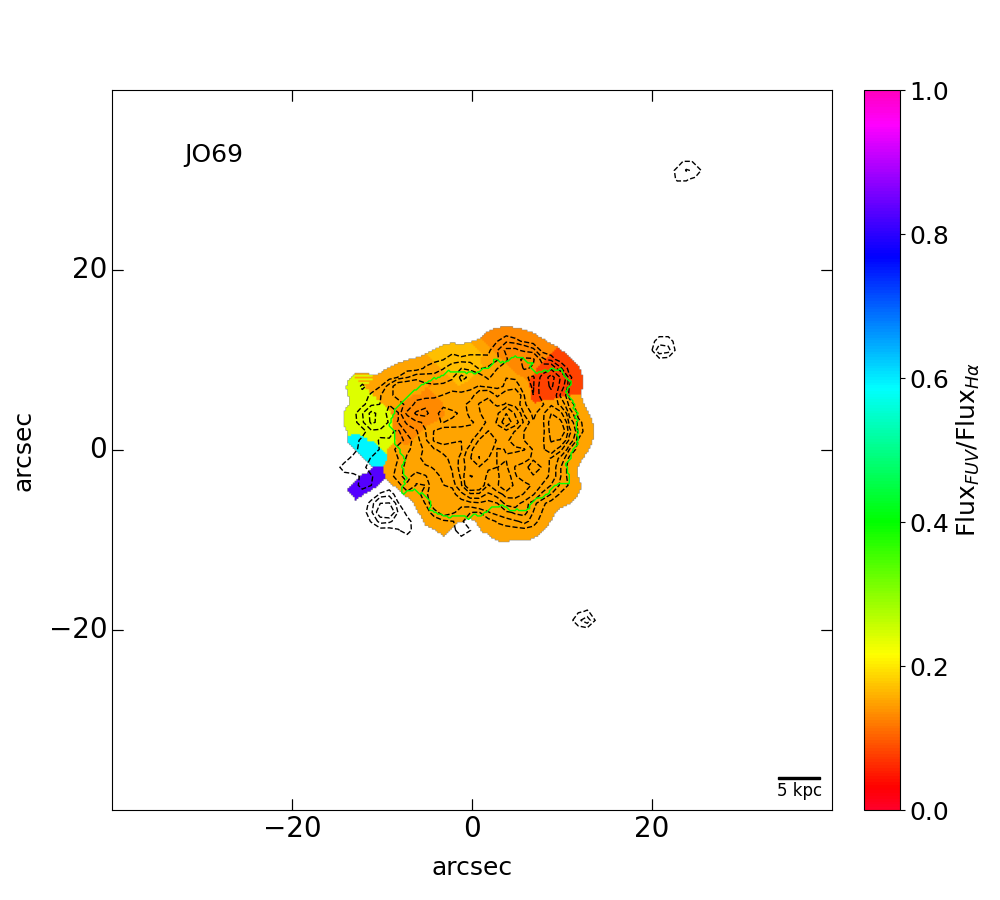}\par 
    \includegraphics[width=6.0cm]{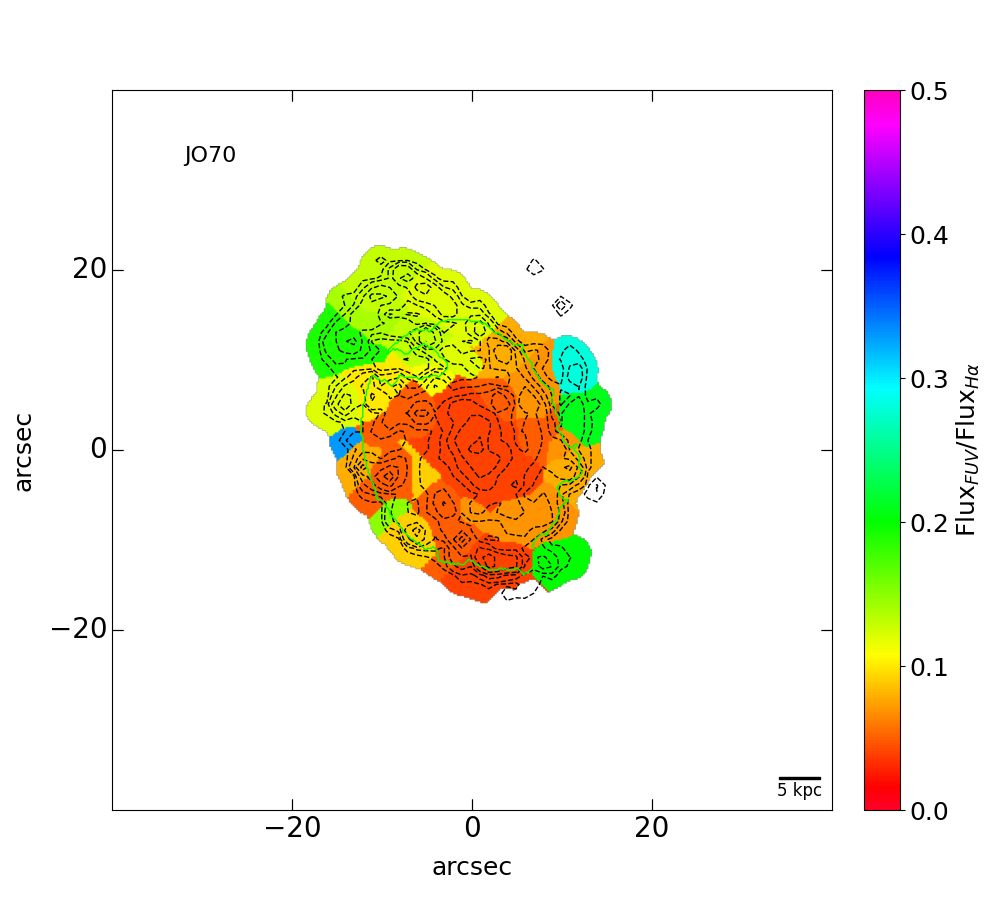}\par 
    \includegraphics[width=6.0cm]{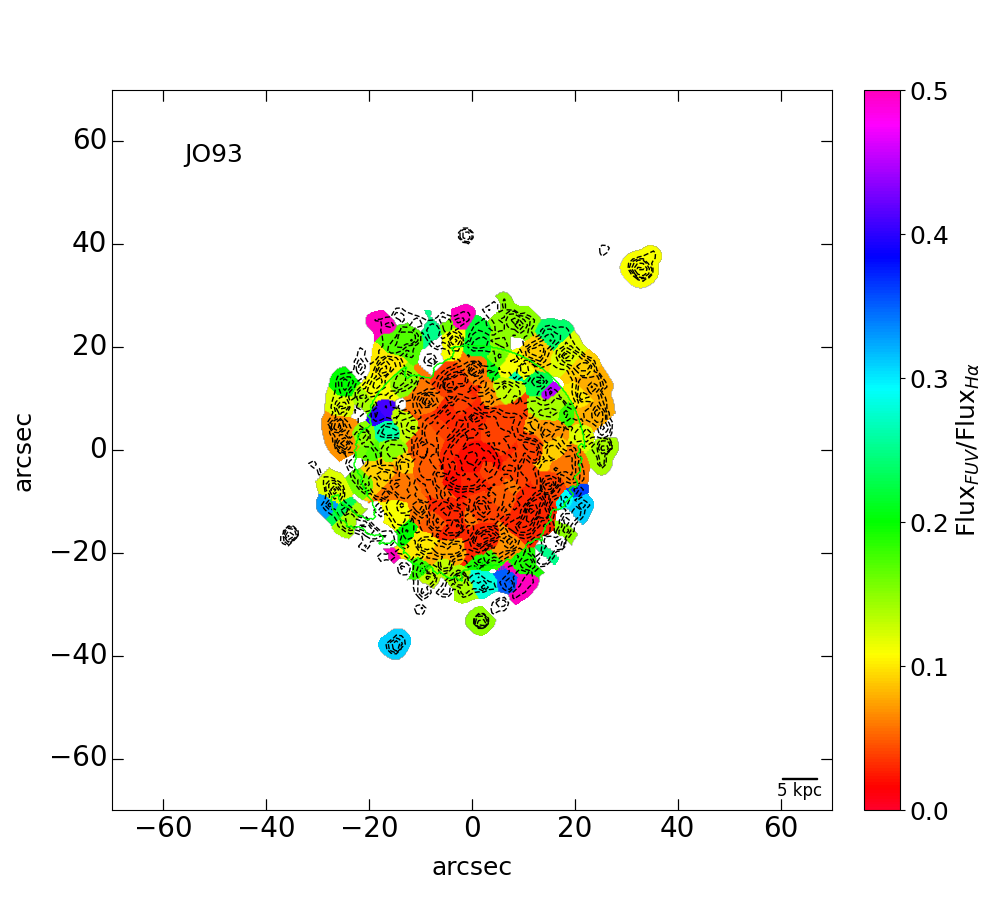}\par
 \end{multicols}
         \caption{Segmentation map for galaxies created from ProFound run over FUV images. The ratio of FUV and $\mathrm{H}{\alpha}$ flux values is used as colour scheme for the segments. The $\mathrm{H}{\alpha}$ flux contours in black and the galaxy main body in green colour are overlaid. Three galaxies shown here and rest are in Appendix.} 
         \label{fig:segmap}
\end{figure*}

\begin{figure}
    {\includegraphics[width=0.5\textwidth]{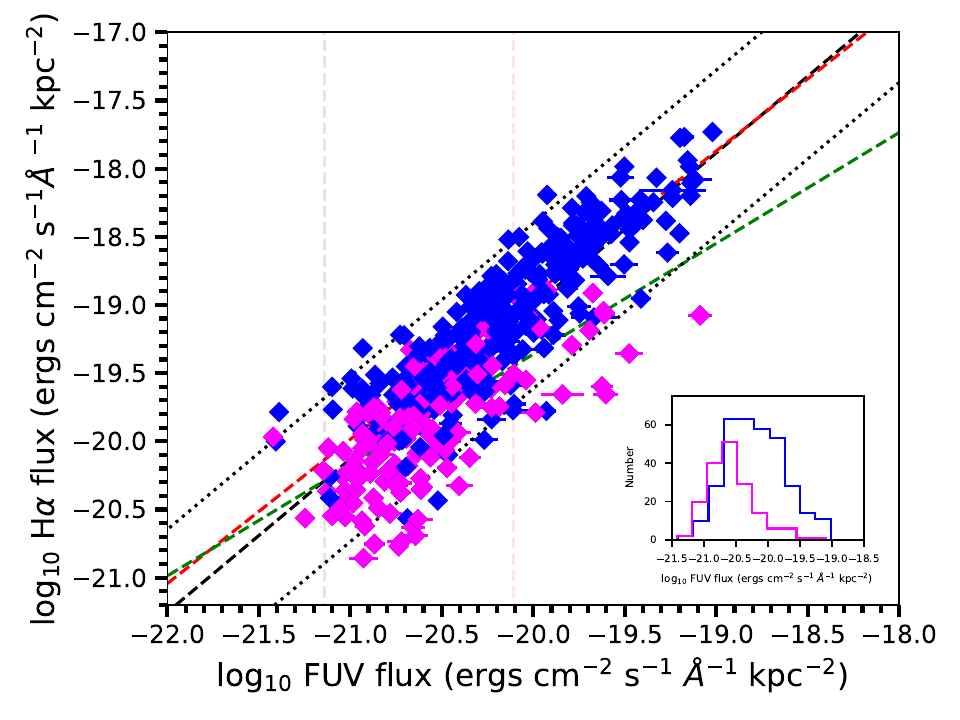}}
    \caption{Far-ultraviolet flux density of the segments detected for 13 galaxies is compared with the $\mathrm{H}{\alpha}$ flux density. The segments from the disc of the galaxy are shown
in blue points and the segments from tails are shown with magenta points. The relation between FUV and H$\alpha$ flux density values of the segments, computed by linear regression, is shown with a black dashed line, with a one-sigma width on each side of the relation indicated by black dotted lines. The relation computed for disc segments are shown in red and the tail segments in green. The FUV limiting flux computed from the full field UVIT imaging of galaxy with highest integration time (JW56) is shown in das black vertical line and for the galaxy with lowest integration time (JO134) is shown in dash red vertical line. The histogram of the FUV flux density of the segments detected from tail and disc of galaxies are shown in the subplot. 
}\label{fig:fuvhalpha}
\end{figure}

 We identify the segments within the galaxy's main body (shown in green in Fig \ref{fig:segmap}) as contributing to the galaxy disc, and the rest as coming from the tails. There are segments detected only in H$\alpha$ with no flux in FUV image. Table 2 gives the details of detected segments on the disc and tails of all 13 galaxies. There are 656 segments with 336 on the disc and 320 on the tails of the 13 galaxies detected from H$\alpha$ imaging. 500 segments are having flux in both H$\alpha$ and FUV images with 170 segments along the tail and 330 segments along the disc. We conducted a detailed analysis of the segments that showed flux in both H$\alpha$ and FUV images to check for a possible correlation. The FUV flux density (flux normalised for the area in kiloparsec$^{2}$) values of the segments are plotted against the H$\alpha$ flux density values for all 13 galaxies in Fig \ref{fig:fuvhalpha}. The grey vertical line represents the FUV limiting flux density for JW56, the galaxy with the highest integration time, while the red vertical line represents the FUV limiting flux density for JO134, the galaxy with the lowest integration time. The distribution of the FUV flux density of segments from the tail and the disc is shown inside the plot. There is a lack of high flux values for the tail segments compared to disc segments. This is due to the disc experiencing more star formation compared to the tails. We note that the disc segments have a larger area, as shown in the segmentation maps in Fig \ref{fig:segmap}.\\

The correspondence or lack of correspondence between the FUV and H$\alpha$ flux density of the segments can constrain the time scales of star formation in the disc and tails of these galaxies. The segments detected from the segmentation map of all 13 galaxies, which have flux in both H$\alpha$ and FUV, exhibit a good correlation between the FUV and H$\alpha$ flux density, as shown in Fig \ref{fig:fuvhalpha}.
The correlation between H$\alpha$ and FUV flux density of the segments implies that star formation over the last 10 Myr is the primary contributor to emission from the segments. The scatter in the FUV and H$\alpha$ flux density values can be attributed to various reasons, including contributions from other emission mechanisms, variations in star formation ages, and internal dust attenuation. We note that we did not perform an extinction correction for these segments. We parameterise the relation by performing a linear regression between FUV and H$\alpha$ flux density of the segments on the tail and disc of the 13 galaxies listed below (also shown in Fig \ref{fig:fuvhalpha}):
\begin{equation}
log_{10} {H\alpha}_{fd}   = (1.12\pm0.03) \times log_{10} {FUV}_{fd} + (3.45\pm0.62).
\end{equation}
%
%
Here, $H\alpha_{fd}$ and $FUV_{fd}$ are the H$\alpha$ and FUV flux density along the tails of the 13 RPS galaxies. This relation will be useful for predicting H$\alpha$ or FUV flux density in other ram-pressure stripped galaxies with FUV or H$\alpha$ flux along the disc and tails. We caution that this relation is only valid for segments with star formation over the last 10 Myr. Few segments deviate significantly from the common trend. This can be seen clearly in the flux ratio scale assigned to the regions on the segmentation map of galaxies in Fig \ref{fig:segmap}. These segments have high FUV flux compared to H$\alpha$, which could be due to the different time scales involved in the star formation process, with integrated star formation in these segments occurring over ages exceeding 10 Myr. We computed the 1$\sigma$ outliers for this relation and consider those segments deviating significantly from the general trend of the segments' H$\alpha$, FUV flux density. There are 30 segments that deviate from this relation, with 20 of them located on the tail and 10 on the disc of the galaxy. The tail segments that are deviant from the relation are from galaxies JO134 (four), JO149 (three), JO171 (one), JO200 (one), JO23 (three), JO69 (two), JO93 (five) with a low H$\alpha$ flux density value with respect to the FUV flux density for these number of segments. JO147 has one segment on the tail that have a high H$\alpha$ flux density value with respect to the FUV flux density. The disc segments that are deviant from the relation are from galaxies JO160, JO171, JO36 (two), JO93 with a low H$\alpha$ flux density value with respect to the FUV flux density for these number of segments. Galaxies JO36 (three), JO85, JO93 have segments on the disc that have a high H$\alpha$ flux density value with respect to the FUV flux density. We don’t see a different emission mechanism (discussed in detail later) contributing to H$\alpha$ in the tails of these galaxies from the emission line maps. This implies these segments may be hosting stellar populations with higher ages (> 10 Myr) that contribute more to the FUV flux compared to H$\alpha$. We note that for all these galaxies FUV imaging was obtained with different integration times, with JO134 having the lowest integration time ($\sim$ 10.25ks) and JW56 the highest integration time ($\sim$ 19ks). We investigated whether the change in integration time affects the detection of FUV flux from the tails of galaxies by calculating the limiting flux from detected objects in the full UVIT imaging of the two galaxy fields. There is a factor of $\sim$ 3 change in limiting FUV flux between the two fields (shown as black and red vertical lines in Fig \ref{fig:fuvhalpha}). This implies there can indeed be a slight shift in the faint end flux values with higher integration time. There are H$\alpha$ segments that lack FUV flux (as shown with magenta arrows) below the FUV detection limits. The limiting FUV flux for the other ten galaxies falls between the red and black vertical lines, and general conclusions about the FUV and H$\alpha$ flux of the segments for all 13 galaxies can be drawn above the red line. We checked for correlation between tail and disc segments separately by performing the linear regression independently for these segments. The slope and intercept values are given in Table 3. The tail segments shows a steeper slope compared to the disc segments.  

Some features in the FUV image may not be present in the H$\alpha$ image of the galaxy. These features are not detected when H$\alpha$ is used as the detection image, but they can be identified by creating a segmentation map from the FUV imaging. We create a segmentation map of each galaxy by running ProFound on the FUV images of 13 galaxies, which identifies various segments on the disc and stripped tails of the galaxy. The positional information of detected segments is then used to obtain the H$\alpha$ flux in the same region from the H$\alpha$ image. FUV imaging of 13 galaxies has detected 311 segments, with 131 located in the tail and 180 in the disc. Details of the number of segments from disc and tail for individual galaxies are given in Table 2. As the table shows, all FUV detected segments, except for JO171, have corresponding H$\alpha$ flux. Two FUV detected segments on the disc of JO171 lack H$\alpha$ flux. We checked for correlation between the FUV and H$\alpha$ flux by performing the linear regression for the total, tail and disc segments. The slope and intercept values are given in Table 3. There are 14 segments that deviate from this relation, with five of them located on the tail and nine on the disc of the galaxy. The deviant segments are from galaxies JO160, JO36, JO69, and JO93 for tail segments, and from galaxies JO160, JO171, JO23, JO36, and JO69 for disc segments. Here, the tail segments exhibit a steeper slope in comparison to the disc segments.

\begin{table*}
\caption{\label{t7} Number of detected segments from the disc and tail of 13 galaxies.}
\centering
\label{galaxy details}
\fontsize{10pt}{10pt}\selectfont
\tabcolsep=0.01cm
\tiny
\begin{tabular}{|l|c|c|c|c|c|c|c|c|c|c|c|c|} 
\hline
Galaxy &  N$_{H\alpha H\alpha disc}$ & N$_{H\alpha H\alpha tail}$ & N$_{H\alpha H\alpha total}$ & N$_{H\alpha FUV disc}$ & N$_{H\alpha FUV tail}$ & N$_{H\alpha FUV total}$ & N$_{FUV FUV disc}$ & N$_{FUV FUV tail}$ & N$_{FUV FUV total}$ & N$_{FUV H\alpha disc}$ & N$_{FUV H\alpha tail}$ & N$_{FUV H\alpha total}$\\
\hline
JO69  &  9     &    12  &  21  & 9   &  6   &  15 &                  5 &    11 &   16  &  5 &    11 &   16\\
JO70  &  25    &    16  &  41  & 25  &  11  &  36 &              6 &    9 &   15 & 6 &    9 &   15 \\
JO93  &  85    &    68  &  153 & 81  &  42  &  123 &      44  &  39   &  83 & 44  &  39   &  83  \\
JO134 & 18 &  15  &  33  & 18  &  10  &  28 &      11  &   5  &  16  & 11  &   5  &  16 \\
JO200 &  58    &    19  &  77  & 56  &  13  &  69 &         24 &    2 &   26  & 24 &    2 &   26\\
JW56  &  2     &    11  &  13  &  2  &  1   &  3 &     1   &   2   &    3  & 1     &   2   &    3 \\
JO85  &  73    &    62  &  135 & 73  &  36  &  109  & 27   &  17   &   44  & 27    &  17   &   44 \\
JO147 &   9    &     33 &  42  & 9   &  4   &  13 &                9    &  5   &  14  & 9    &  5   &  14  \\
JO149 &  11     &    25 &  36  & 11  &  21  &  32 &            5  &  15   &  20   & 5  &  15   &  20 \\
JO160 &  4     &    12  &  16  &  4  &  1   &  5 &     5   &  11   &   16  & 5     &  11   &   16\\
JO171 &  30    &   37   &  67  & 30  &  19  &  49   & 18   &  6    &   24  & 16    &  6    &   22  \\
JO36  &  11    &     7  &  18  & 11   &  3   &  14 &              22  &   8  &   30  & 22  &   8  &   30\\
JO23  &   1    &     3  &  4   & 1   &  3   &  4 &                         4 &    1 &   5 & 4 &    1 &   5 \\
\hline
\end{tabular}
\tablefoot{Key to columns: N$_{H\alpha H\alpha disc,tail,total}$, number of H$\alpha$ detected disc, tail, and total segments. N$_{H\alpha FUV disc,tail,total}$), number of H$\alpha$ detected disc, tail, and total segments with FUV detections.  N$_{FUV FUV disc,tail,total}$, number of FUV detected disc, tail, and total segments. N$_{FUV H\alpha disc,tail,total}$), number of FUV detected disc, tail, and total segments with H$\alpha$ detections.}
\end{table*}

\begin{table}
\caption{\label{t7} Slope and intercept values of the best-fit relation between the flux density of the segments with both FUV and H$\alpha$ flux detected in the H$\alpha$ and FUV images of 13 galaxies.
}
\centering
\label{galaxy details}
\tabcolsep=0.18cm
\begin{tabular}{|l|c|c|c|c|} 
\hline
          &   Slope$_{H\alpha}$  & Intercept$_{H\alpha}$ &  Slope$_{FUV}$  & Intercept$_{FUV}$ \\
\hline          
Tail      &   0.81$\pm$0.06  & -3.10$\pm$1.34 &  0.81$\pm$0.10 & -3.14$\pm$1.96 \\
Disc      &   1.06$\pm$0.03  & 2.22$\pm$0.64 &  0.84$\pm$0.08 & -2.15$\pm$1.53 \\
All       &   1.12$\pm$0.03  & 3.45$\pm$0.62 &  0.83$\pm$0.06 & -2.56$\pm$1.22 \\
\hline
\end{tabular}
\end{table}

\begin{figure*}
\centering
\begin{multicols}{3}
    \includegraphics[width=6.0cm]{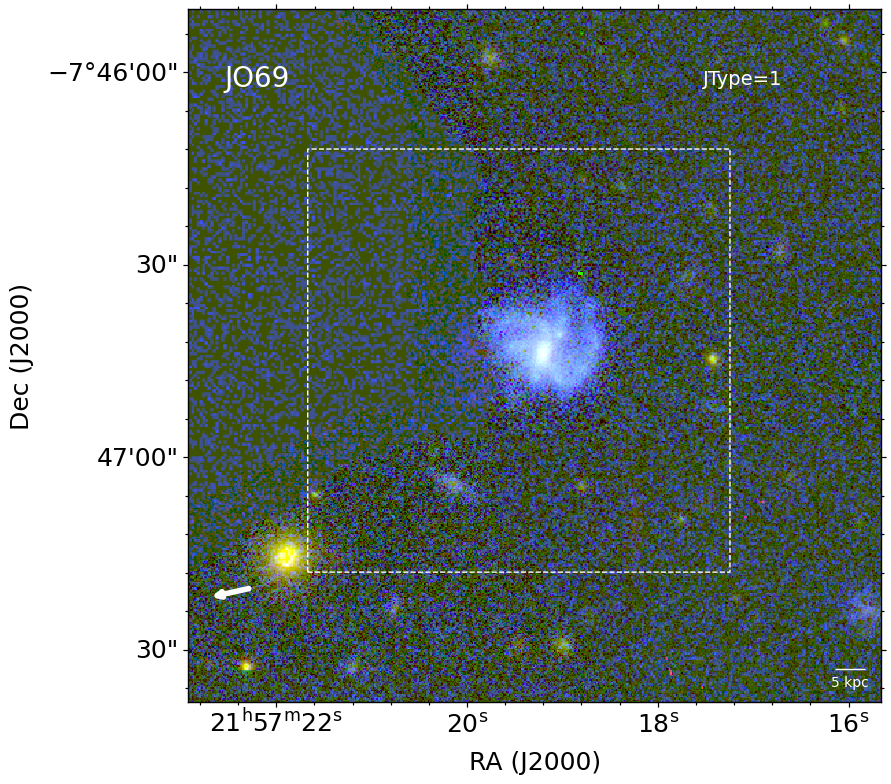}\par 
    \includegraphics[width=6.0cm]{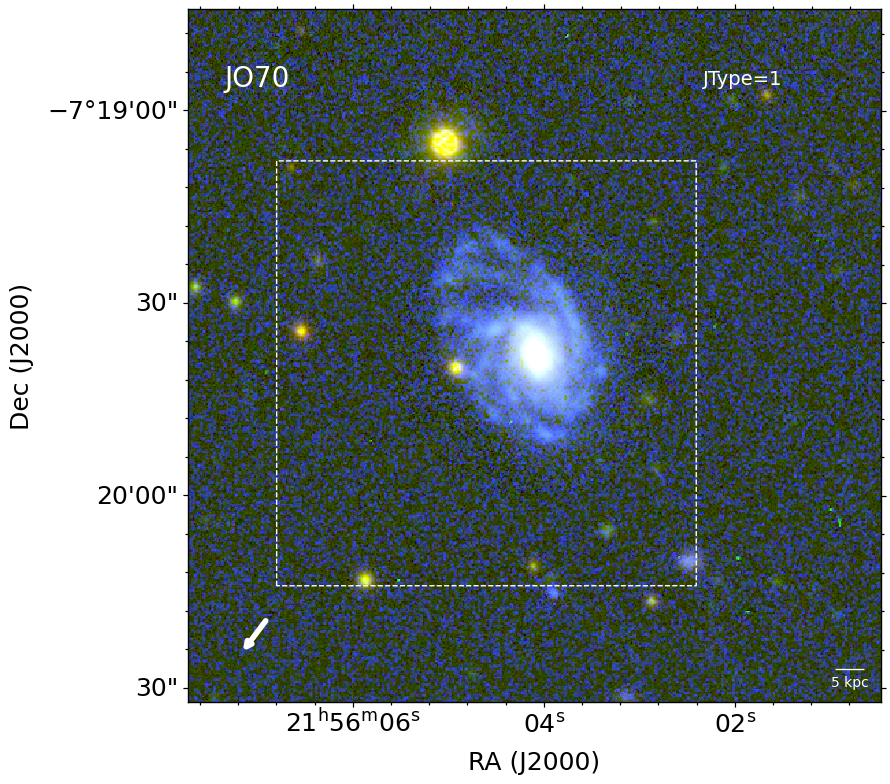}\par 
    \includegraphics[width=6.0cm]{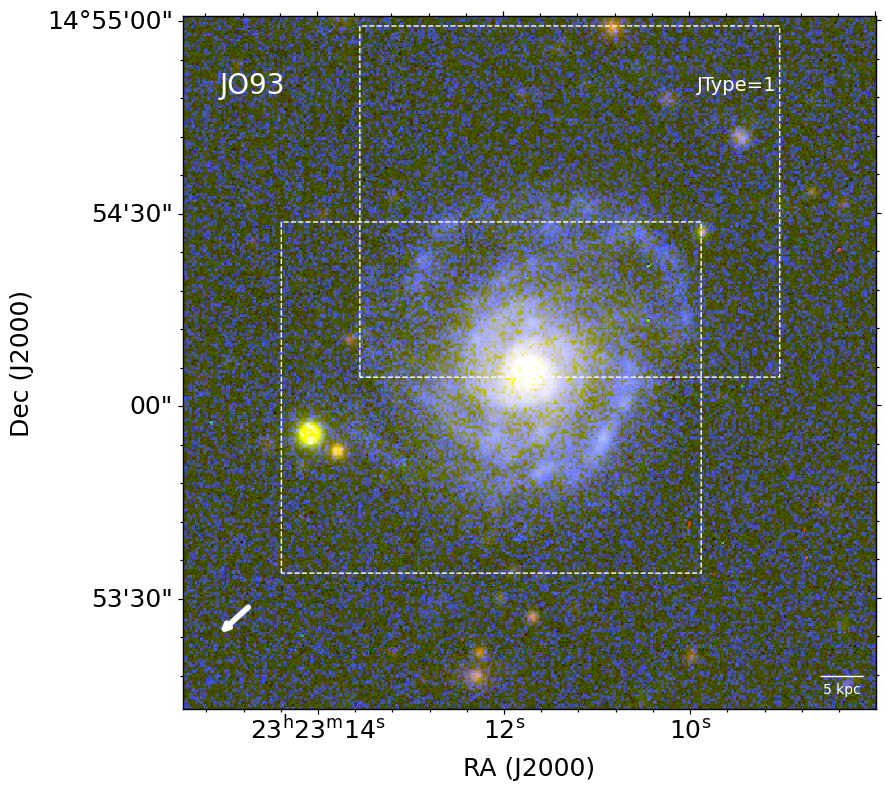}\par
 \end{multicols}
\caption{Colour composite image of galaxies created by combining the FUV image from UVIT with optical B,V imaging from the WINGS/OmegaWINGS survey with 1 arcmin $\times$ 1 arcmin field of view of VLT/MUSE pointings  marked
with a white-dashed line box. The image was made such that the galaxy is at the centre. The direction towards the BCG is shown with a white arrow. The image measures 1.8 arcmin $\times$ 1.8 arcmin. Three galaxies shown here and rest are in Appendix}\label{fig:rgb}
\end{figure*}

\begin{figure*}
\centering
\begin{multicols}{3}
    \includegraphics[width=6.0cm]{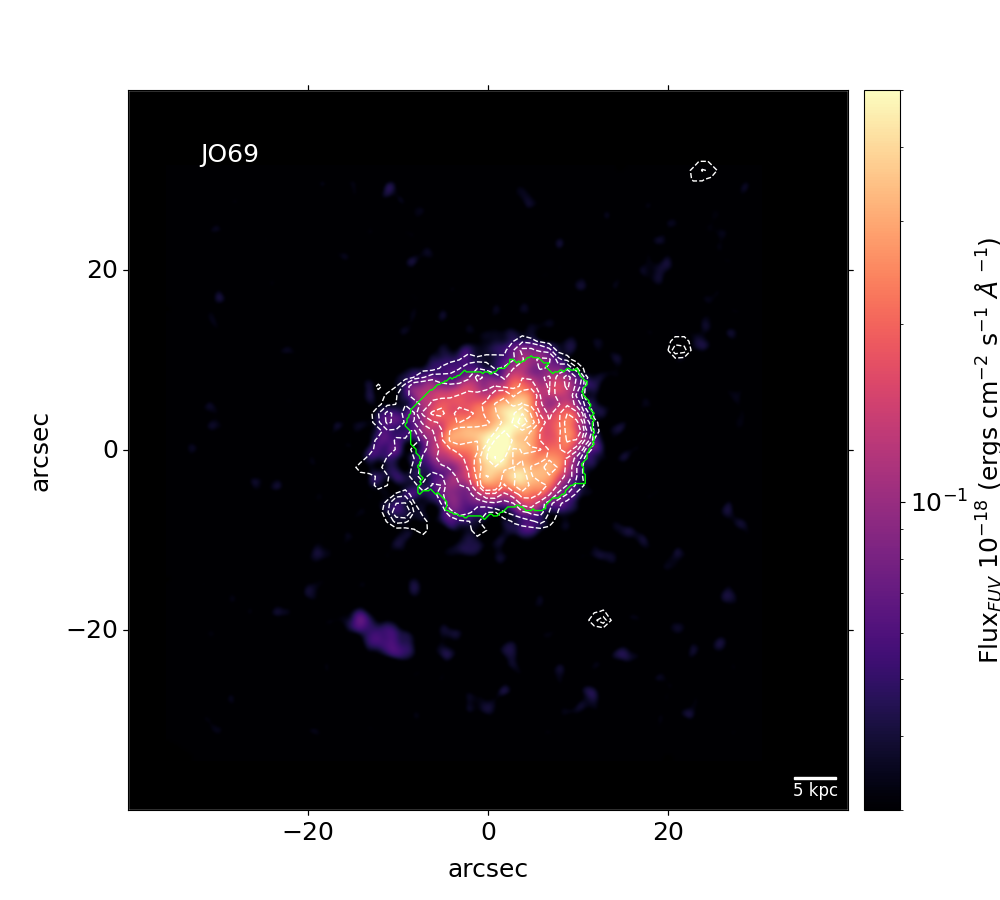}\par 
    \includegraphics[width=6.0cm]{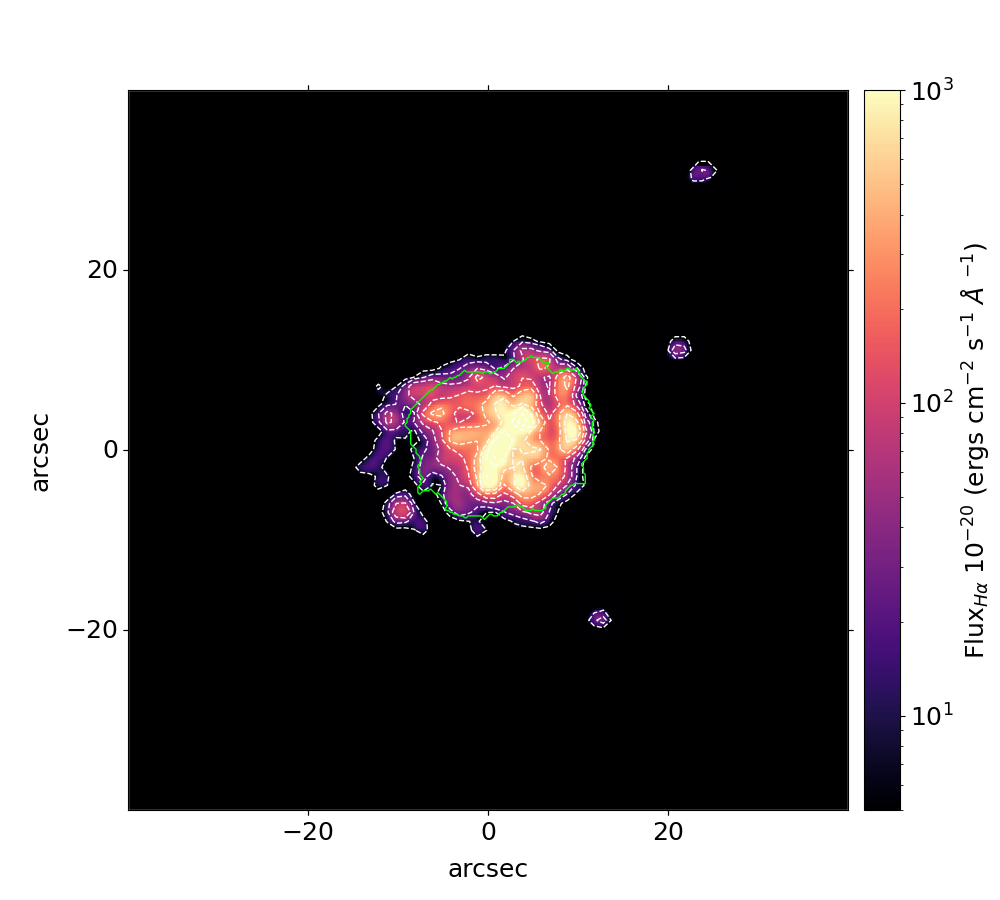}\par 
    \includegraphics[width=6.0cm]{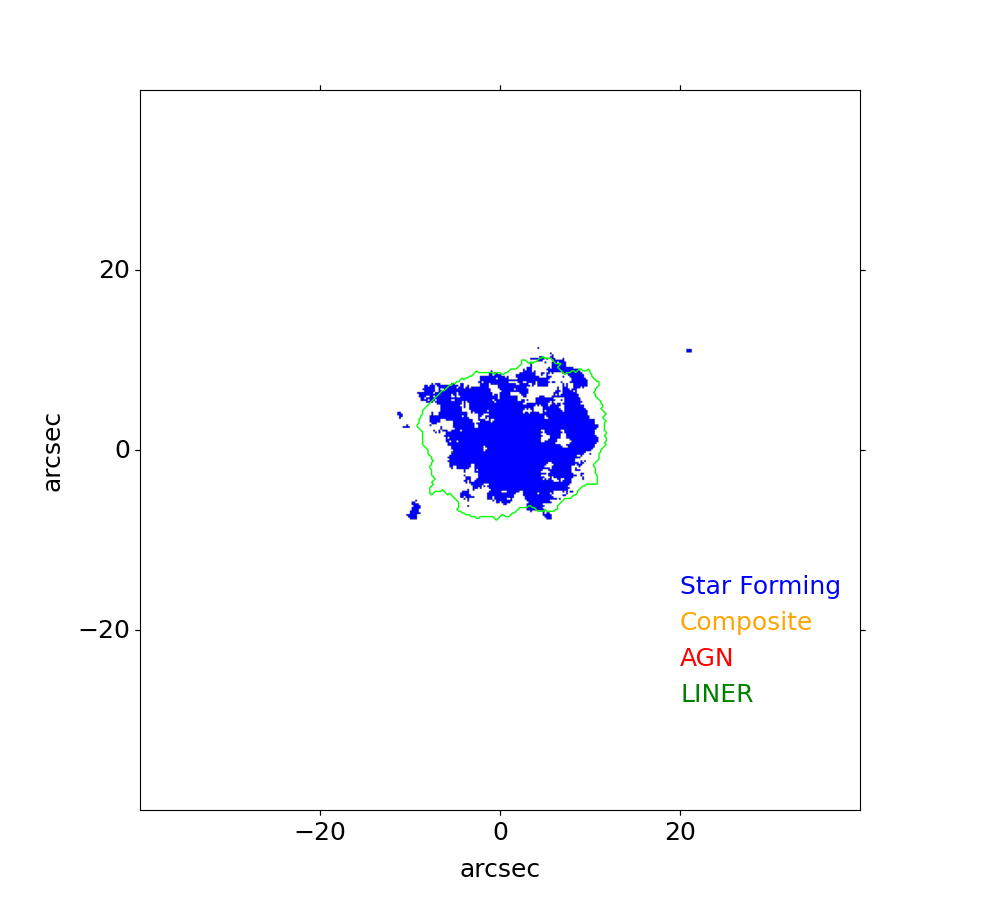}\par
 \end{multicols}
  \caption{Far-ultraviolet, H$\alpha$ flux image, and emission line diagnostic map of galaxies. The $\mathrm{H}{\alpha}$ flux contours in white colour are overlaid over FUV and H$\alpha$ images. The green line defines the galaxy's main body. Regions covered due to emission from LINER, composite (AGN+SF) and star
formation are assigned with different colours in the emission line map. Galaxy JO69 shown here and rest are in Appendix}
 \label{fig:fuvhalphabptmap}
\end{figure*}

\subsection{Star-forming regions}

\subsubsection{Ultraviolet, $\mathrm{H}{\alpha}$ imaging, and emission line region maps}

We investigate whether regions with UV flux on the disc and tails of the galaxy coincide with star-forming regions in the emission line maps. Fig \ref{fig:rgb} displays the colour composite images of galaxies created from FUV and optical B, V imaging data. Fig \ref{fig:fuvhalphabptmap} presents the FUV, H$\alpha$ flux maps, and emission line region maps. The emission line region map shows the regions corresponding to LINER (green), AGN (red), Composite (orange), and star formation (blue). Although the low signal-to-noise ratio of the emission lines along the stripped tails leads to differences in detection, the FUV-emitting regions align well with the star-forming regions in the galaxies' emission line maps of all 13 galaxies. This is a further confirmation that the FUV flux from these regions is due to star formation. Details of a morphological analysis of each galaxy, based on our visual inspection of FUV, H$\alpha$ flux, and emission line region maps can be seen in Appendix.

We examine the star-forming properties of the discs and tails of 13 galaxies, as revealed by FUV imaging observations combined with H$\alpha$ information. We categorise galaxies into two groups: those with FUV detection along the tail and significant H$\alpha$ detection, and galaxies with H$\alpha$ detection along their tails but no significant FUV emission. We have detected significant FUV flux from the stripped tails of seven galaxies (JO171, JO85, JO93, JO200, JO70, JO69, JO149). FUV imaging of these galaxies reveals clumpy star formation in their tails and discs, which is well matched with the H$\alpha$ emitting regions. All of these galaxies belong to strong and extreme cases (JType 1 and 2) of ram-pressure stripping. There are six galaxies with no significant FUV flux (JO160, JW56, JO36, JO23, JO134, JO147) but with H$\alpha$ along the tails. Two of these galaxies have truncated discs.\\

\subsubsection{Truncated disc}

Compared to galaxies of similar mass in the field environment, galaxies in clusters tend to have smaller gaseous discs \citep{Giovanelli_1985}. This truncation in the size of the gaseous disc is attributed to the after-effect of ram-pressure stripping. Spiral galaxies have loosely bound halo gas and tightly bound disc gas. The loosely bound halo gas is the first to strip as a galaxy falls into the cluster. As the galaxy moves into the dense regions of the cluster, the stripping progresses from the outside to the inner regions, and after a few cluster crossings, only the gas anchored to the gravitational potential in the centre regions of the disc remains \citep{Kulier_2023}. Star formation occurs in this cold gas, and as a result, the ongoing star formation in the galaxy’s disc follows the gas stripping pattern. Star formation gradually progresses from covering the whole disc, including the outskirts, to just the inner regions, resulting in an outside-in quenching. In ultraviolet imaging observations, this can be seen as a truncated disc. Two galaxies, JO36 and JO23, in our sample have been assigned JType 3, which is designated for galaxies with truncated ionised gas discs. This implies these two galaxies have gas contained at the central regions of the disc and are probably in an advanced stage of stripping. It is highly likely that these galaxies went through one of the other JType(s) in the past. The case is clear for JO23, as shown in Fig \ref{fig:fuvhalphabptmap}. Truncation is prominent with the FUV flux confined to the very central regions of the disc. This is also the case for JO36, as Fig \ref{fig:fuvhalphabptmap} illustrates, where we observe some flux outside the H$\alpha$ contour, both at the top side of the disc and on the tail. The FUV imaging clearly reveals that star formation is displaced to one side of the disc in two galaxies, JO160 and JW56. These galaxies may be forming a truncated disc, with star formation in the disc being quenched from the outside in.\\

\subsubsection{Special case of JO147}

JO147 is a galaxy classified as undergoing extreme stripping, as evident from JType=2. However, we observe that the FUV flux along the stripped tail of this galaxy does not align with the JType assigned to the galaxies (JType is assigned based on H$\alpha$ morphology of the galaxy.) The reason why the tail of galaxy JO147, which contains significant H$\alpha$ flux, lacks FUV flux is not clear. The H$\alpha$ tail of the galaxy appears unresolved, with clumpy features detected only near the galaxy's disc. 29 segments have been detected with H$\alpha$ flux outside the stellar disc with no corresponding flux in the FUV image. The tail segments highest H$\alpha$ flux density is 1.2 $\times$ 10$^{-19}$ erg cm$^2$ s$^{-1}$ $\AA^{-1}$ kpc$^{-2}$ and median H$\alpha$ flux density value = 6.20 $\times$ 10$^{-21}$ erg cm$^2$ s$^{-1}$ $\AA^{-1}$ kpc$^{-2}$. We used the form of Eq.1 with the coefficients taken for the tail segments from Table.3 to compute the expected FUV flux density for these H$\alpha$ flux density values and found them to be 3 $\times$ 10$^{-20}$ erg cm$^2$ s$^{-1}$ $\AA^{-1}$ kpc$^{-2}$ for the highest flux density and 7.6 $\times$ 10$^{-22}$ erg cm$^2$ s$^{-1}$ $\AA^{-1}$ kpc$^{-2}$ for the median. As shown in Fig \ref{fig:fuvhalpha}, the highest value is well above the detection limits and is expected to be detected with an FUV integration time of $\sim$ 18.7 ks for this galaxy. In general, the segments detected in H$\alpha$ is very diffuse and a corresponding detection in FUV is not possible for all the segments. However if there has been star formation in the regions responsible for the highest H$\alpha$ flux density over the last 10 Myr, we should detect FUV flux  from these regions. A lack of FUV flux likely indicates that other mechanisms are responsible for the detected H$\alpha$ flux from the tail of this galaxy. In the emission line region map for JO147, as shown in Fig \ref{fig:fuvhalphabptmap}, there are also indications of composite emission, although the signal-to-noise ratio is too low to make a robust claim. \\

When interpreting the lack of FUV flux along the tails of these galaxies, one caveat is the possible role of dust attenuation. This is especially true for galaxies viewed slightly edge-on, where dust can cause more attenuation due to a large optical depth \citep{GildePaz_2007}. The flux maps presented here have not undergone attenuation correction. Dust attenuation is expected to affect both H$\alpha$ and FUV flux, resulting in a greater loss of flux for FUV emission compared to H$\alpha$. In regions along the tails of GASP galaxies, the median dust extinction is A$_{V}$ = 0.5 mag for star-forming clumps \citep{Poggianti_2019b}. There can be FUV flux loss due to dust extinction along the clumpy star-forming regions within stripped tails. The H$\alpha$ morphology of the detected features along the tails of galaxies JW56 and JO160, as well as the special case of JO147, displays rather diffuse emission, implying that significant dust content and FUV flux attenuation are unlikely along the stripped tails of these galaxies.

The FUV imaging observations of 13 galaxies presented here provide direct evidence of ongoing star formation in the stripped tails of galaxies. This is pronounced in the case of extreme and strong cases of ram-pressure stripping where the tails display regions with significant FUV flux. We detected FUV flux along the tails where H$\alpha$ detection is clumpy. We note that using H$\alpha$ as the detection image yields a higher number of segments detected from the tail and disc of all 13 galaxies compared to using the FUV image. One common feature of all galaxies is that no counterpart is detected in FUV imaging in the regions of the tail with unresolved H$\alpha$ emission. This suggests that the unresolved H$\alpha$ emission along the tail (with an expected FUV flux above the UVIT detection limit) may be caused by processes other than star formation.

\subsection{Phase space diagram}

\begin{figure}
\centerline{\includegraphics[width=0.5\textwidth]{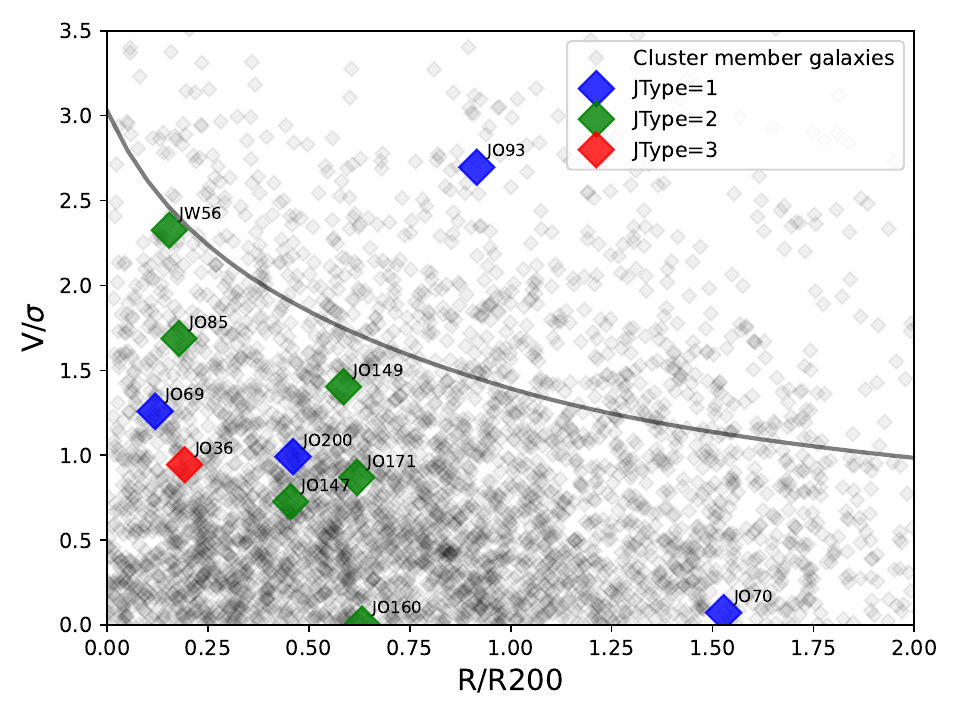}}
\caption{Phase space diagram for 11 jellyfish galaxies is shown with coloured diamond points. We group the galaxies with JType and assign the GASP name of the galaxies to the corresponding points in the plot. Member galaxies in ten galaxy clusters are shown with grey colour diamond points. The normalised, projected distance from the cluster centre (in units of R200) against the galaxy line-of-sight velocities with respect to the cluster mean, also normalised to cluster velocity dispersion. The three-dimensional (unprojected) escape velocity in an NFW halo with concentration $c $= 6 is shown by the grey curve for reference.}\label{fig:vbsigma}
\end{figure}

Galaxies that fall into galaxy clusters with high velocities in radial orbits undergo strong ram-pressure stripping \citep{Jaffe_2018}. These galaxies are observed in elongated orbits at large radii compared to other cluster member galaxies \citep{Biviano_2024,Salinas_2024}. The orbital histories of galaxies undergoing ram-pressure stripping can be studied using the location of galaxies in the projected position versus radial velocity phase-space diagram. A phase-space diagram can help clarify any connection between the observed FUV properties of galaxies and their orbital history. Fig \ref{fig:vbsigma} shows the distance of the galaxies from the centre plotted against the ratio of the differential velocity of the galaxy to the cluster velocity dispersion ($\Delta v_{cl}$/ $\sigma_{cl}$) colour coded for the JType. We normalised the distance to $r_{200}$, which is the radius within which the mean density is 200 times the critical density of the Universe and can be approximated to the cluster virial radius. Galaxies with low velocities relative to the cluster (i.e. $\Delta v_{cl}$/$\sigma_{cl}$ $<1$ ) are expected to have completed at least one cluster crossing, where as galaxies on their first infall are preferentially found with higher velocities across a range of cluster radii  \citep{Jaffe_2015,Jaffe_2018,Jaffe_2019}. Table 1 provides details of the galaxies, including $\Delta v_{cl}$/ $\sigma_{cl}$ and distance from the centre (BCG$_{sep}$). We do not have the $\Delta v_{cl}$/ $\sigma_{cl}$ information for JO23. Galaxies JO134 and JO190 have extremely high $\Delta v_{cl}$/ $\sigma_{cl}$ values and do not belong to the clusters. JO134 is most likely a member of a group structure \citep{Vulcani_2021}. The remaining 11 galaxies are used in the phase space diagram analysis. The galaxies are found close to the centre (except two) in the phase space diagram. At the cluster core, we find the galaxy JO36, which has a truncated disc and may have already completed one cluster centre crossing, stripping away most of its loosely bound disc gas. The remaining gas anchored to the galaxy is forming stars at the central regions as seen in FUV imaging. Located at the outskirts, JO70 and JO93 could be currently undergoing infall to the cluster. JO93 is infalling at high velocities with respect to the cluster, as evident from the high $\Delta v_{cl}$/ $\sigma_{cl}$ value. The remaining eight galaxies are confined within the cluster and could be undergoing ram-pressure stripping prior to cluster centre  crossing. Galaxies undergoing strong and extreme stripping (JType 1 and 2) are experiencing their first infall, with intense star formation occurring in the stripped tails, which explains the observed FUV and H$\alpha$ emission. The galaxies with truncated discs (JType 3) observed in FUV and H$\alpha$ have likely experienced extreme gas stripping from the outer regions of the disc in the past, leaving gas only in the centre. The stripping process removes most of the gas from the outer regions of the galaxy disc, facilitating an outside-in quenching of star formation in these galaxies.

\section{Discussion}

Normally, star formation in galaxies occurs in the disc, where gas loses angular momentum and collapses to form star-forming regions. In ram-pressure stripped galaxies with star formation detected in the tails within the hot ICM, the scenario is entirely different. These galaxies show an enhancement in star formation along the tails (and in the disc) due to the action of ram-pressure \citep{Vulcani_2020}. The process by which stripped gas cools in such environments to form new stars is not well understood. The star formation happening in the stripped tails therefore provides a unique opportunity to understand star formation happening outside the disc within a hot, dense medium. Analyzing star formation progression along the tails and discs of galaxies in a large, statistically significant sample, such as the one presented here, provides insight into the star formation process in galaxies at various stages of stripping, across a range of masses and environments. The detection of FUV emission from galaxies serves as direct evidence of ongoing star formation \citep{Kennicutt_2012}. In principle, comparing FUV emission with H$\alpha$ can confirm the time scales of star formation, but this is complicated by factors such as stochasticity, the shape of the initial mass function, and the recent star formation history, including whether it was bursty, continuous, or occurred over specific time scales \citep{Boselli_2009,Goddard_2010,Barnes_2011,Koda_2012,Hermanowicz_2013,Boquien_2015,Flores_2021,Lomaeva_2022}. Internal attenuation due to dust, along with these effects, can contribute to the scatter observed in the FUV and H$\alpha$ flux comparison shown in Fig \ref{fig:fuvhalpha}. We do not have a clear understanding of how these effects influence the stripped tails of galaxies, and we have assumed they are negligible when comparing H$\alpha$ with FUV flux along these tails.\\

The galaxies presented here have a range of stellar mass, from JO149 (M$_{\star}$=10$^{8.87}$ M$_\odot$), to JO200 and JO36 (M$_{\star}$=10$^{10.82}$ M$_\odot$). These galaxies are drawn from ten different clusters and could thus have experienced different environments. This sample of 13 galaxies, which covers a wide range of stellar masses and cluster conditions, is ideal for exploring the various stages and features imprinted by ram-pressure stripping. Five out of the 13 galaxies presented here, namely JO85, JO93, JO200, JO70, and JO69, display a pattern of unwinding of the spiral arms in FUV imaging. This scenario has been studied in detail based on optical and H$\alpha$ imaging of GASP galaxies by \citet{Bellhouse_2021}, which reveals that the pattern is due to the stripping process. The gas from the outer regions of the spiral arm is first stripped which then collapses and forms new stars at higher orbits as seen in FUV imaging. The pitch angle from the disc is increased in this process which gives the appearance of a pattern of spiral arm unwinding. These galaxies are undergoing strong and extreme stripping, as evident from their JType 1 and 2 classification. This phenomenon can therefore be a common feature in strong and extreme cases of ram-pressure stripped galaxies \citep{Vulcani_2022}. \\

Two galaxies (JO36 and JO23) are displaying strong truncation with FUV flux confined to the central regions of the disc. This could be considered direct evidence for star formation happening only in the central regions of the galaxy. The observed disc truncation is a clear case of outside-in quenching, where the fuel for star formation, outer halo gas, is progressively stripped as the galaxy moves through the cluster's ICM following the first infall. This triggers a shift in star formation from the outer regions to the inner disc, and after several cluster crossings, the galaxy can be completely depleted of gas with a global quenching of star formation. We have observed H$\alpha$ flux emanating from the disc regions of these galaxies, which exhibit FUV emission only at their centres. The emission line map reveals regions with star-formation-related emission lines at the centres of these galaxies. This implies there is ongoing star formation (age less than 10 Myr) in the central region of the galaxy. The H$\alpha$ emission from the disc is more extended compared to FUV emission for JO36 and JO23. The emission line diagnostic map shows regions with composite emission in these outer regions for these two galaxies. This could hint at sources other than star formation responsible for H$\alpha$ emission in the outer regions of the truncated disc. We note that three more galaxies (JO160, JW56, JO147) in our sample can be considered on the way to disc truncation. These galaxies have different levels of stripping as demonstrated from the JType in Table 1. In the cases of JO160, JW56, and JO147, stripping is now occurring on one side of the disc, while the other side lacks FUV flux. All these five galaxies (including JO36 and JO23) are seen edge-on. We note that the observed disc truncation depends on the orientation of the galaxy with respect to the observer. Face-on galaxies undergoing ram-pressure stripping in the direction of the observer can show the unwinding of the spiral arms, but the truncation of the disc cannot be clearly observed. In edge-on galaxies, stripping creates a more pronounced truncation due to the reduction in disc size. The stripping process will remove most of the gas from the galaxy disc, facilitating an outside-in star formation quenching in these galaxies.\\ 

The 13 galaxies undergoing ram-pressure stripping show a good correspondence between the FUV and H$\alpha$ flux emitting regions. We derived a correlation between the H$\alpha$ and FUV flux of segments detected in the tails and discs of these galaxies, indicating ongoing star formation is responsible for these segments with ages under 10 Myr. Eight galaxies were observed to have segments on their tails that deviate from the established relation between FUV and H$\alpha$ flux. The FUV flux in these segments from the tails is high compared to H$\alpha$, which may be due to a lack of star formation within the segments over the last 10 Myr. We note that this is seen in other GASP galaxies studied in FUV and H$\alpha$ \citep{George_2018,George_2023,Poggianti_2019b,Tomicic_2024}.  The star-forming regions in the emission line maps of all galaxies show a good match with regions of FUV flux. The absence of FUV emission from the tails in six galaxies (JO160, JW56, JO36, JO23, JO134, JO147) can be attributed to a lack of star formation. We note that the FUV integration time for two galaxies (JO23, JO134) is shallow, but even with deep integration, little FUV flux is detected along the tails of JO147 and JW56. This is particularly true for JO147, which hosts significant H$\alpha$ emission (mostly unresolved) along the tails, with not much FUV flux. The star formation efficiency might differ along the tails of these galaxies.\\

A common finding in all these galaxies is that no FUV counterpart has been detected to match the unresolved emission observed in H$\alpha$ imaging. There can be regions detected in H$\alpha$ with FUV flux below the detection limits, but for those above the limits, this suggests that mechanisms beyond star formation may also contribute to the unresolved H$\alpha$ emission. One possible reason for the H$\alpha$ emission in such cases could be the thermal ionisation of stripped gas in the tail as it interacts with the hot ICM \citep{Poggianti_2019a,Poggianti_2019b,Sun_2021,Campitiello_2021,Bartolini_2022,Vollmer_2024}. This scenario is appealing because it explains why FUV emission is not observed along the stripped tails, particularly in JO147, where unresolved H$\alpha$ is detected and FUV is expected to be above detection limits. This implies that we should detect FUV emission if the detected H$\alpha$ is caused by star formation. The composite emission regions present in the line diagnostic map, located further outside the disc in the direction of the tail, are likely caused by processes other than star formation or AGN/LINER. The low signal-to-noise ratio along the tail is preventing us from making a robust claim. In this scenario, the observed H$\alpha$ emission along the stripped tail can be explained by thermal conduction between the hot ICM and the relatively cold stripped gas which  depends on local ICM conditions, such as temperature and density, which can vary locally and with distance from the cluster's centre (JO147 is located at $\sim$ 0.5R$_{200}$). Our study shows that only one out of 13 galaxies displays this process across the full extent of stripped tail, which is expected to be prominent at the boundary between the stripped gas and the hot ICM. Notably, this effect is pronounced in edge-on galaxies undergoing ram-pressure stripping, as observed in JW100 (see \citet{Poggianti_2019a}) and JO147 in the GASP survey. Similar to D100 in the Coma cluster \citep{Cramer_2019} and ESO 137–002 in the Abell 3627 cluster \citep{Laudari_2022}, there are other galaxies undergoing ram-pressure stripping that can display similar characteristics in H$\alpha$ and UV, with very low star formation efficiency along the tails. FUV and H$\alpha$ observations of more such galaxies can provide further insights into how thermal conduction inhibits star formation along the stripped tails.

\section{Summary}

We studied the star formation properties of 13 galaxies undergoing different strengths of ram-pressure stripping, using FUV imaging observations from UVIT on board Astrosat. These galaxies were selected from ten galaxy clusters based on signs of ram-pressure stripping in optical B-band imaging data, which were observed as part of the GASP survey and have extensive multi-wavelength data available. Based on H$\alpha$ imaging data, a more refined classification scheme grouped 13 galaxies into strong (five), extreme (six), and truncated disc (two) cases of ram-pressure stripping. We excluded one galaxy, initially identified as a strong stripping case, from our analysis since its status was later classified as unknown in the latest scheme.
We analysed the star-forming regions along the tails and discs of 13 galaxies, comparing the detected FUV flux with the H$\alpha$ flux. We examined how the properties of star formation in the stripped tails are influenced by various stages of stripping. Our main results can be summarised as follows:

\begin{itemize}

\item FUV emission is detected in the tails of galaxies, indicating that star formation is occurring in the stripped tails. Seven galaxies (JO171, JO85, JO93, JO200, JO70, JO69, JO149) show significant FUV and H$\alpha$ emission along the stripped tails. \\
 
\item There are six galaxies with no significant FUV flux (JO160, JW56, JO36, JO23, JO134, JO147) but with unresolved H$\alpha$ along the tails. The unresolved H$\alpha$ emission along the tail may be caused by processes other than star formation, such as thermal conduction.\\

\item We identified 656 segments in 13 galaxies using H$\alpha$ imaging, with 336 located on the disc and 320 along the tails. Among the 500 segments with flux in both H$\alpha$ and FUV images, 170 are situated along the tail and 330 along the disc. The FUV and H$\alpha$ flux density of these segments shows a correlation, which can be described by a global relation derived from 13 galaxies. The correlation becomes steeper when only segments from the tail of the galaxies are considered. This correlation suggests that the primary source of emission in these segments is star formation that has occurred within the last 10 Myr. The stripped tails of eight galaxies (JO134, JO149, JO171, JO200, JO23, JO69, JO93, and JO147) show 20 segments that deviate from the H$\alpha$, FUV flux density correlation, possibly due to different time scales (>10Myr) involved in the star formation within these segments.\\

\item We detected a lack of FUV and H$\alpha$ emission along the disc region of galaxy JO171, with a nearby region exhibiting only FUV flux and no H$\alpha$, indicating star formation with ages > 10 Myr. This could be due to the propagation of star formation quenching within the disc as the galaxy moves within the cluster.\\

\item Two galaxies, JO36 and JO23, display FUV emission confined to the central region of the disc, which appears smaller than the H$\alpha$ region. These galaxies are experiencing truncated star formation, with extended H$\alpha$ emission suggesting that a process other than star formation is at play.\\

\item There is a lack of FUV emission along the tail of galaxy JO147, which hosts an extended ionised gas tail in H$\alpha$. The presence of H$\alpha$ along the tail, without FUV emission, suggests that the emission is likely not caused by star formation, but rather by other processes like thermal conduction. \\

The present study demonstrates that FUV emission from the galaxies studied here is the result of star formation in galaxies at different stripping stages. Galaxies with significant FUV and H$\alpha$ emission along their tails (JO171, JO85, JO93, JO200, JO70, JO69, and JO149) are undergoing recent star formation in the stripped tails and are likely recent infalls. There are six galaxies (JO160, JW56, JO36, JO23, JO134, and JO147) that exhibit minimal star formation in their tails, with JO36 and JO23 being two cases where star formation is limited to the central regions and their discs are truncated. Galaxies with truncated discs confine star formation to the centre, likely because they have completed a cluster crossing that depleted most of their outer gaseous disc. Galaxies with low FUV flux along their tails, such as JO147, display unresolved H$\alpha$ emission in extended tail regions where no FUV emission is present. The unresolved H$\alpha$ emission along the tail may result from processes other than star formation, such as thermal conduction. A detailed follow-up observation of galaxies with H$\alpha$ emission but no FUV emission can provide more insight into the underlying process driving this emission.

\end{itemize}

\begin{acknowledgements}
This paper uses the data from the AstroSat mission of the Indian Space Research  Organisation  (ISRO),  archived  at  the  Indian  Space  Science  Data Centre (ISSDC). Based on observations collected by the European Organisation for Astronomical Research in the Southern Hemisphere under ESO program 196.B-0578 (MUSE).
This project has received funding from the European Research Council (ERC) under the European Union's Horizon 2020 research and innovation programme (grant agreement GASP n. 833824. J.F. acknowledges financial support from the UNAM-DGAPA-PAPIIT IN111620 grant, Mexico. We acknowledge support from the INAF GO grant 2023 "Identifying ram pressure induced unwinding arms in cluster galaxies" (PI Vulcani). Y.L.J. acknowledges support from the Agencia Nacional de Investigaci\'on y Desarrollo (ANID) through Basal project FB210003, FONDECYT Regular projects 1241426 and 123044, and  Millennium 
Science Initiative Program NCN2024\_112.

\end{acknowledgements}


%
   
%


\begin{appendix}
\counterwithin*{figure}{part}
\stepcounter{part}
\renewcommand{\thefigure}{A.\arabic{figure}}

\begin{figure*}
\centering
\begin{multicols}{3}
    \includegraphics[width=6.0cm]{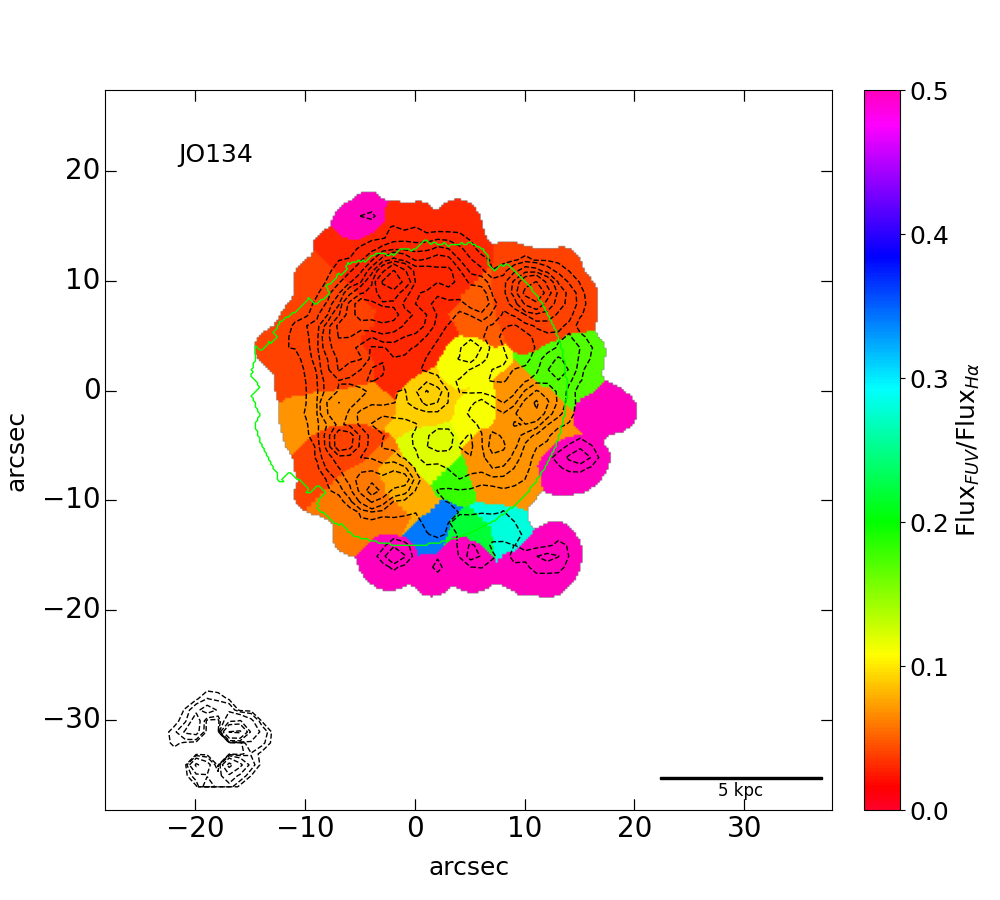}\par 
    \includegraphics[width=6.0cm]{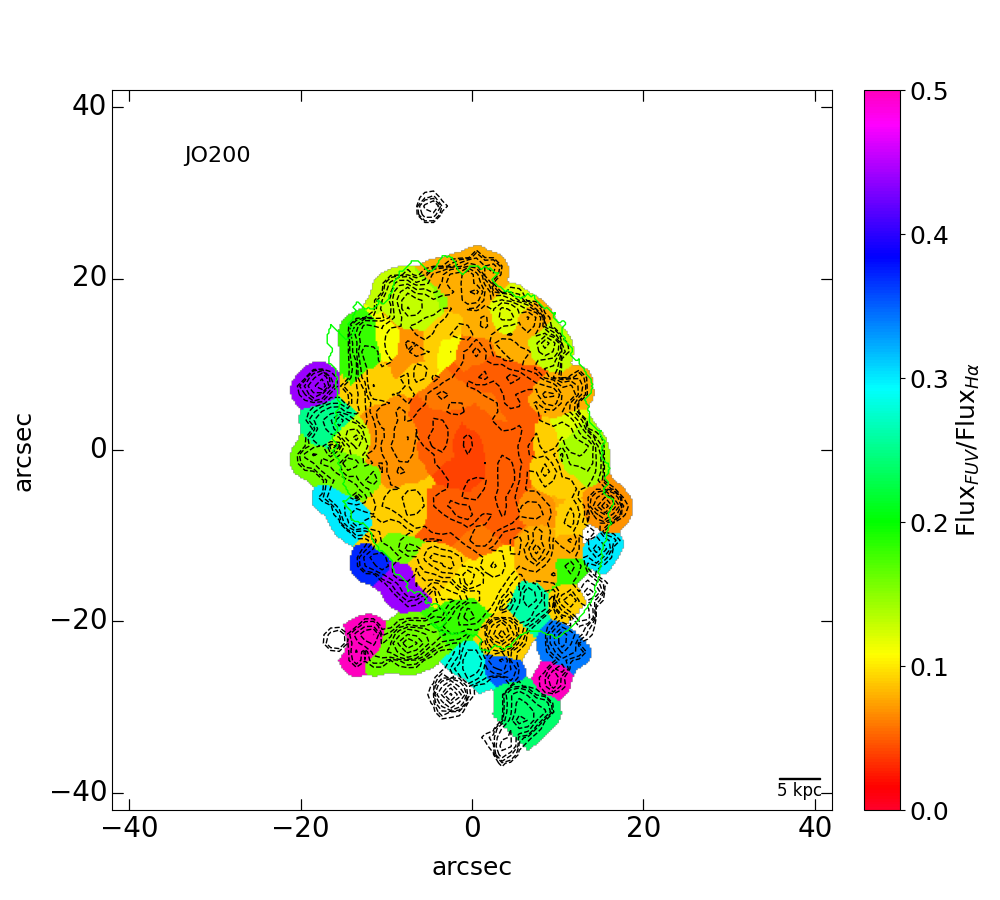}\par 
    \includegraphics[width=6.0cm]{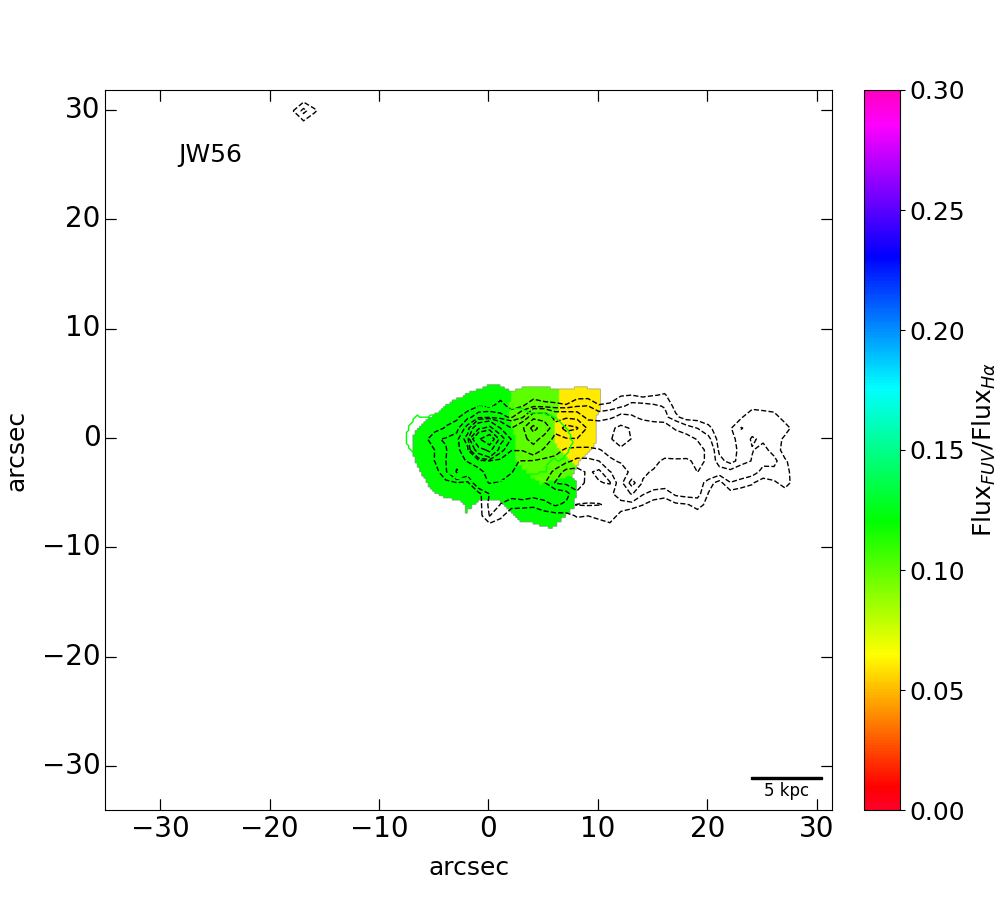}\par
 \end{multicols}
 \begin{multicols}{3}
    \includegraphics[width=6.0cm]{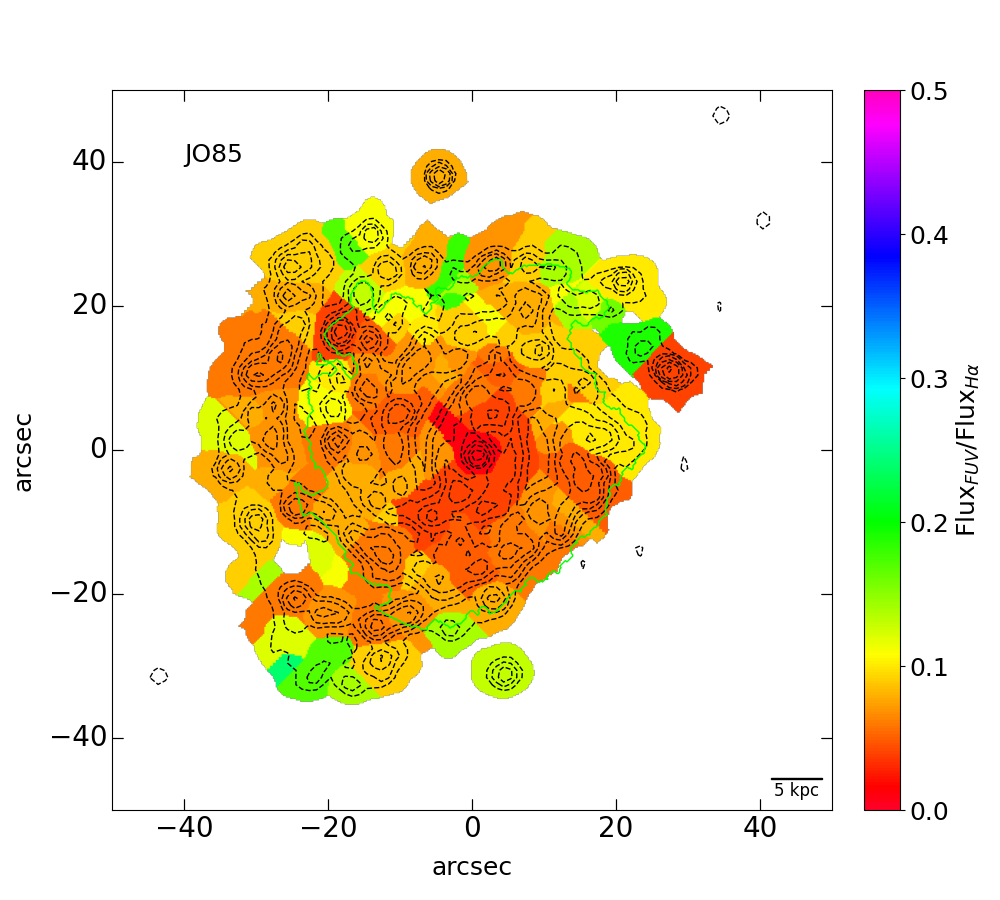}\par 
    \includegraphics[width=6.0cm]{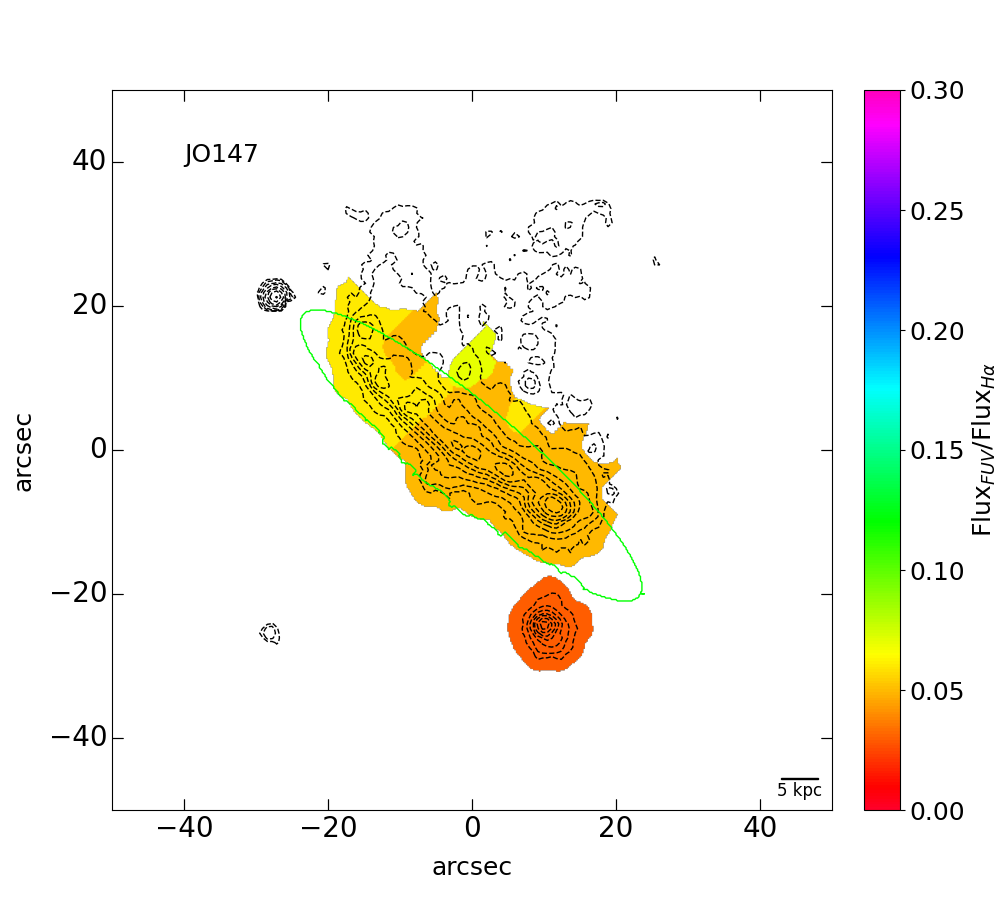}\par 
    \includegraphics[width=6.0cm]{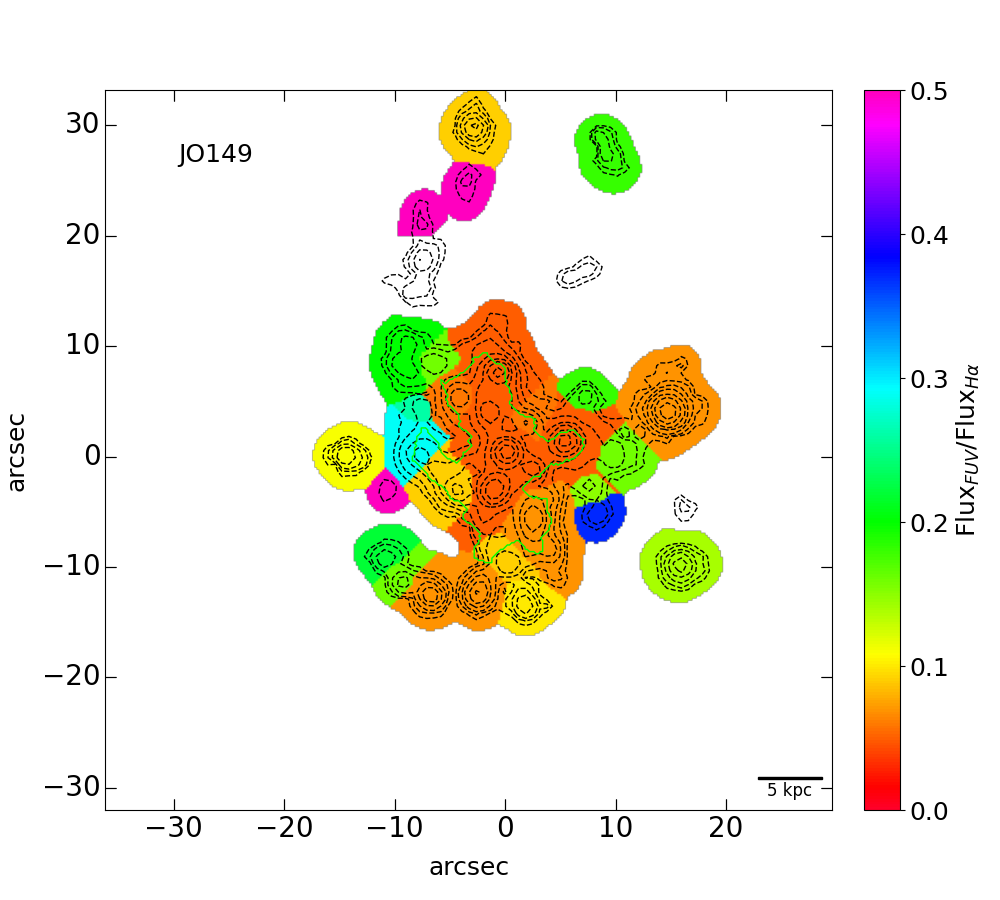}\par
 \end{multicols}
  \begin{multicols}{3}
    \includegraphics[width=6.0cm]{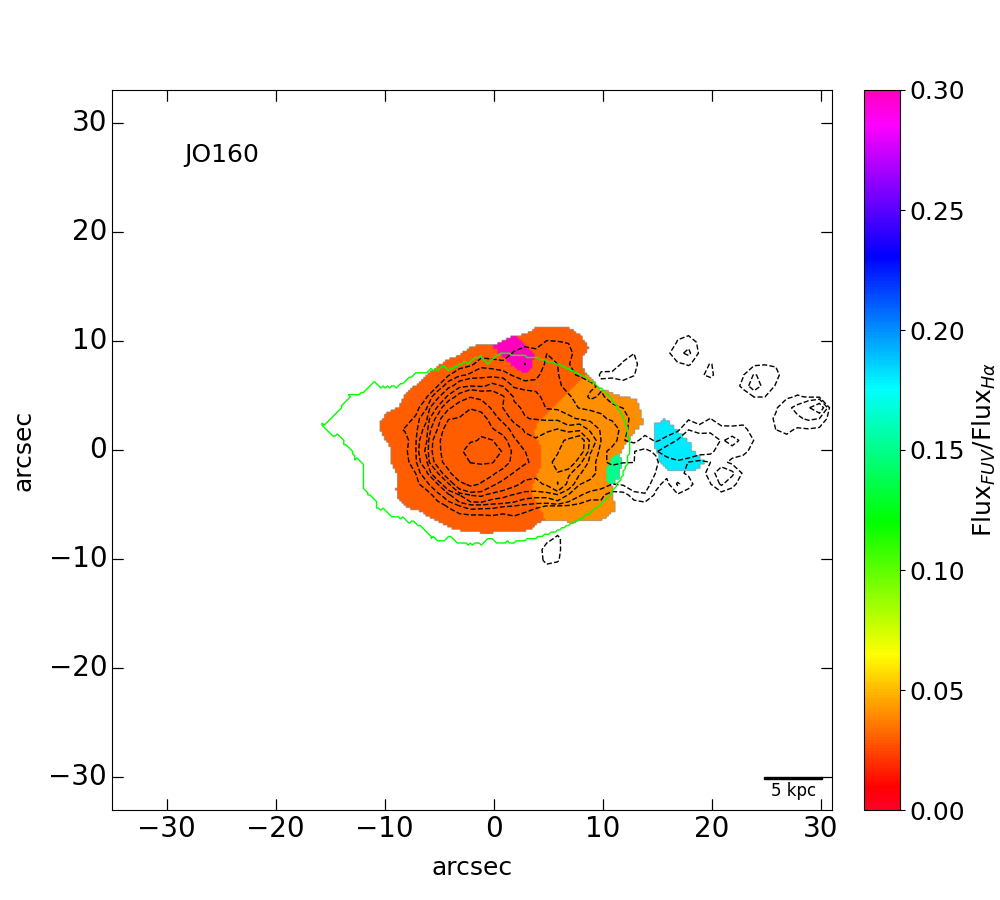}\par 
    \includegraphics[width=6.0cm]{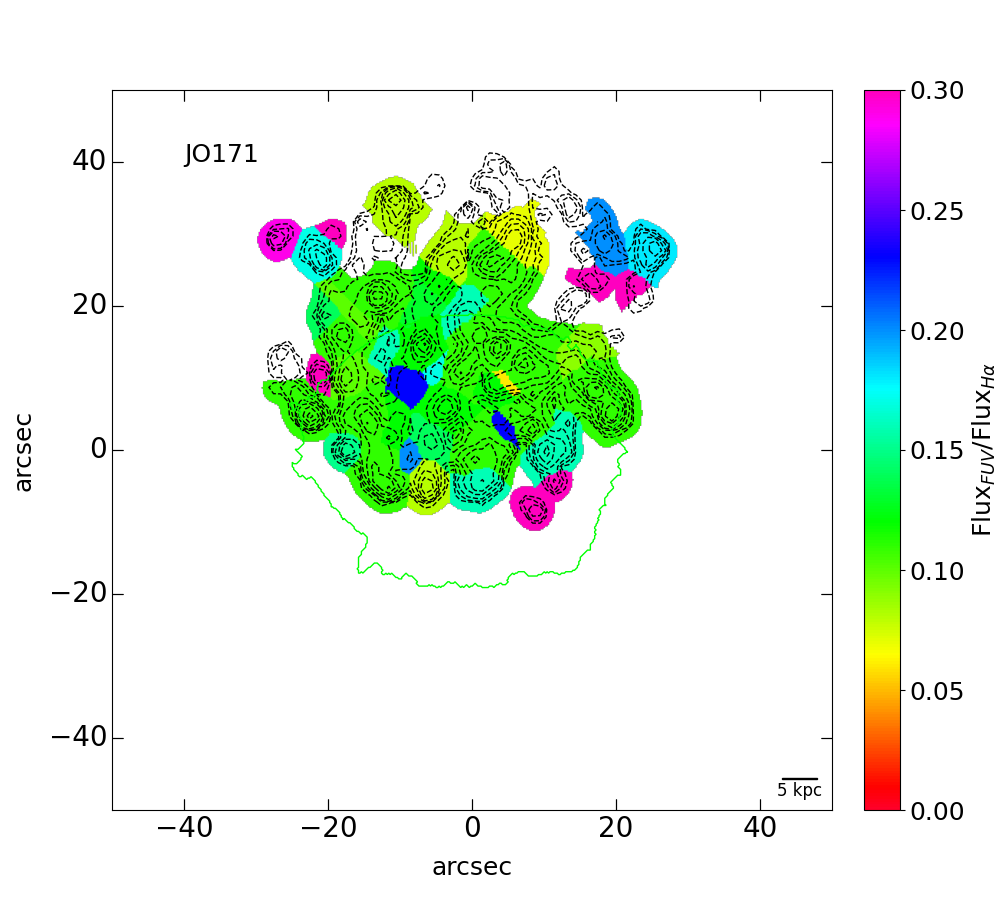}\par 
    \includegraphics[width=6.0cm]{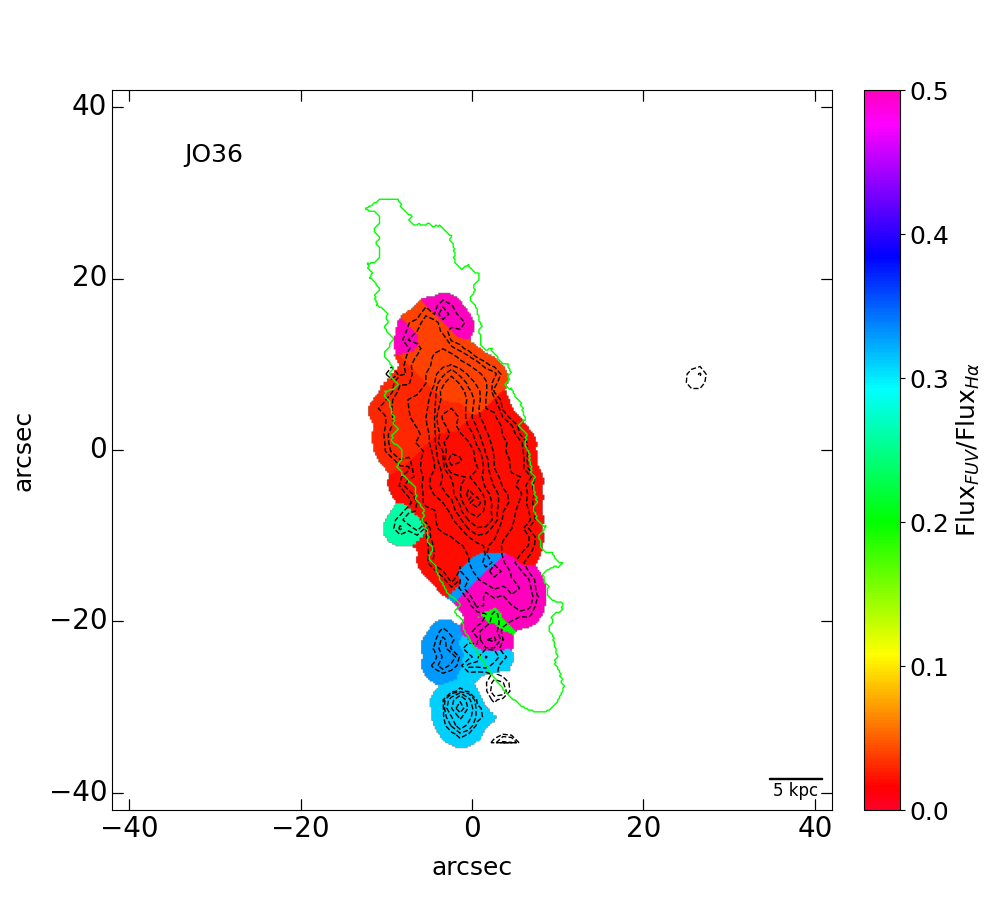}\par
 \end{multicols}

 \begin{multicols}{3}
    \includegraphics[width=6.0cm]{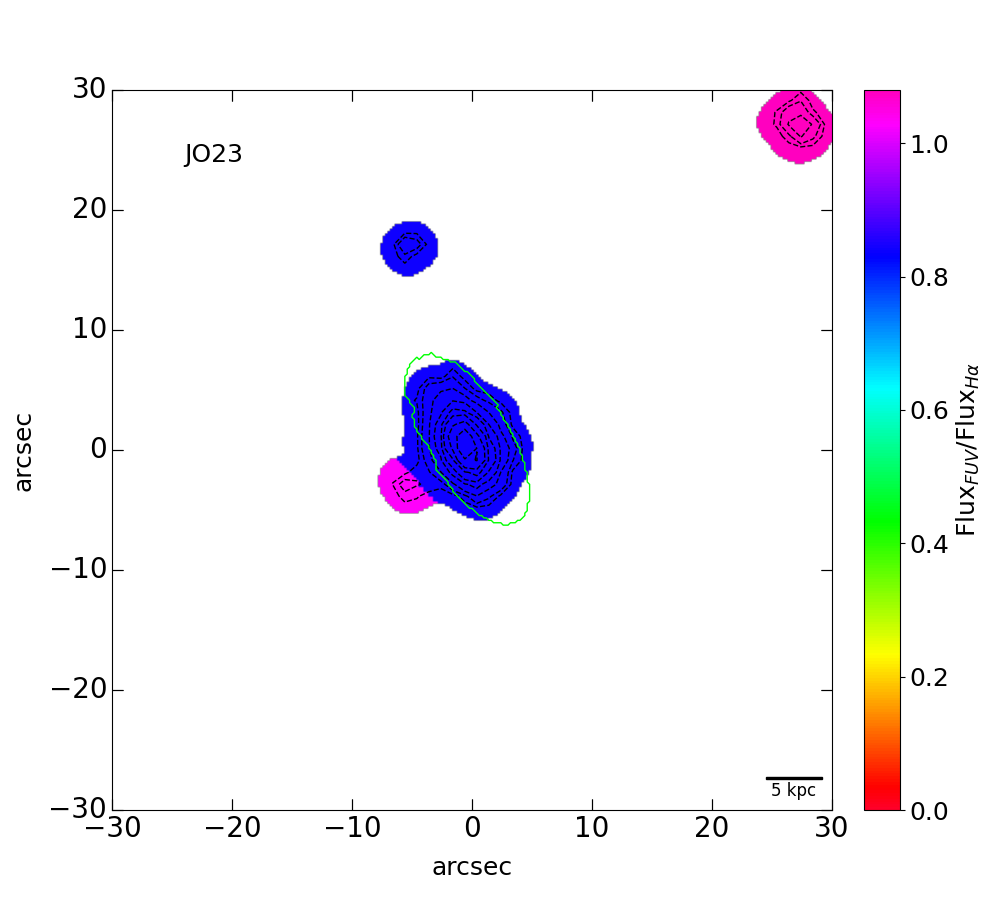}\par 
     \end{multicols}
         \caption{Segmentation map for galaxies created from ProFound run over FUV images. The ratio of FUV and $\mathrm{H}{\alpha}$ flux values is used as colour scheme for the segments. The $\mathrm{H}{\alpha}$ flux contours in black and the galaxy main body in green colour are overlaid.} 
         \label{fig:segmapA1}
\end{figure*}

\begin{figure*}
\centering
\begin{multicols}{3}
    \includegraphics[width=6.0cm]{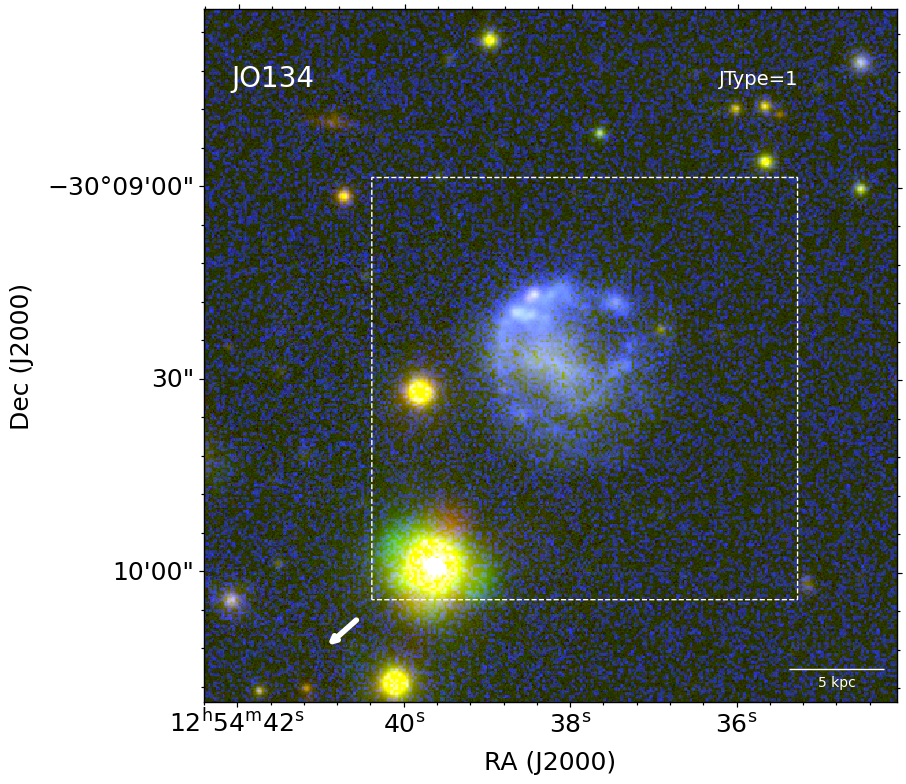}\par 
    \includegraphics[width=6.0cm]{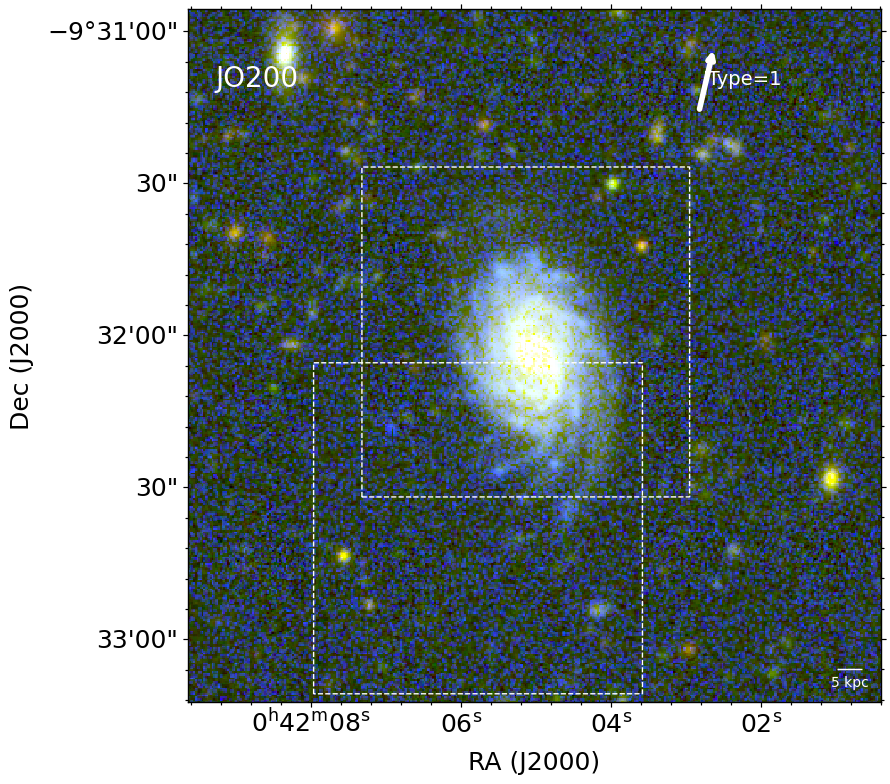}\par 
    \includegraphics[width=6.0cm]{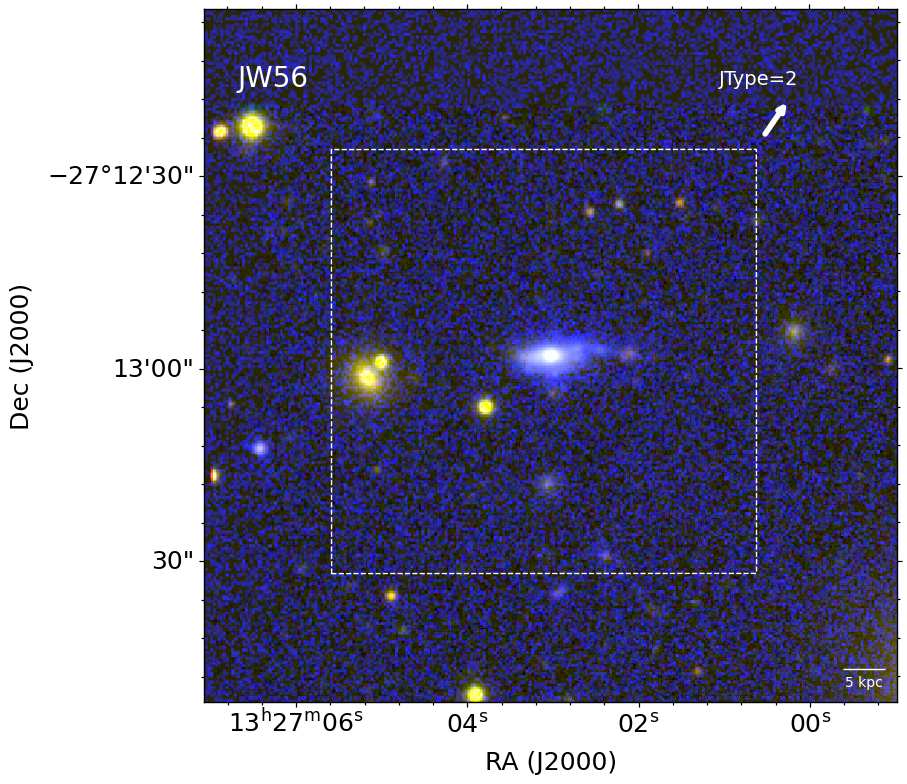}\par
 \end{multicols}
 \begin{multicols}{3}
    \includegraphics[width=6.0cm]{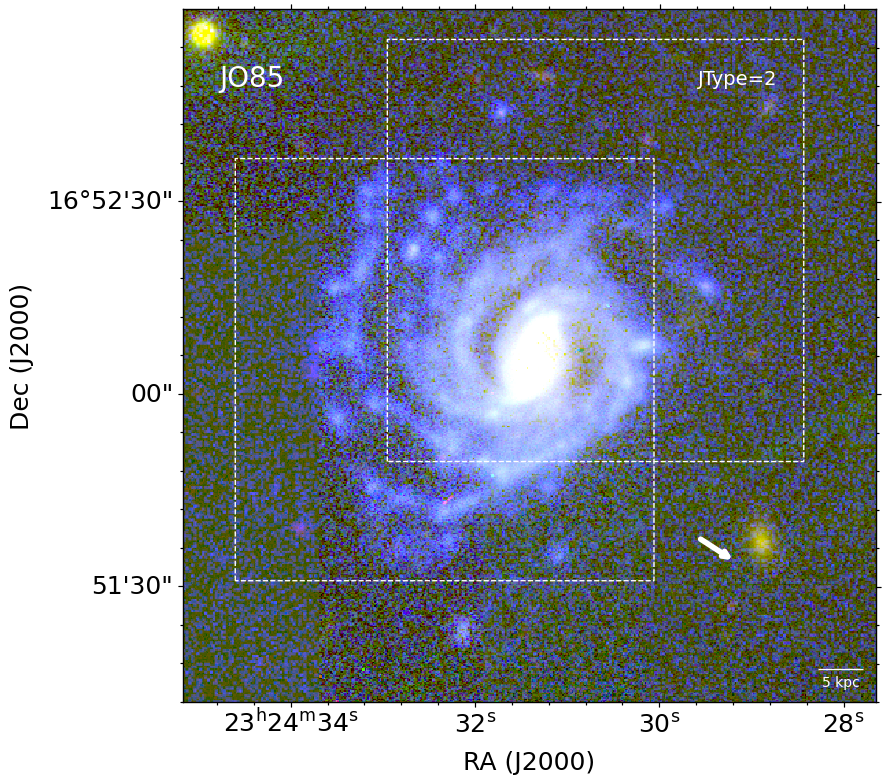}\par 
    \includegraphics[width=6.0cm]{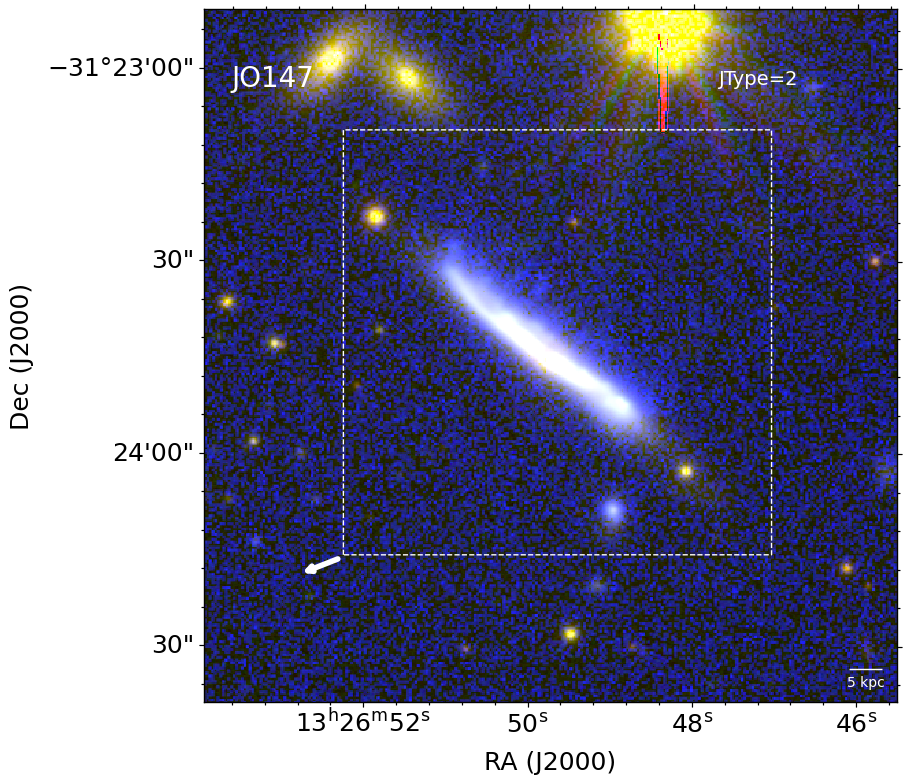}\par 
    \includegraphics[width=6.0cm]{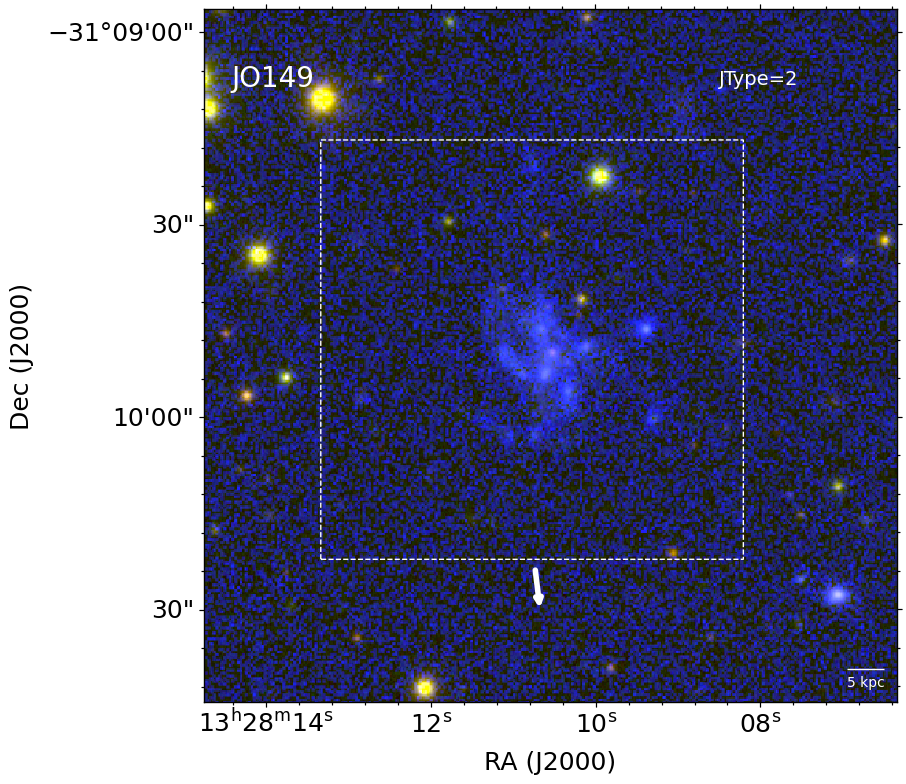}\par
 \end{multicols}
  \begin{multicols}{3}
    \includegraphics[width=6.0cm]{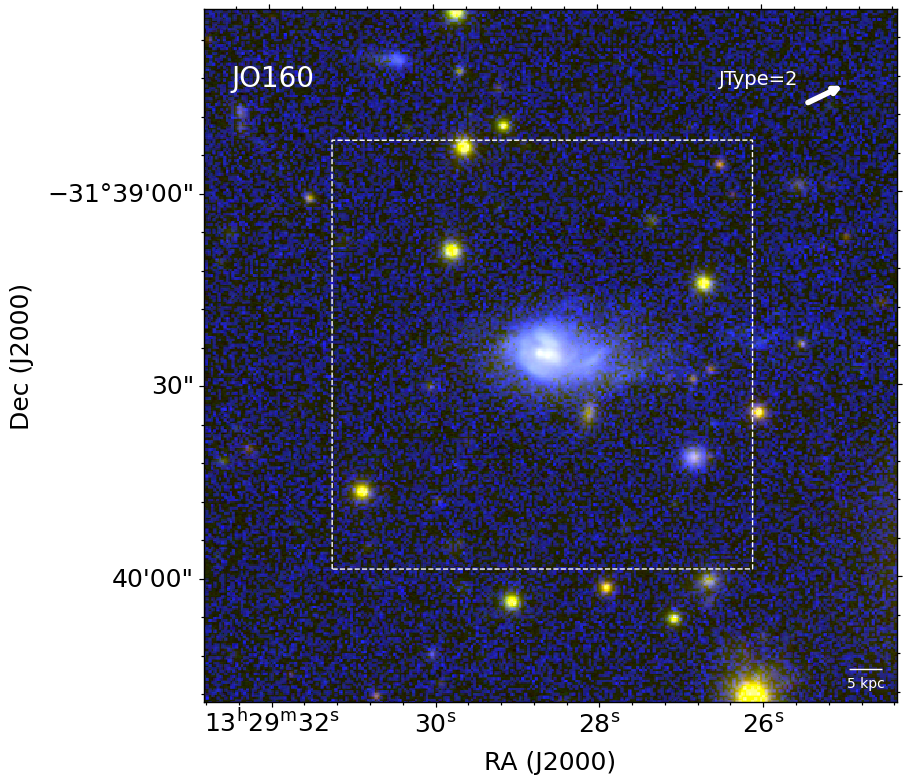}\par 
    \includegraphics[width=6.0cm]{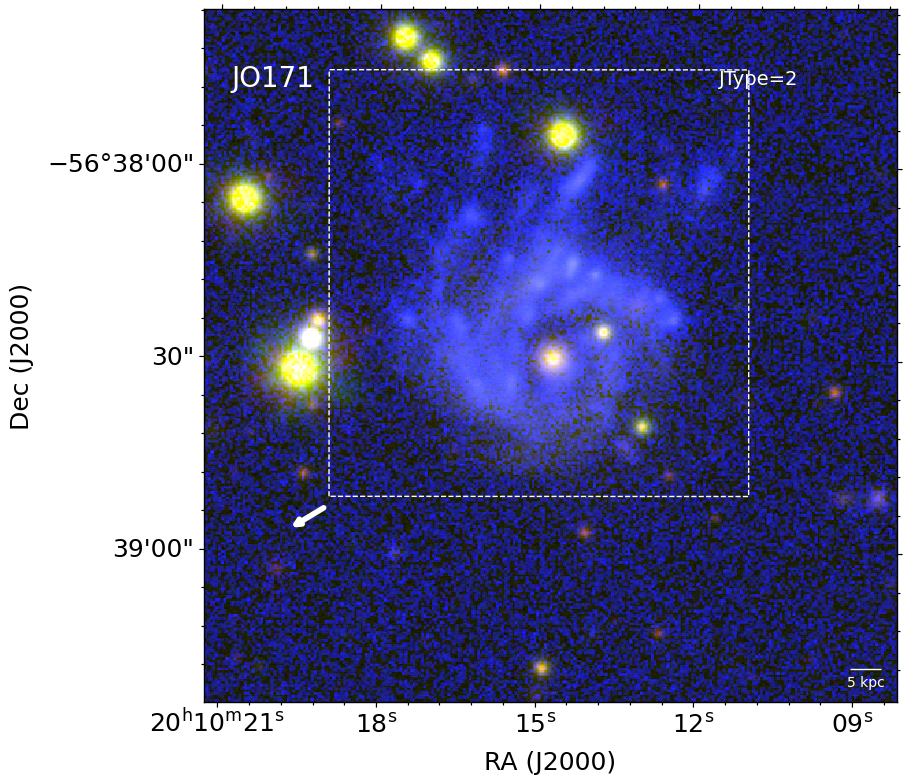}\par 
    \includegraphics[width=6.0cm]{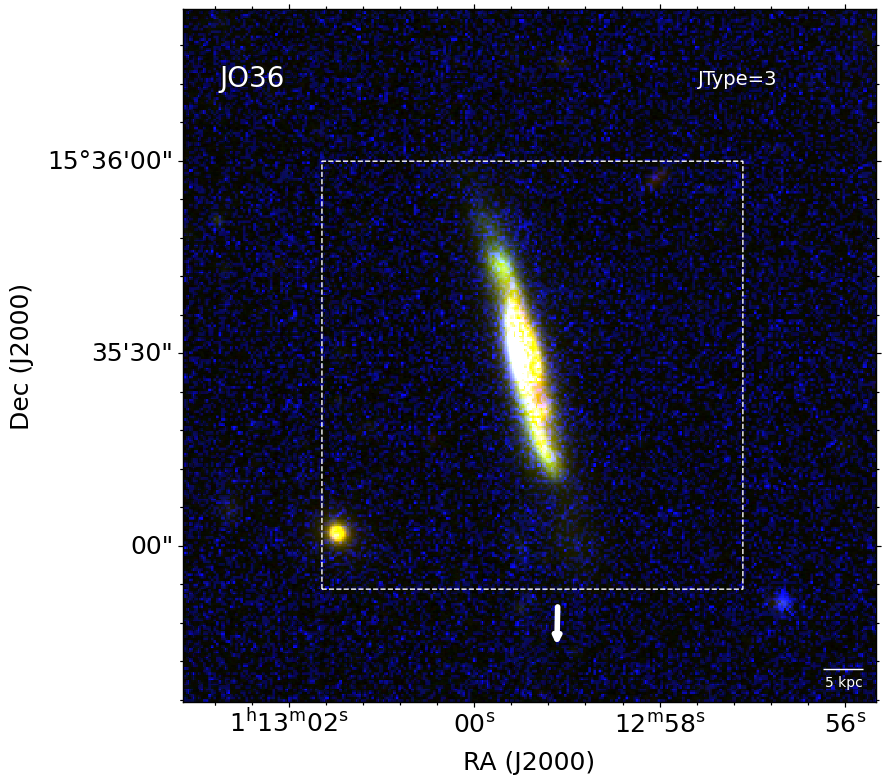}\par
 \end{multicols}
  \begin{multicols}{3}
    \includegraphics[width=6.0cm]{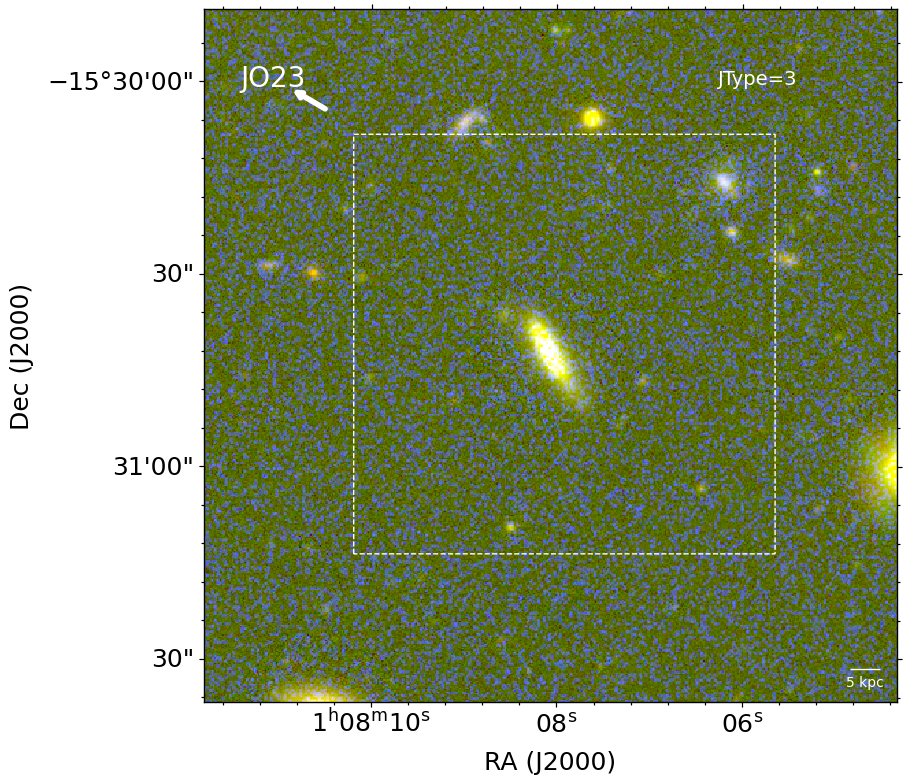}\par 
    \includegraphics[width=6.0cm]{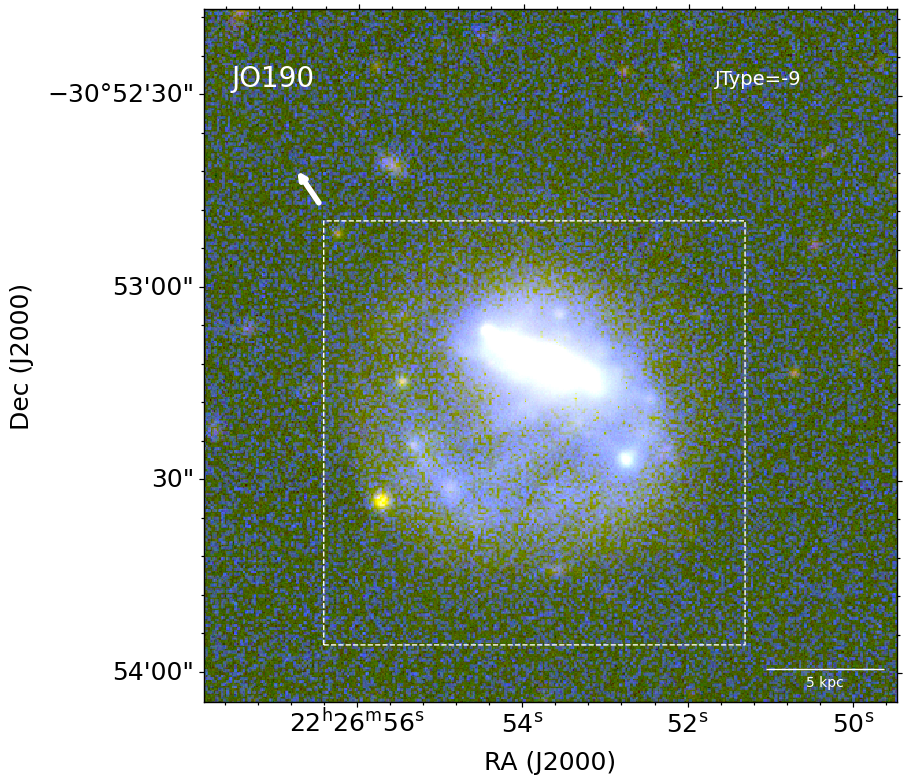}\par 
  \end{multicols}
\caption{Colour composite image of galaxies created by combining the FUV image from UVIT with optical B,V imaging from the WINGS/OmegaWINGS survey with 1 arcmin $\times$ 1 arcmin field of view of VLT/MUSE pointings  marked
with a white-dashed line box. The image is made such that the galaxy is at the centre. The direction towards the BCG is shown with a white arrow. The image measures 1.8 arcmin $\times$ 1.8 arcmin.}\label{fig:rgbA1}
\end{figure*}

\begin{figure*}
\centering
\begin{multicols}{3}
    \includegraphics[width=6.0cm]{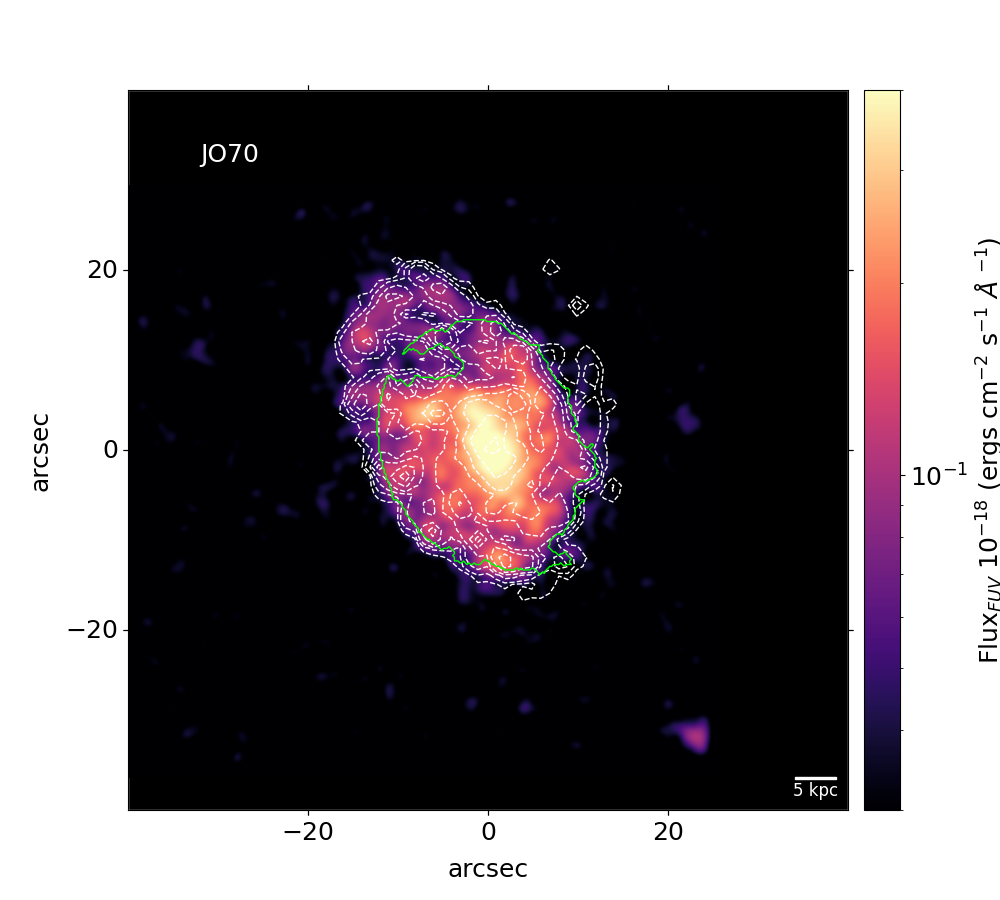}\par 
    \includegraphics[width=6.0cm]{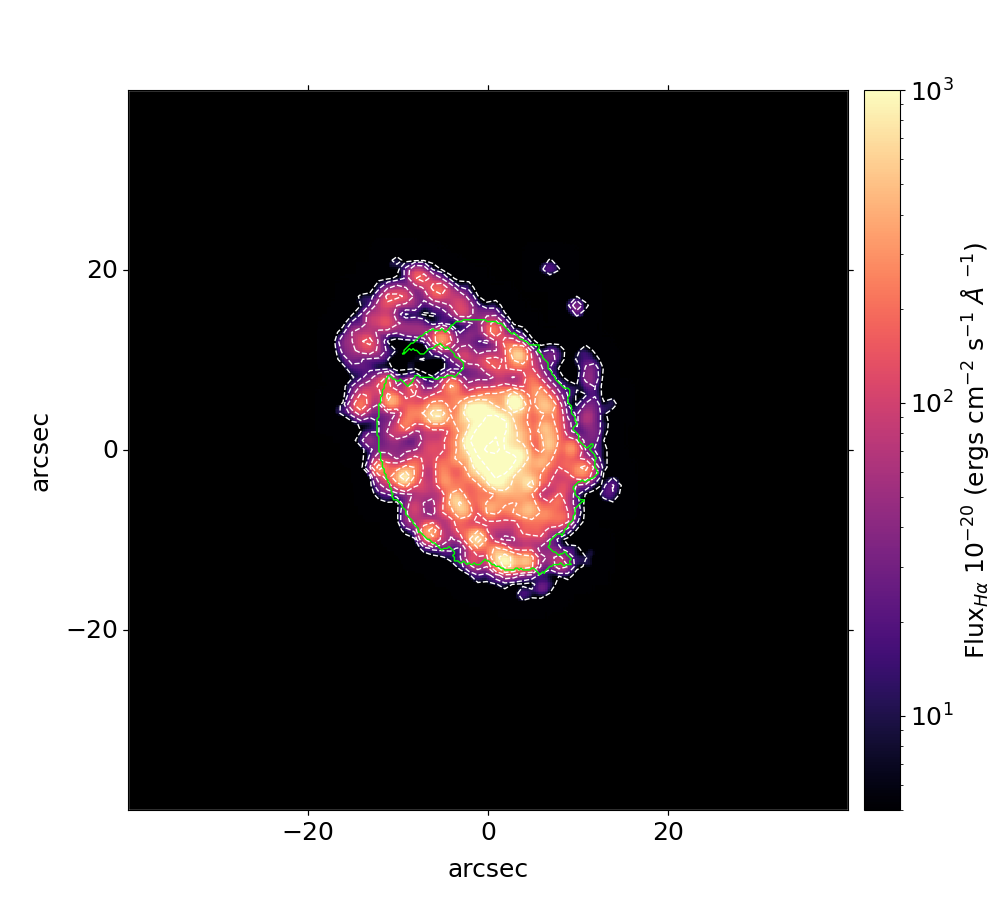}\par 
    \includegraphics[width=6.0cm]{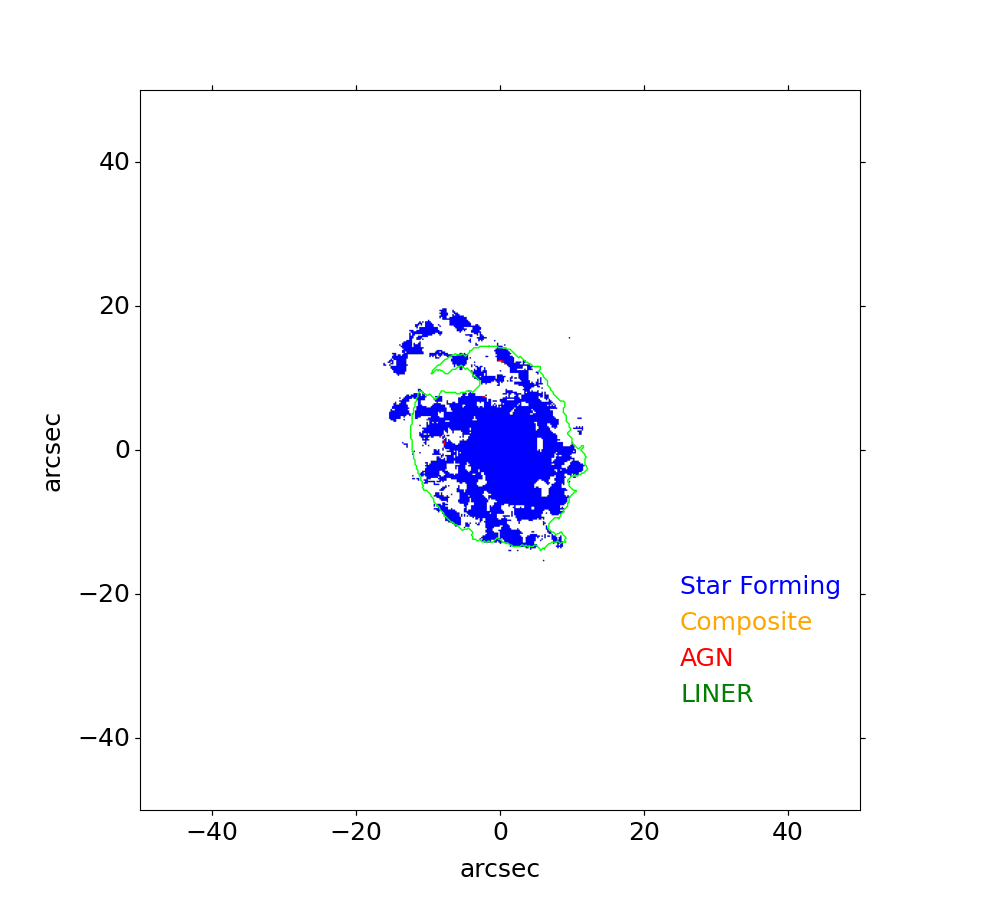}\par
 \end{multicols}
 \begin{multicols}{3}
    \includegraphics[width=6.0cm]{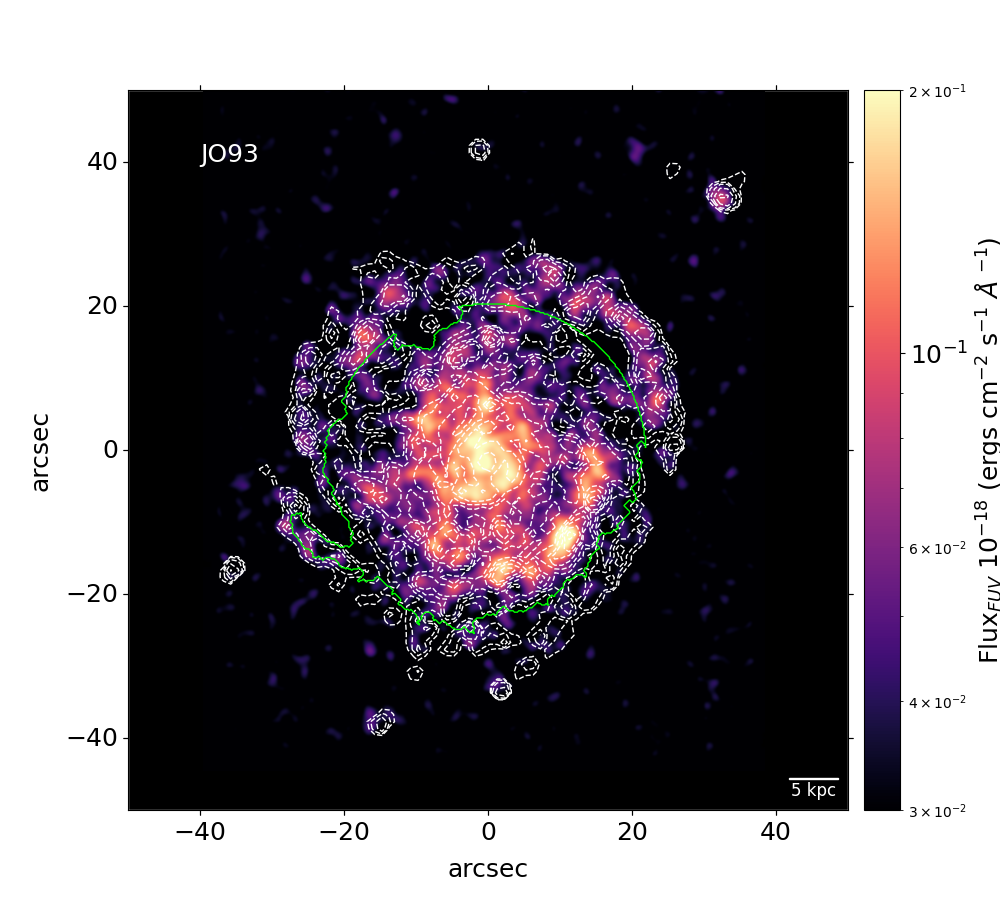}\par 
    \includegraphics[width=6.0cm]{JO93_fuv.png}\par 
    \includegraphics[width=6.0cm]{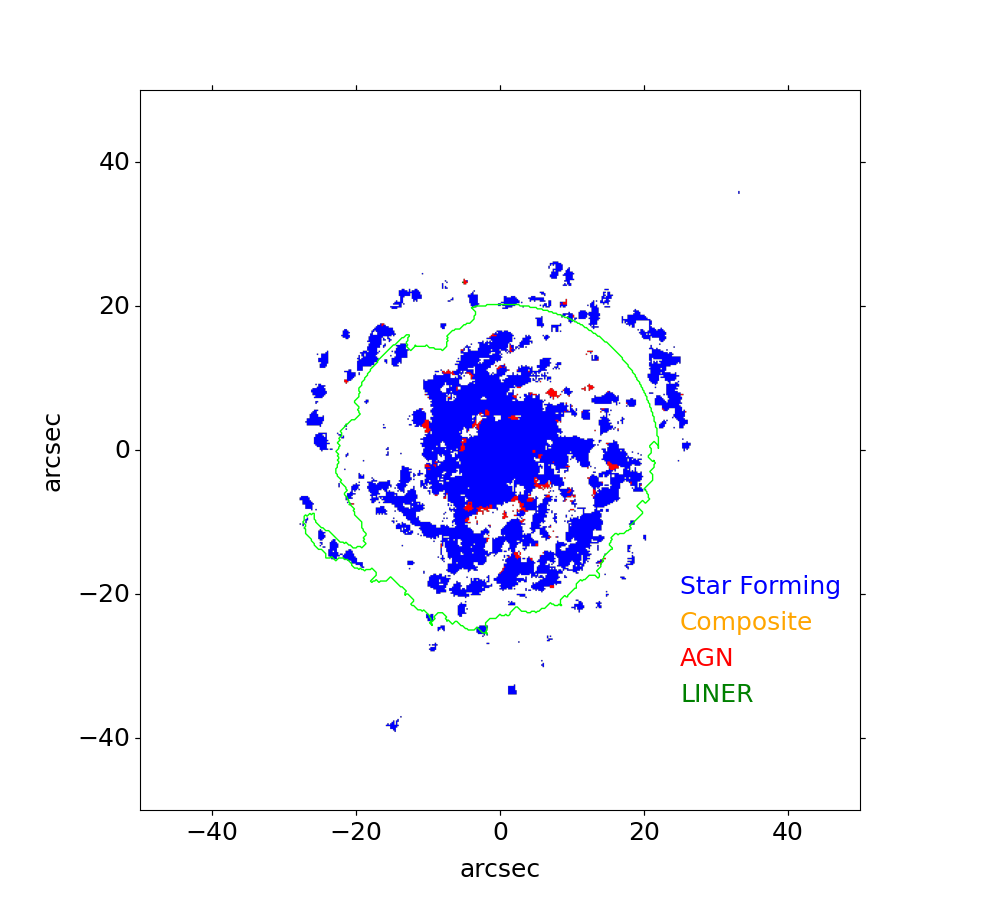}\par
 \end{multicols}
  \begin{multicols}{3}
    \includegraphics[width=6.0cm]{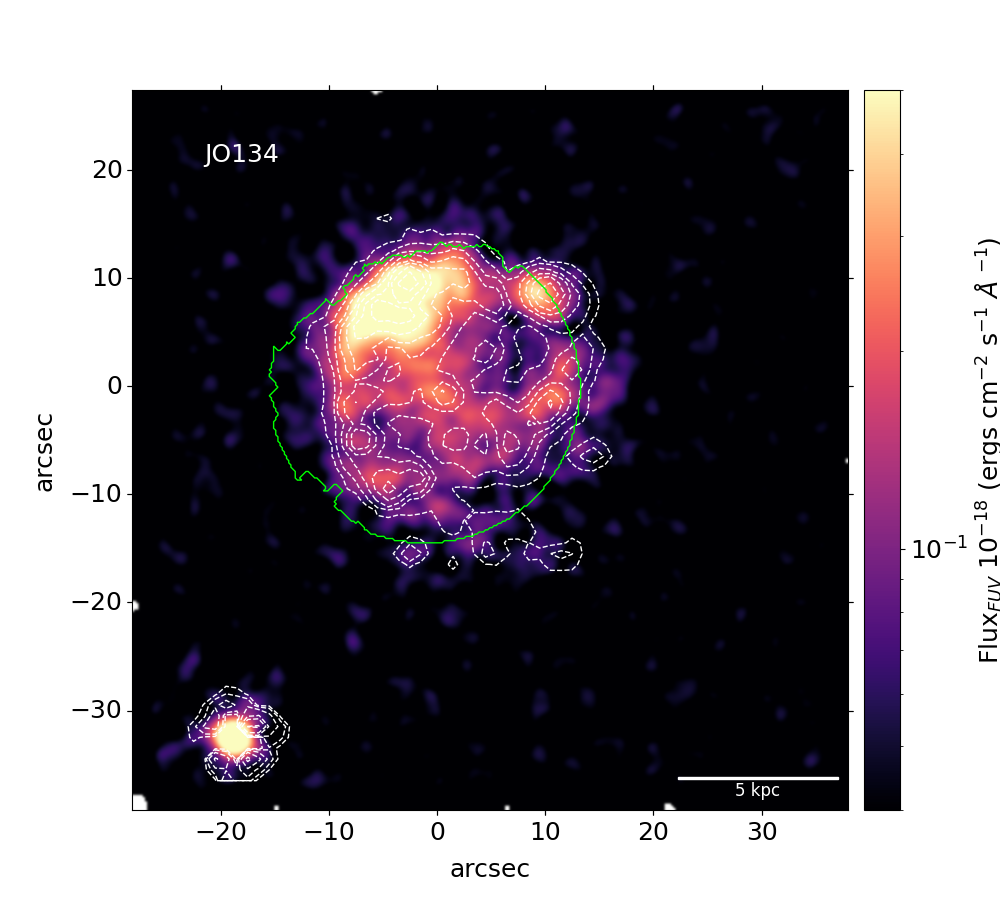}\par 
    \includegraphics[width=6.0cm]{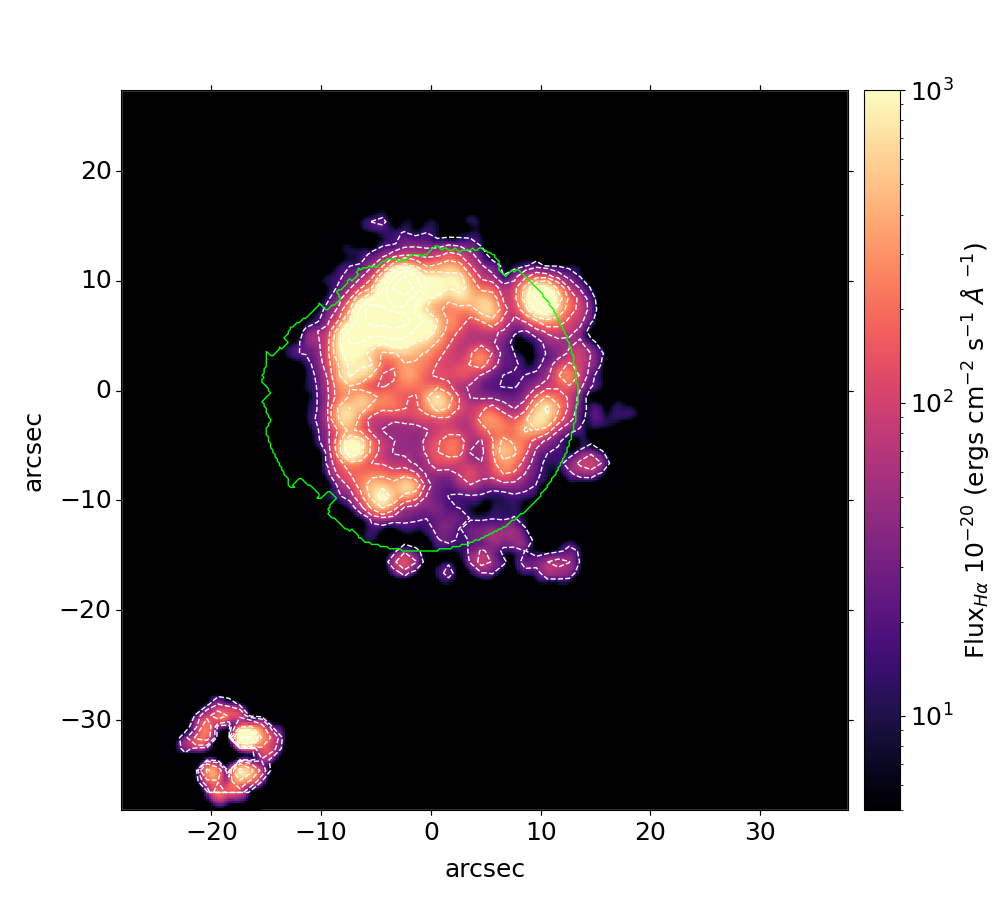}\par 
    \includegraphics[width=6.0cm]{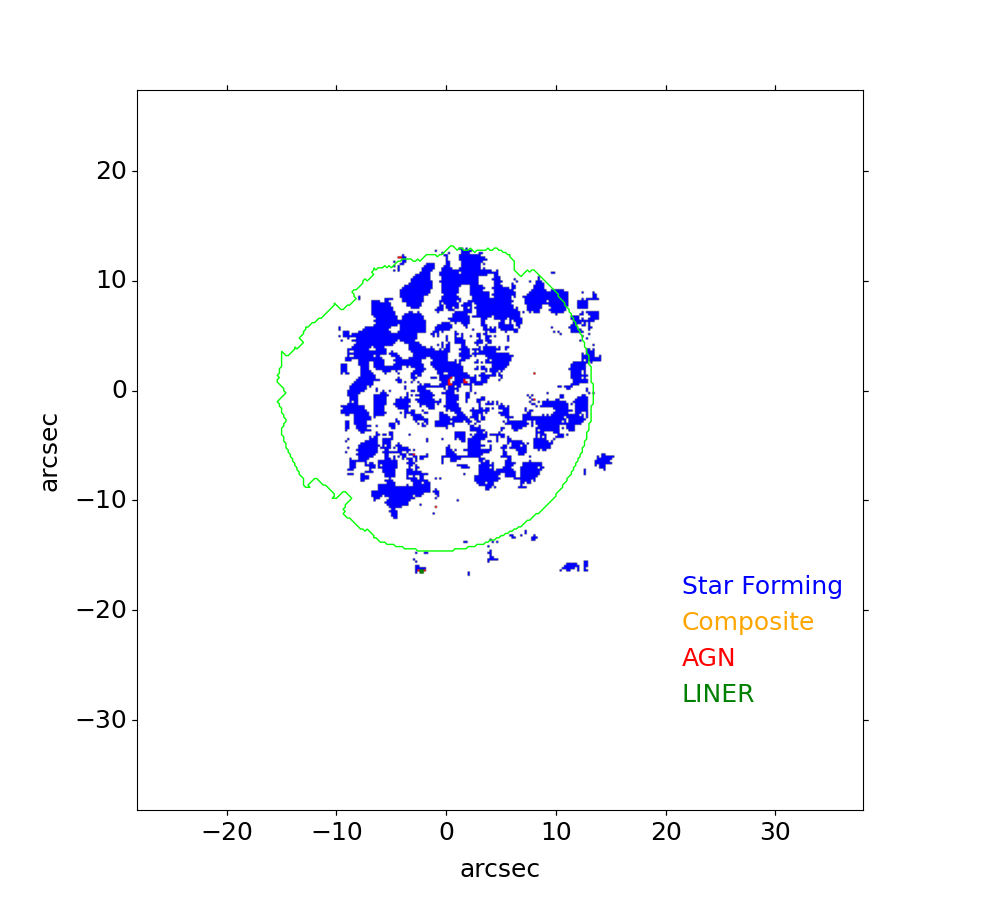}\par
 \end{multicols}
  \begin{multicols}{3}
    \includegraphics[width=6.0cm]{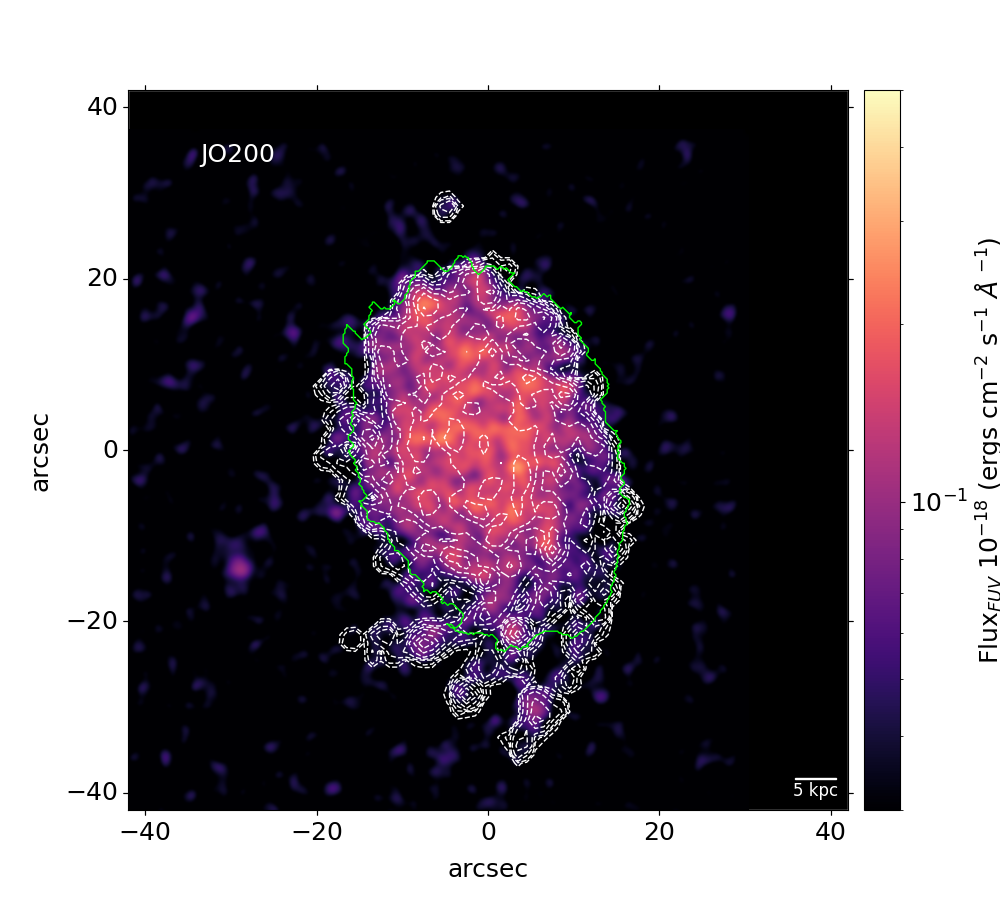}\par 
    \includegraphics[width=6.0cm]{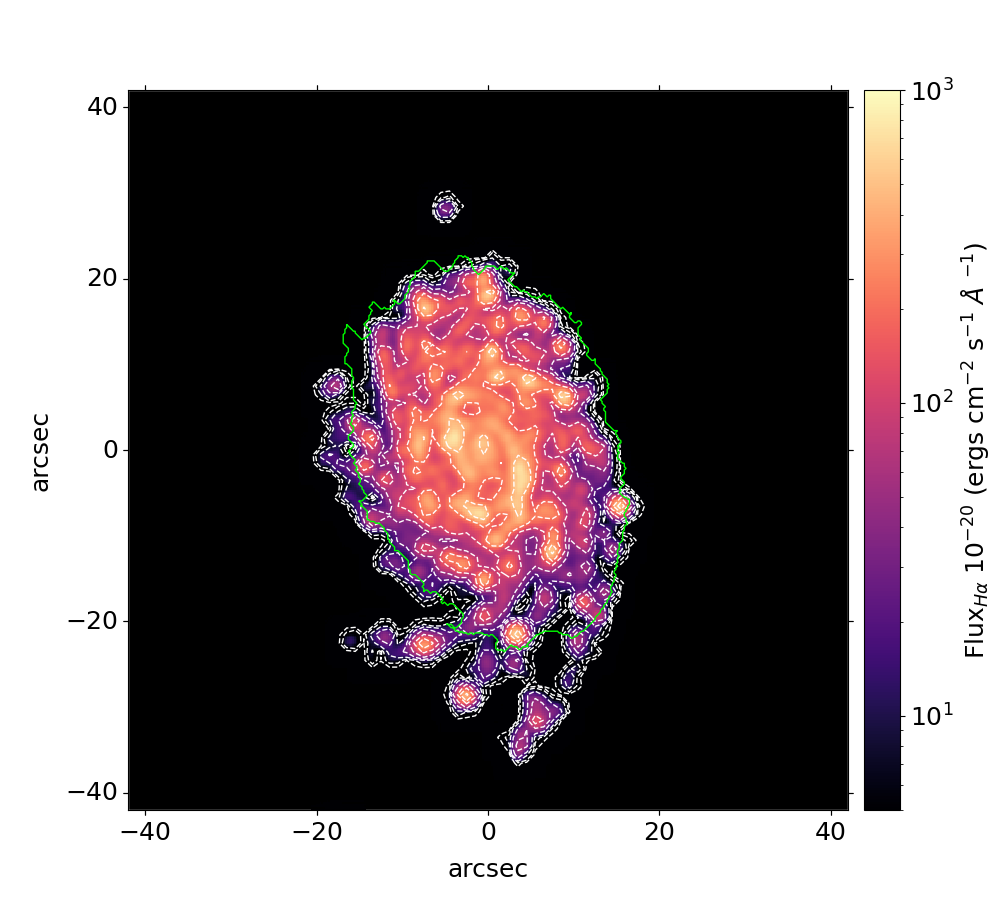}\par 
    \includegraphics[width=6.0cm]{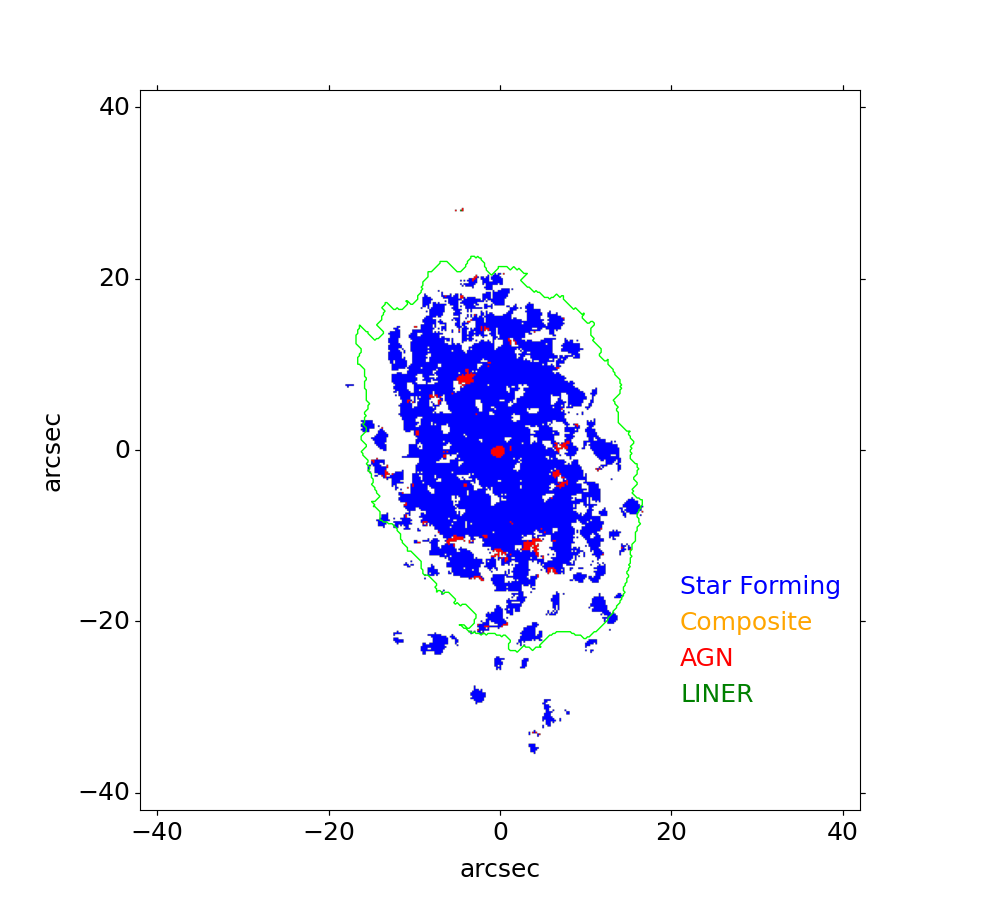}\par
 \end{multicols}
 \end{figure*}
 \begin{figure*}
 \begin{multicols}{3}
    \includegraphics[width=6.0cm]{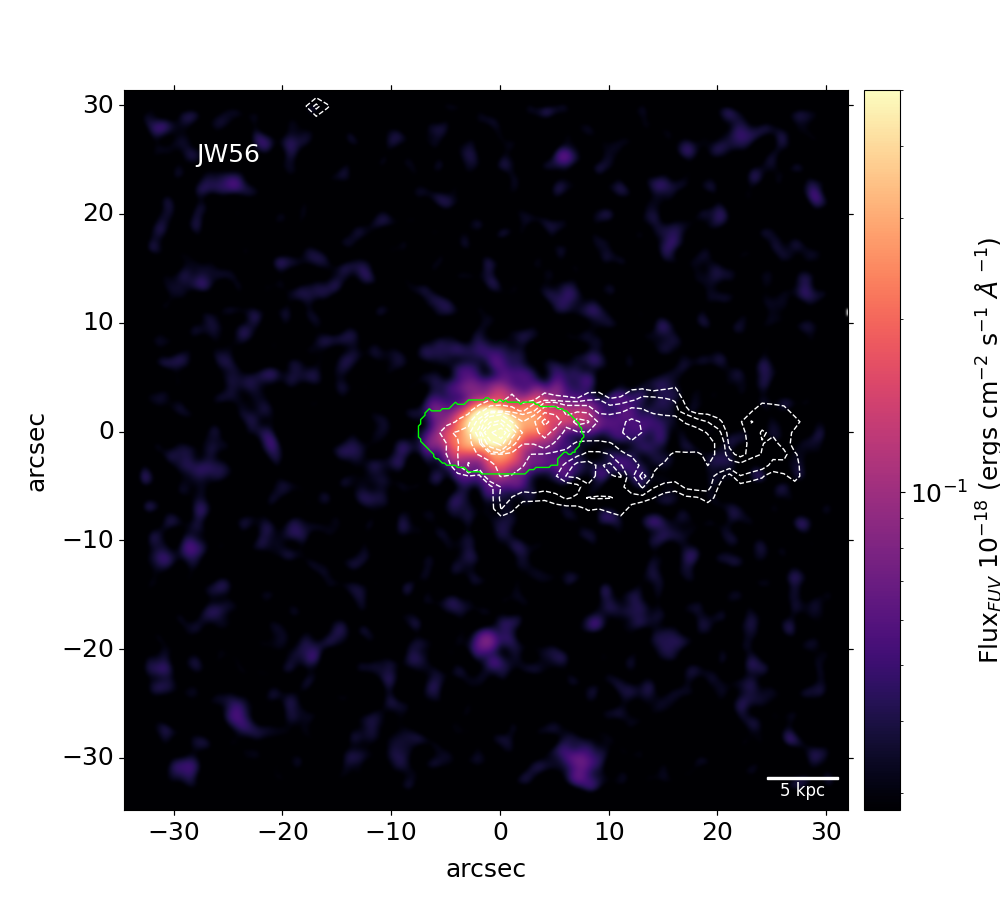}\par 
    \includegraphics[width=6.0cm]{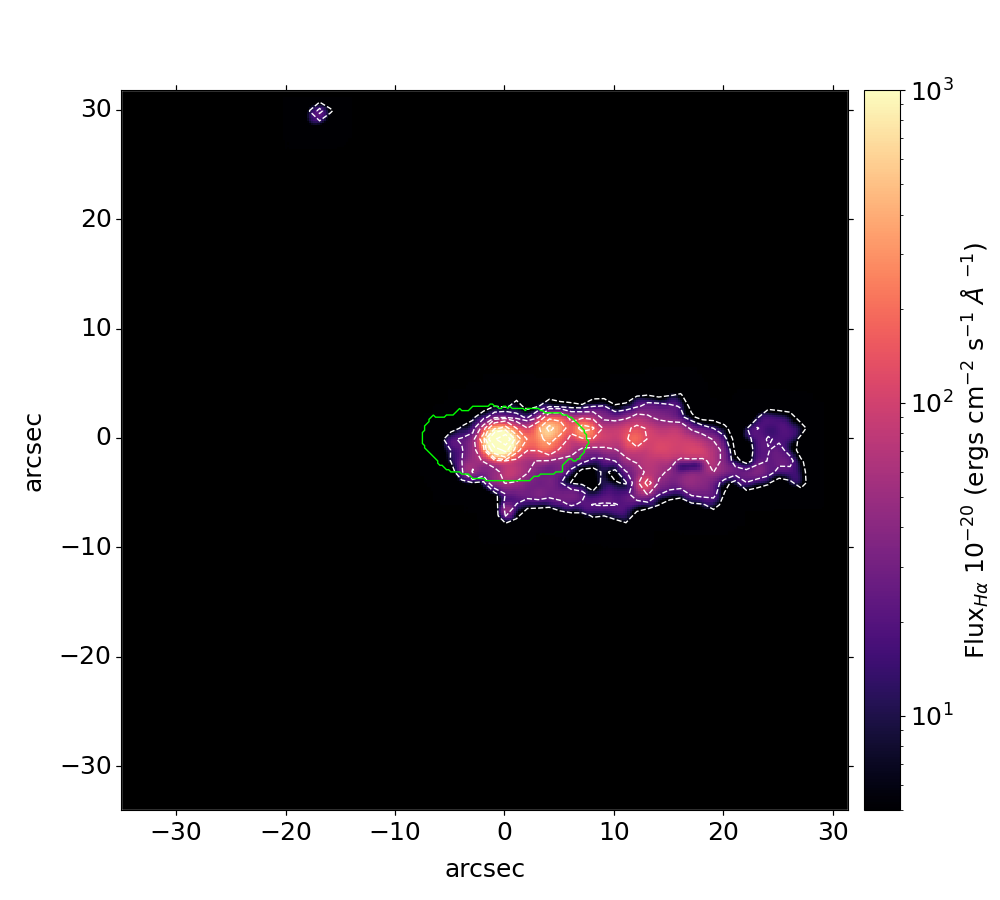}\par 
    \includegraphics[width=6.0cm]{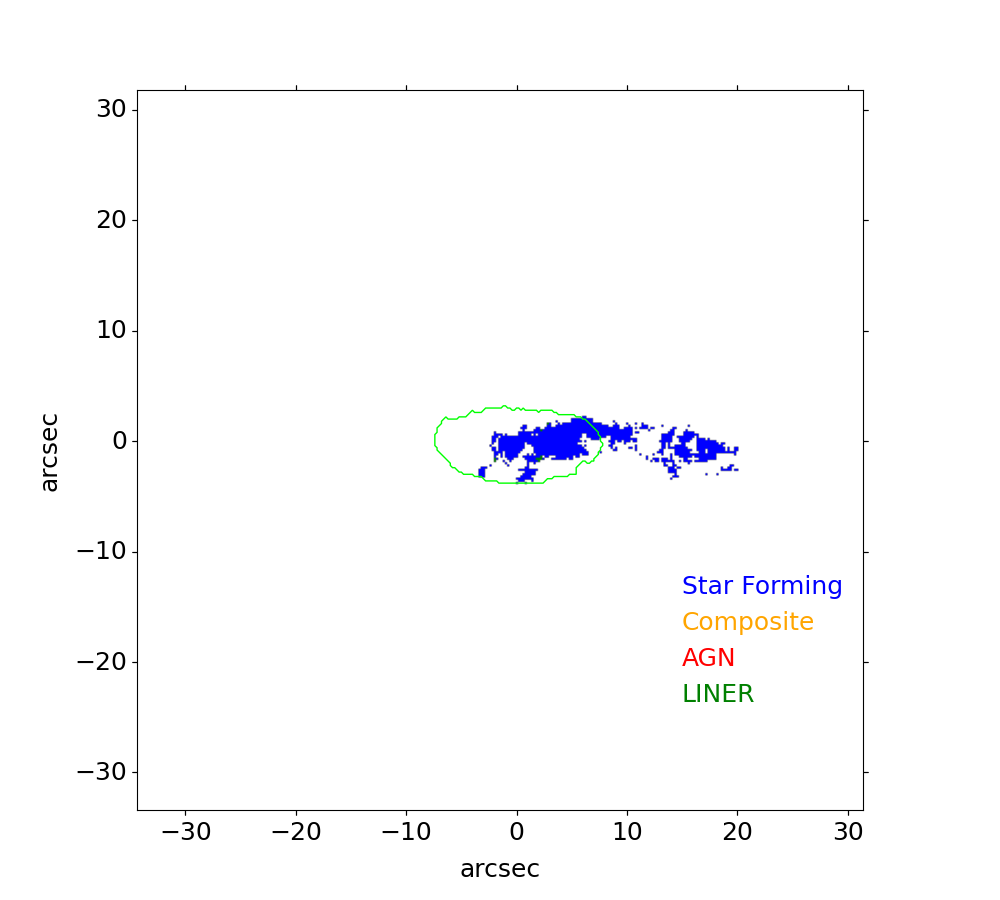}\par
 \end{multicols}
 \begin{multicols}{3}
    \includegraphics[width=6.0cm]{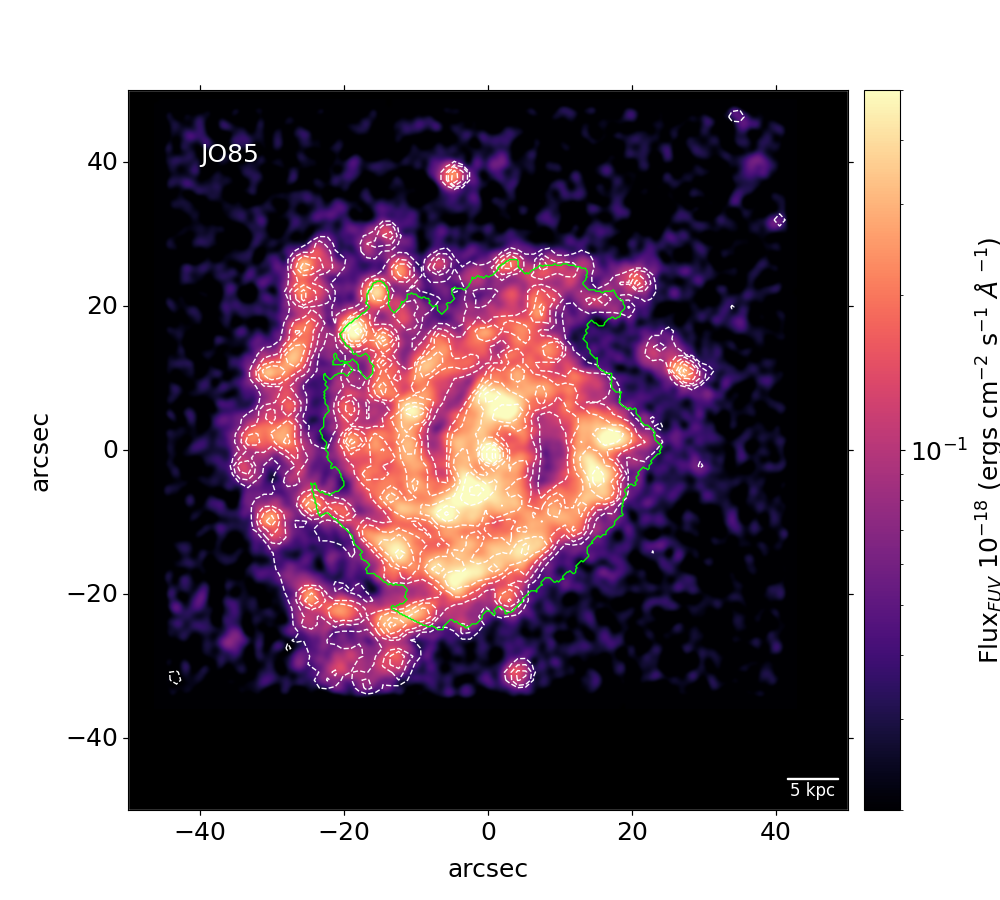}\par 
    \includegraphics[width=6.0cm]{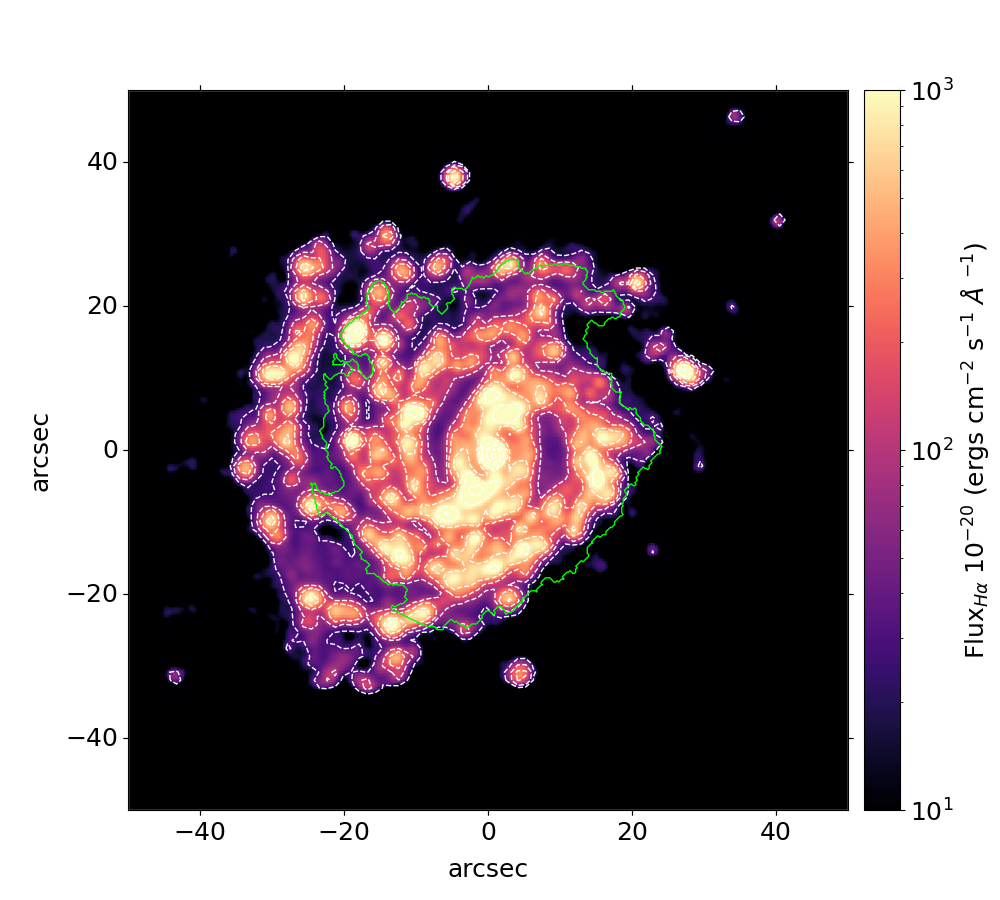}\par 
    \includegraphics[width=6.0cm]{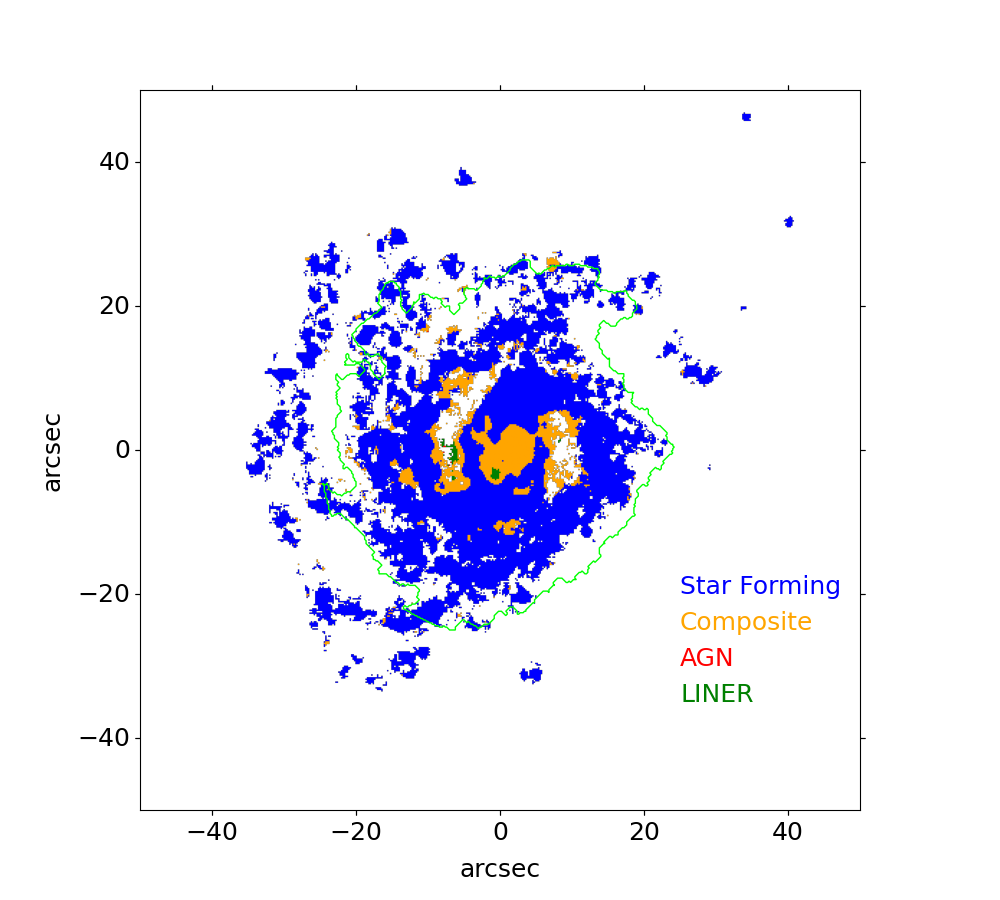}\par
 \end{multicols}
  \begin{multicols}{3}
     \includegraphics[width=6.0cm]{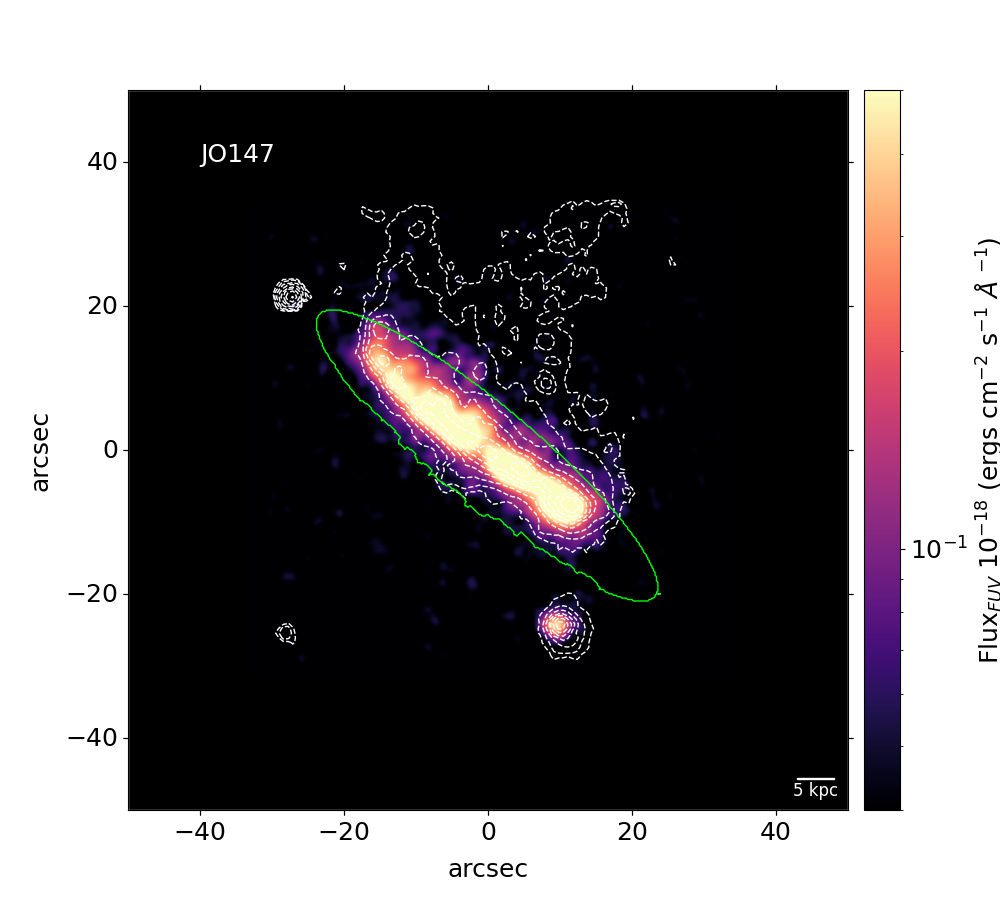}\par 
    \includegraphics[width=6.0cm]{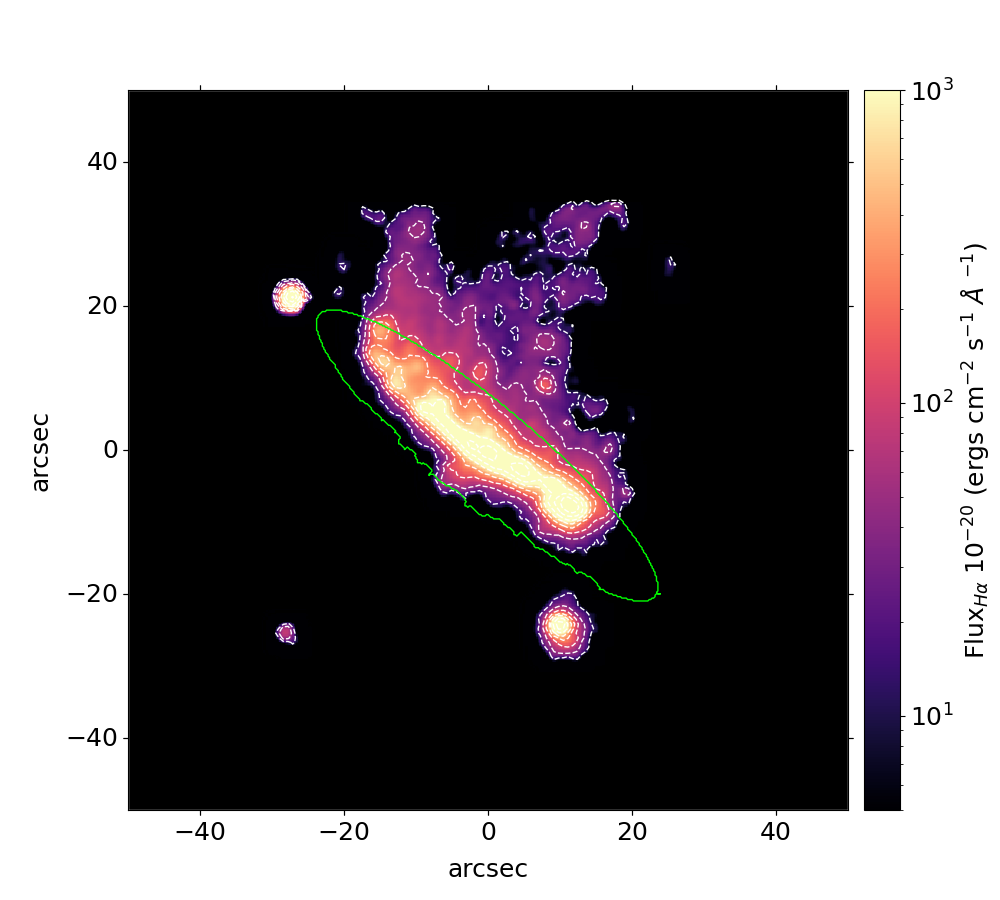}\par 
    \includegraphics[width=6.0cm]{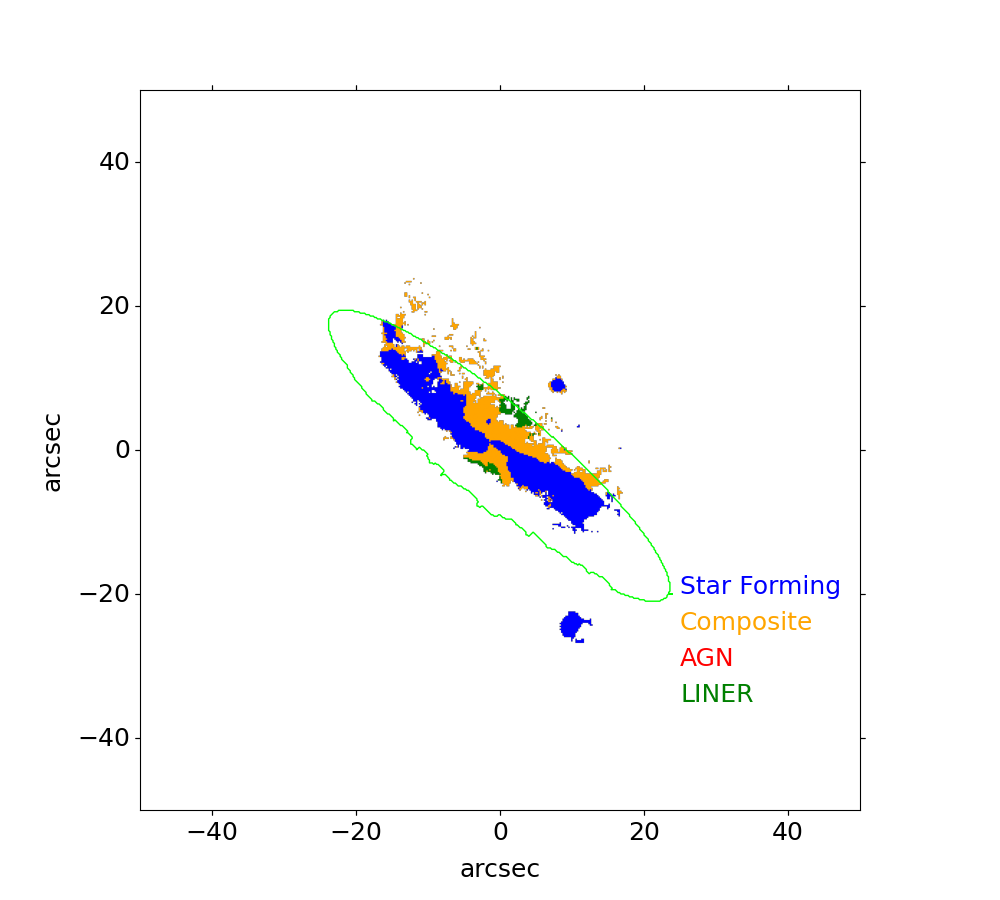}\par
 \end{multicols}
 \begin{multicols}{3}
    \includegraphics[width=6.0cm]{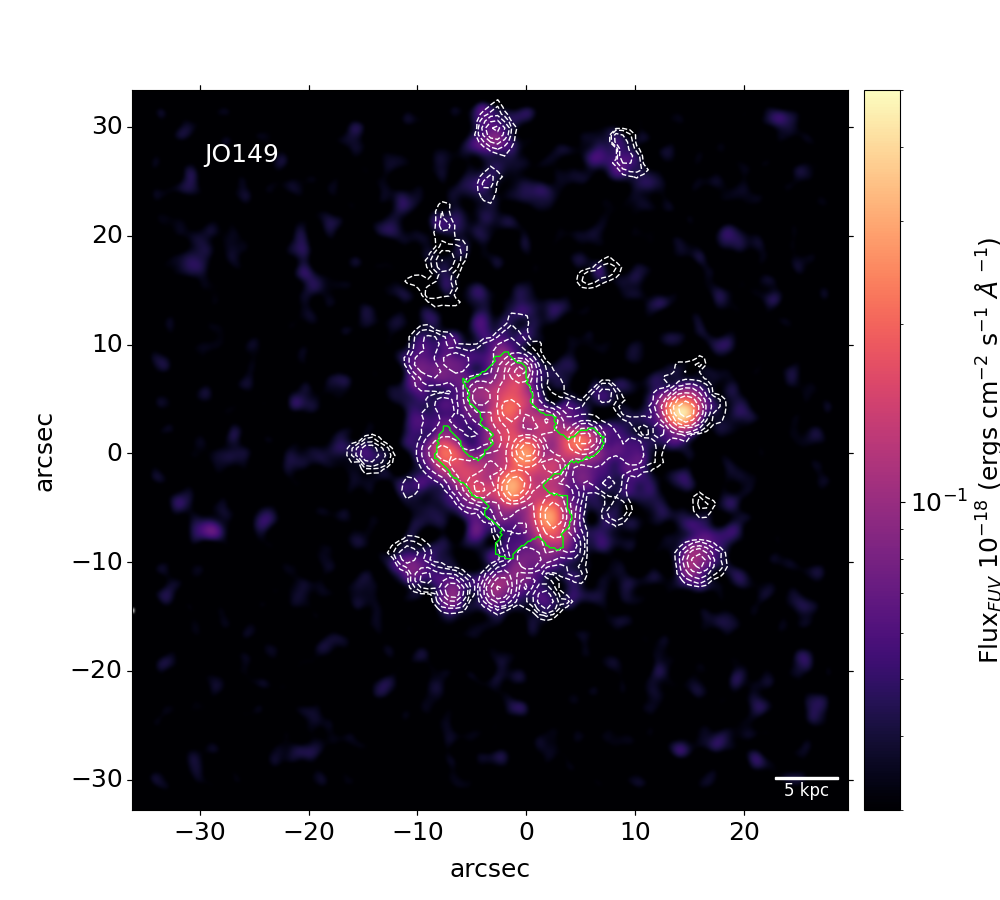}\par 
    \includegraphics[width=6.0cm]{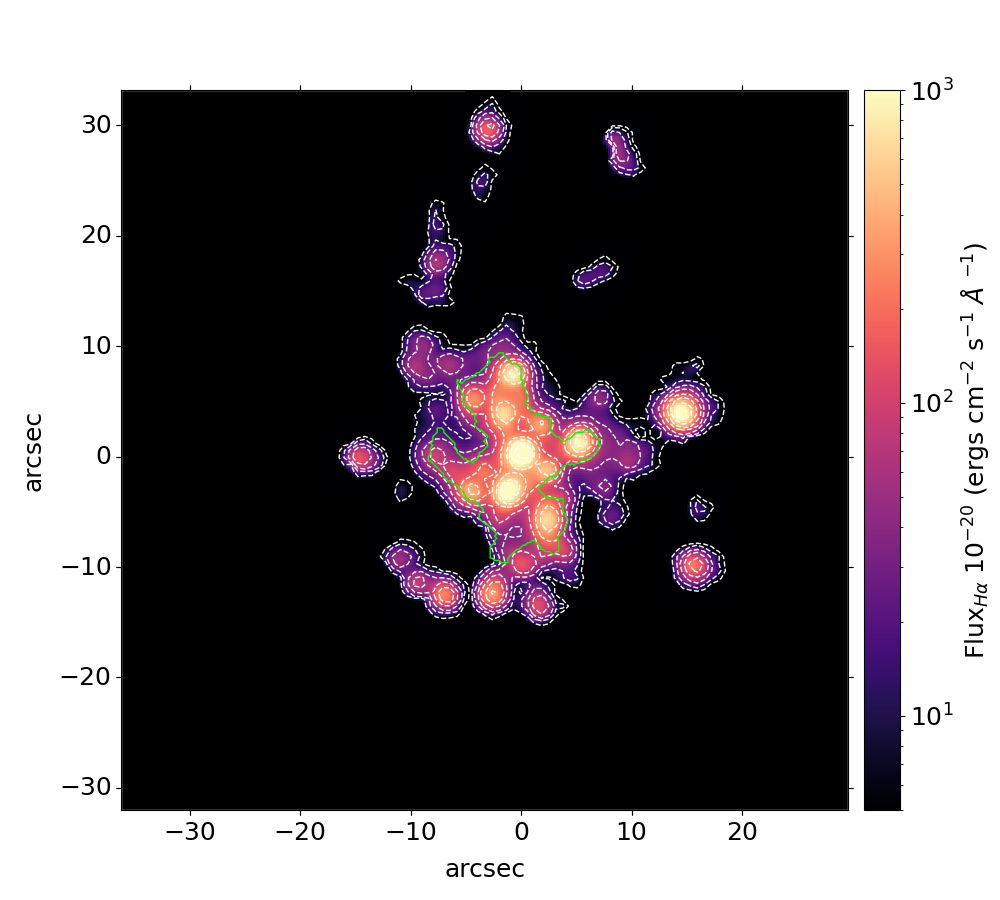}\par 
    \includegraphics[width=6.0cm]{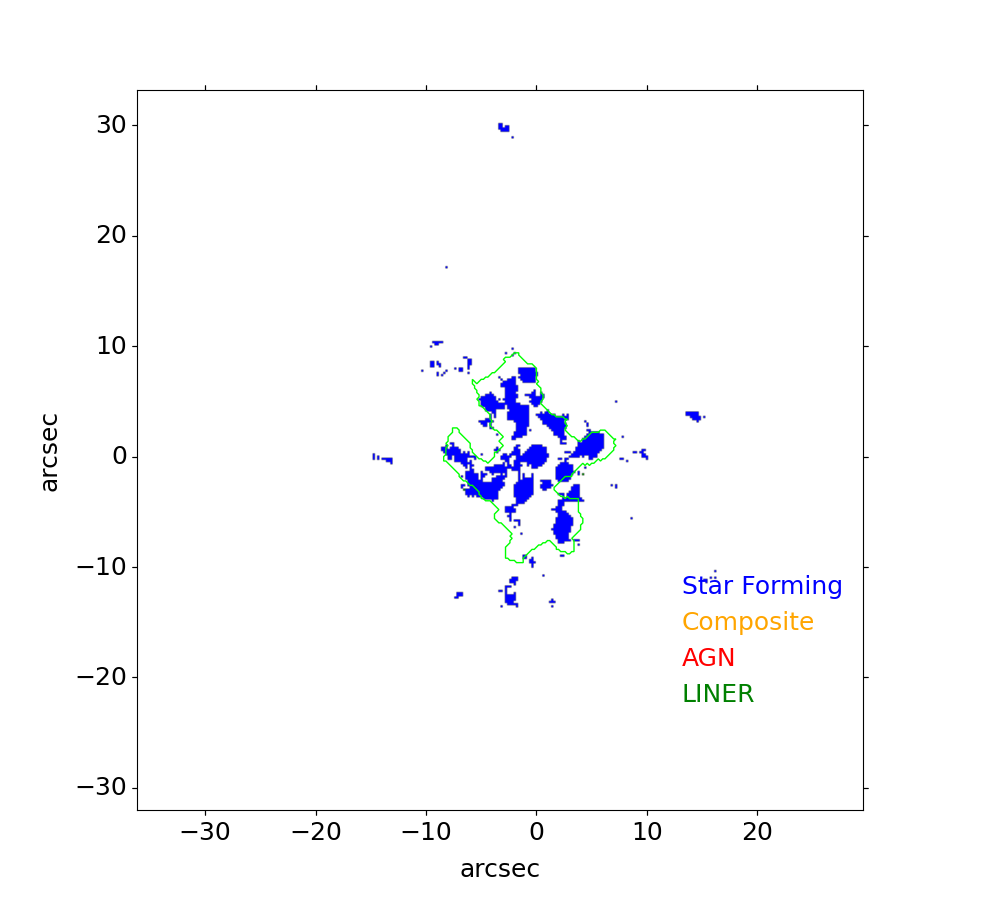}\par
 \end{multicols}
 \end{figure*}
 \begin{figure*}
 \begin{multicols}{3}
    \includegraphics[width=6.0cm]{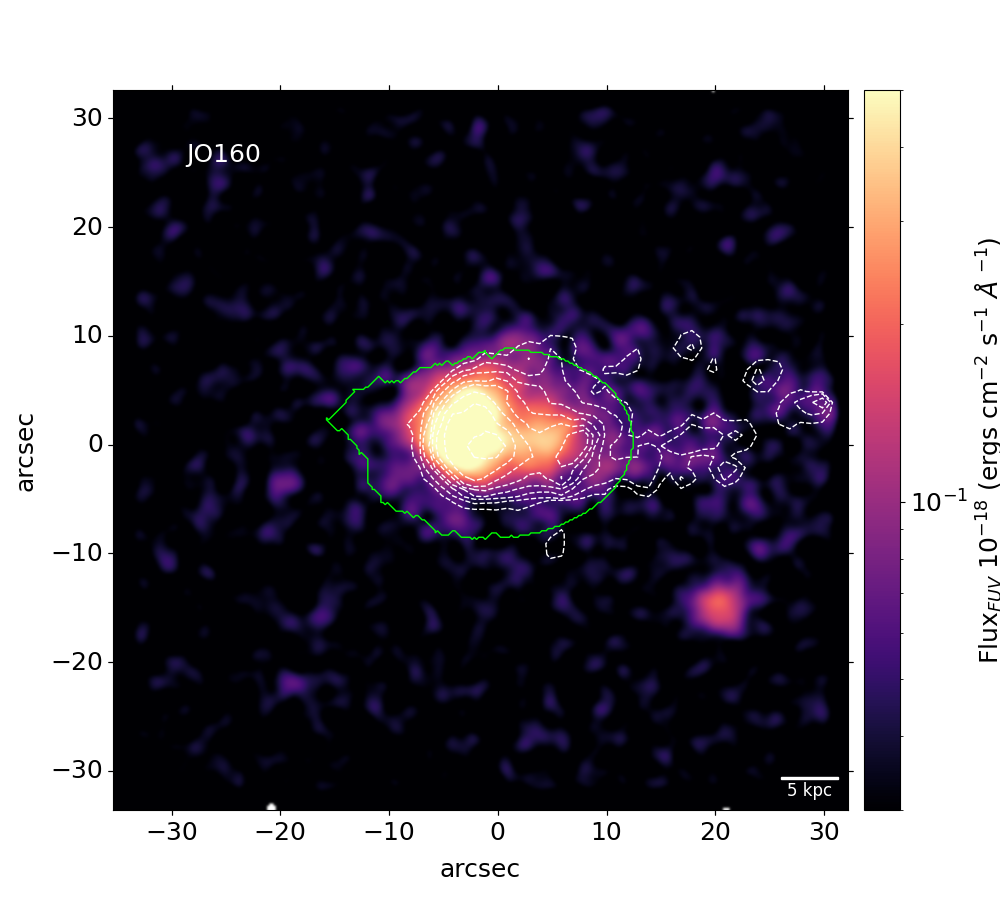}\par 
    \includegraphics[width=6.0cm]{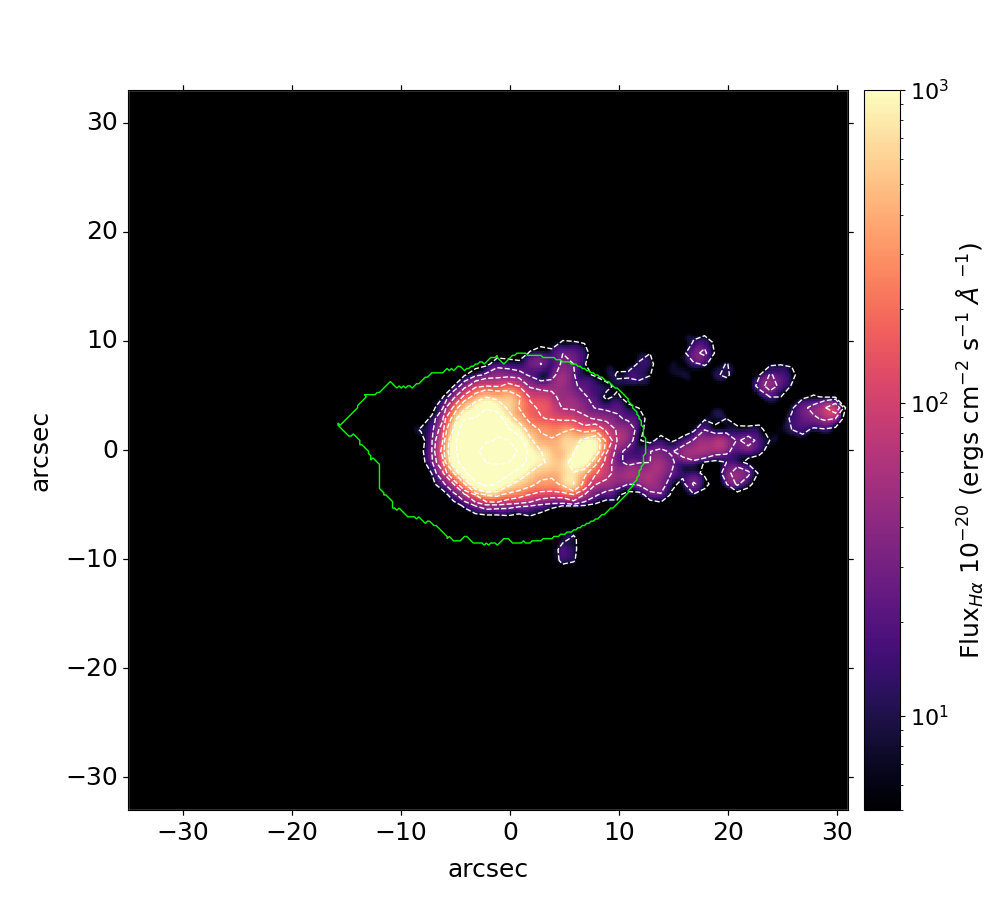}\par 
    \includegraphics[width=6.0cm]{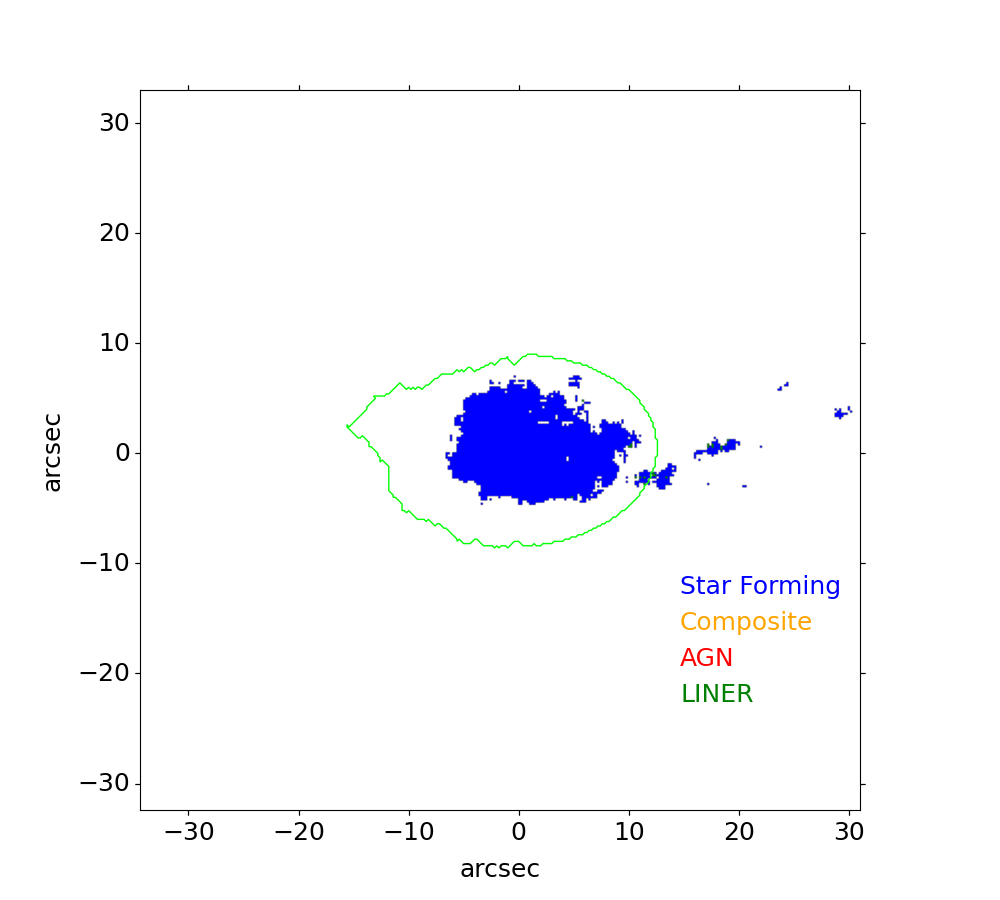}\par
 \end{multicols}
 \begin{multicols}{3}
    \includegraphics[width=6.0cm]{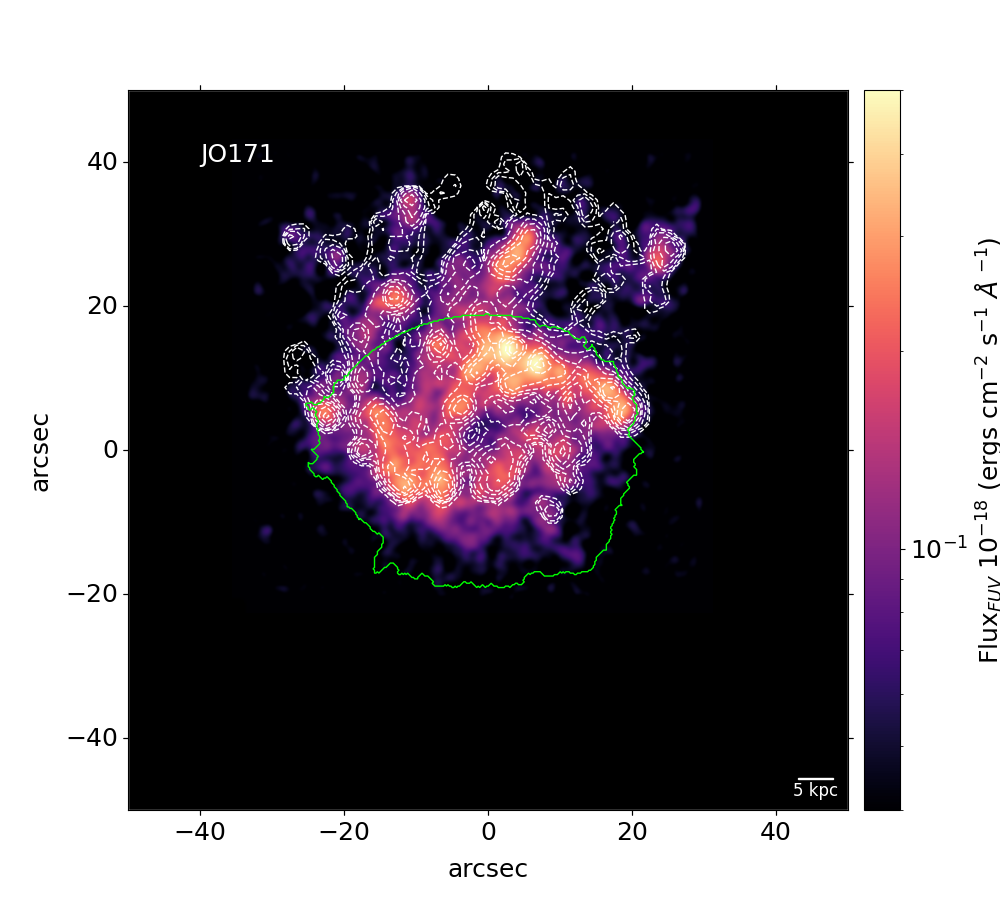}\par 
    \includegraphics[width=6.0cm]{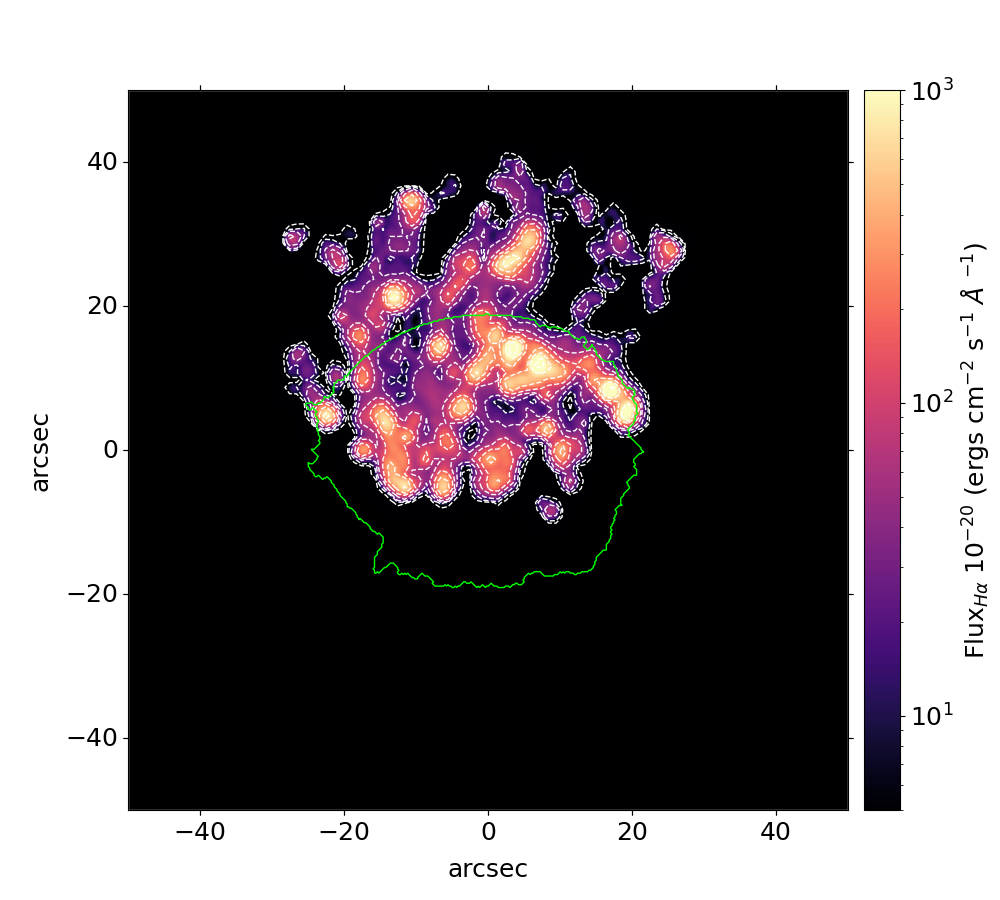}\par 
    \includegraphics[width=6.0cm]{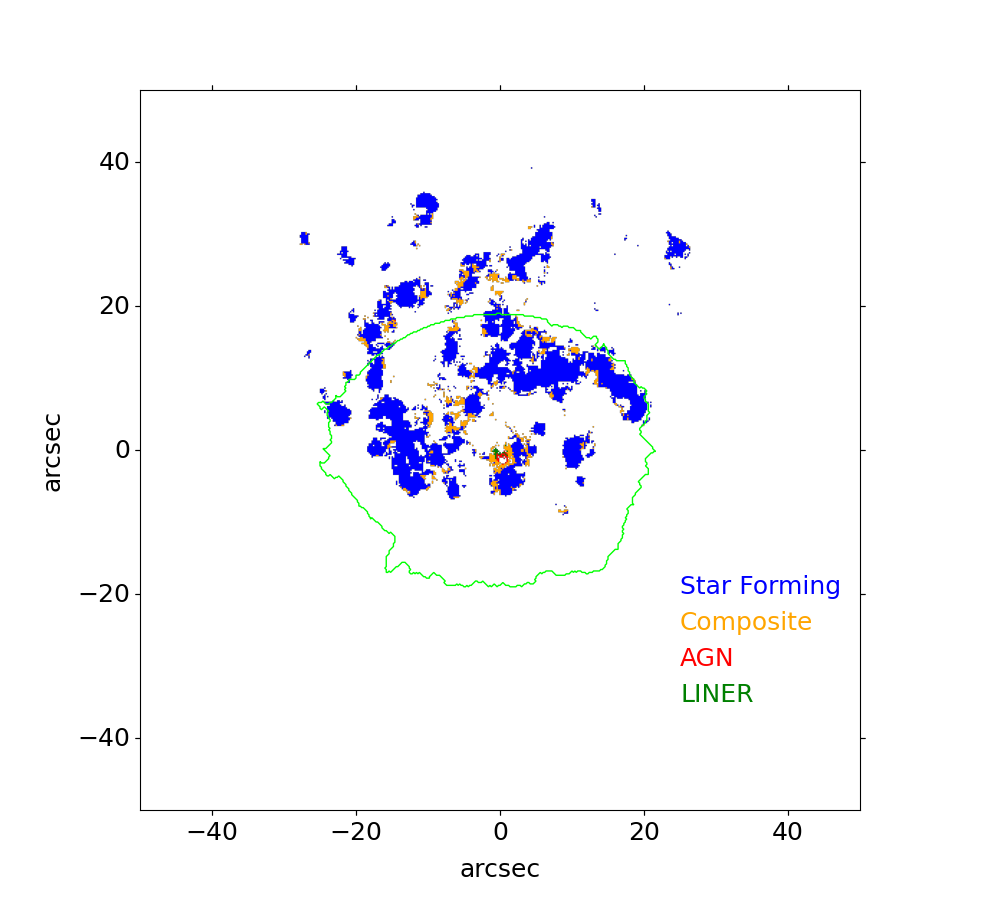}\par
 \end{multicols}
  \begin{multicols}{3}
    \includegraphics[width=6.0cm]{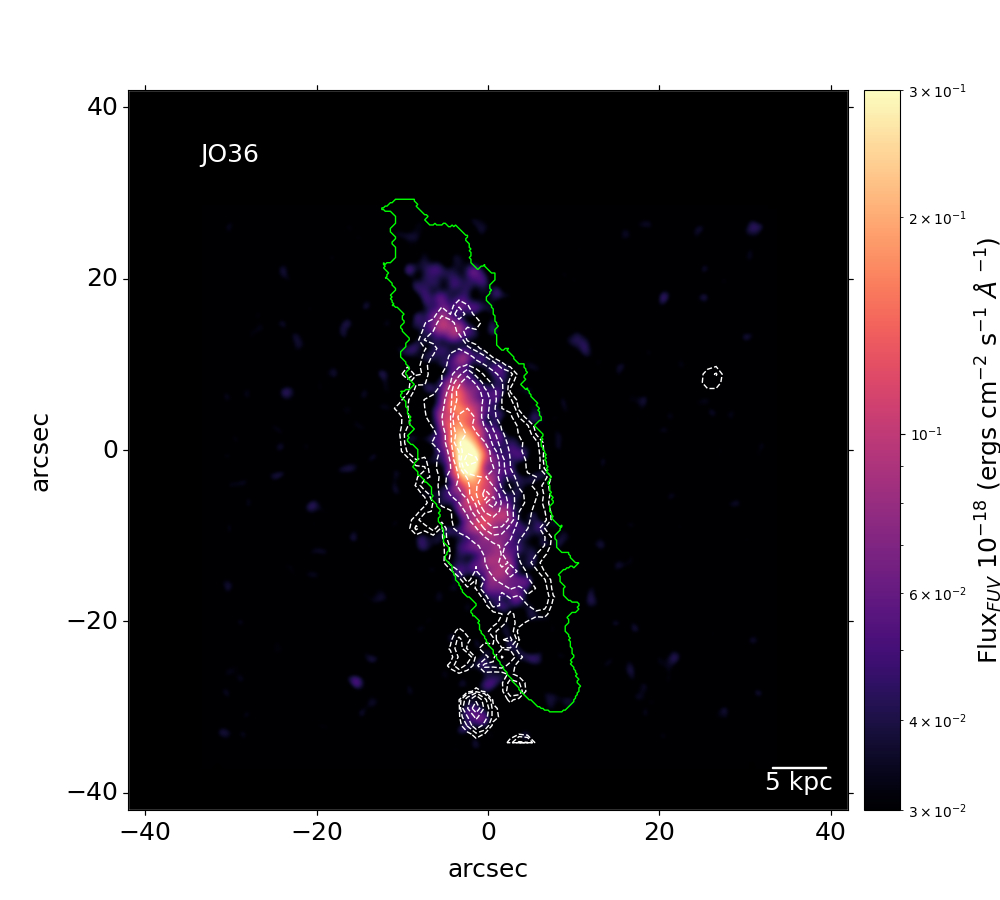}\par 
    \includegraphics[width=6.0cm]{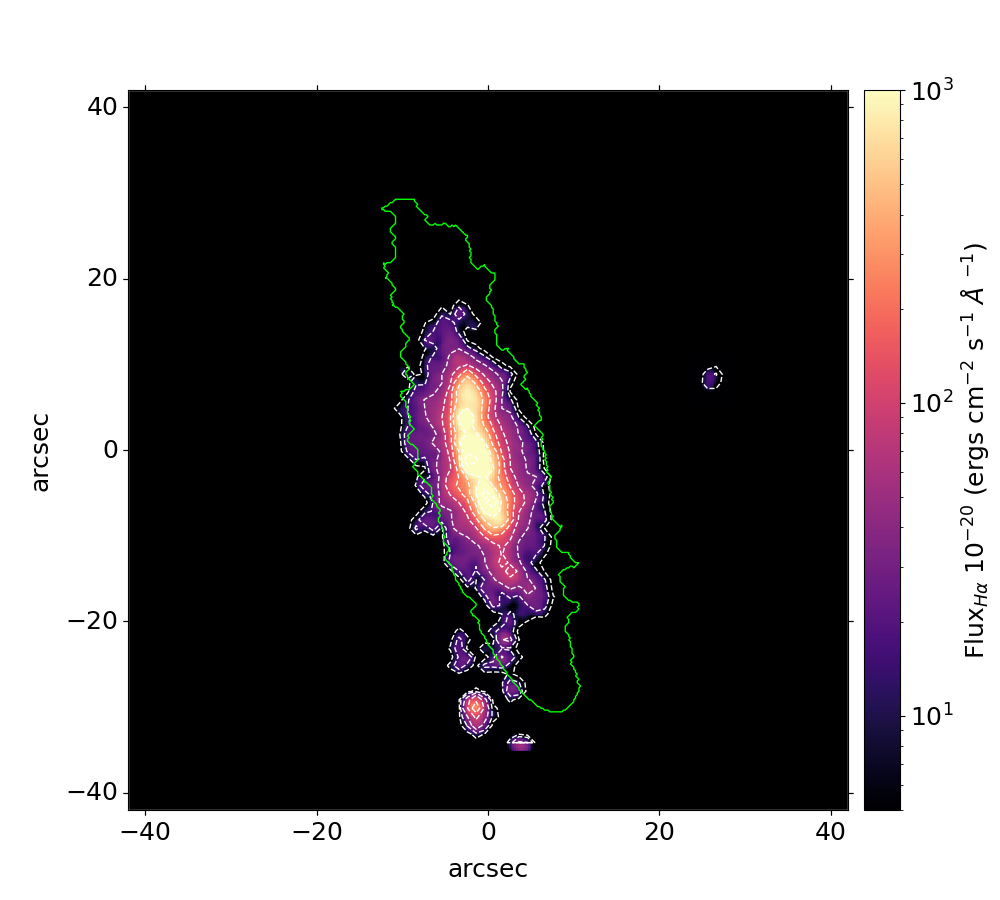}\par 
    \includegraphics[width=6.0cm]{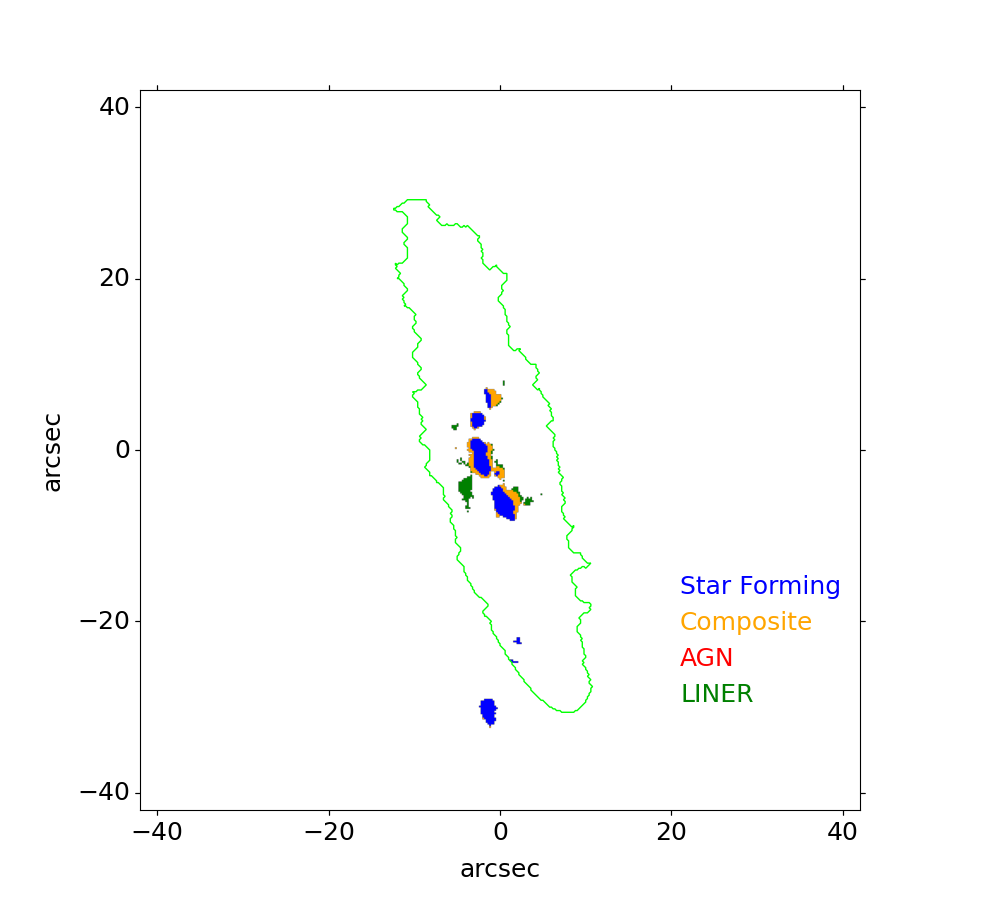}\par
 \end{multicols}
   \begin{multicols}{3}
    \includegraphics[width=6.0cm]{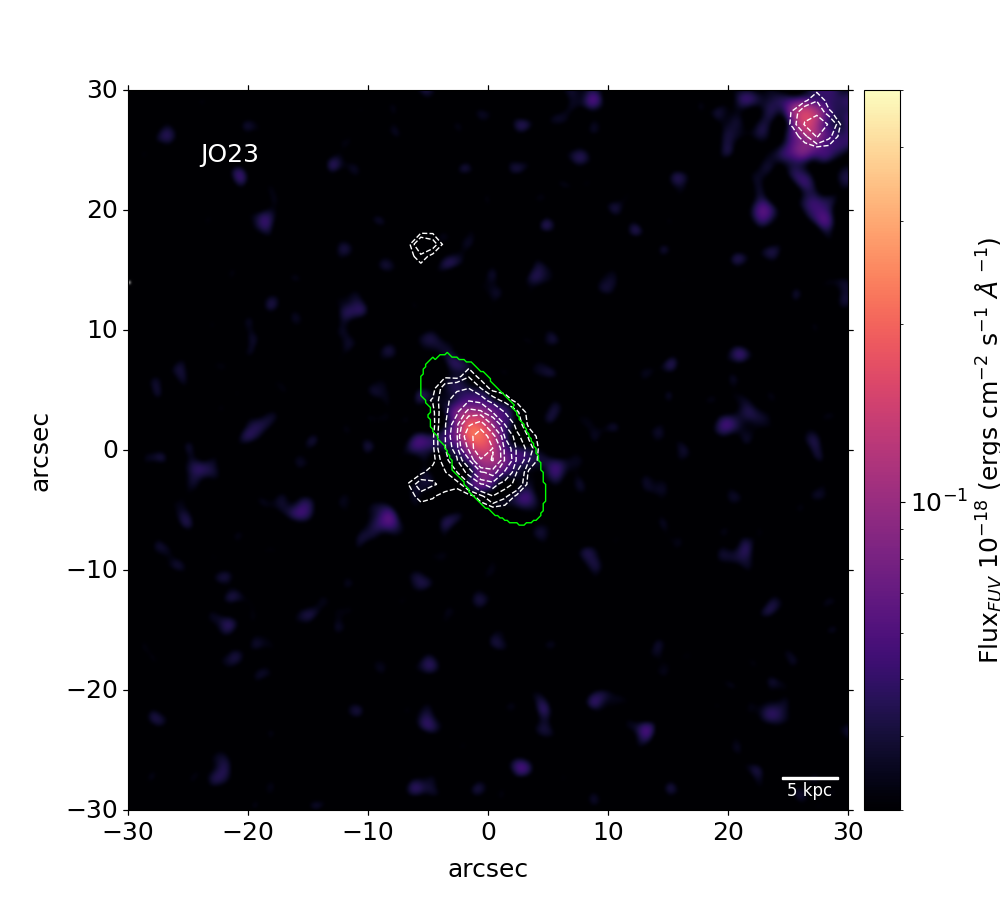}\par 
    \includegraphics[width=6.0cm]{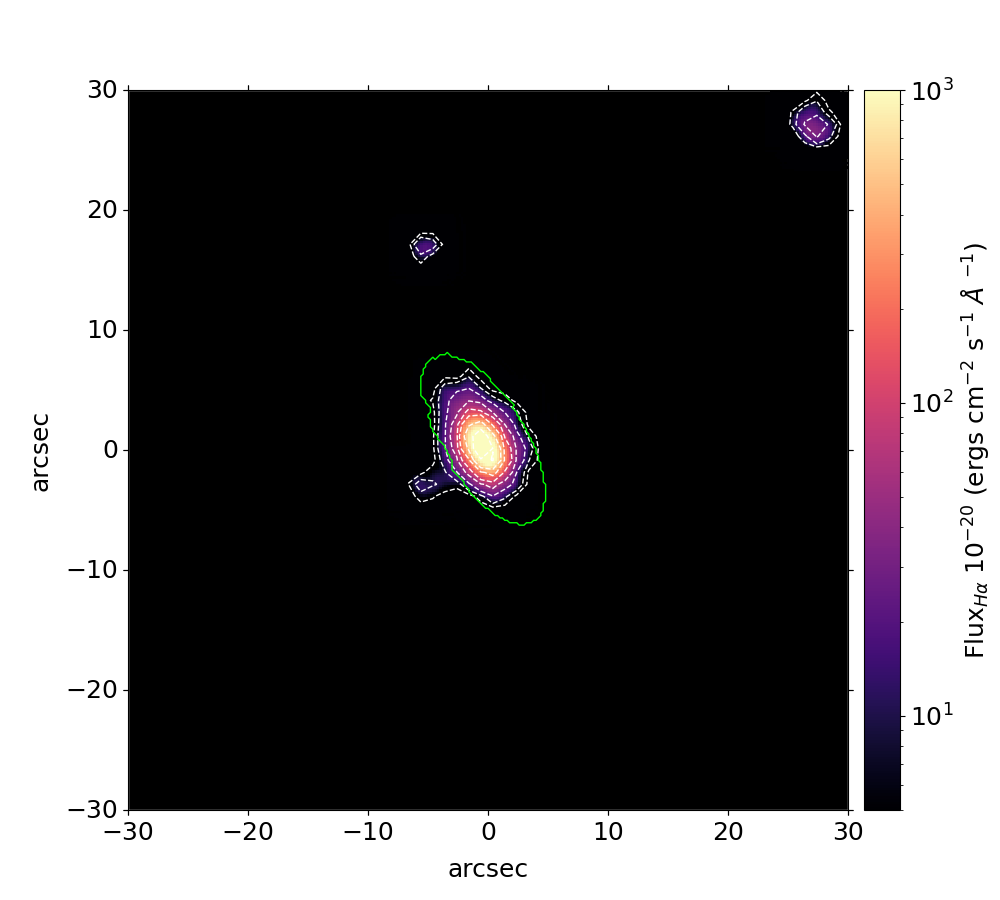}\par 
    \includegraphics[width=6.0cm]{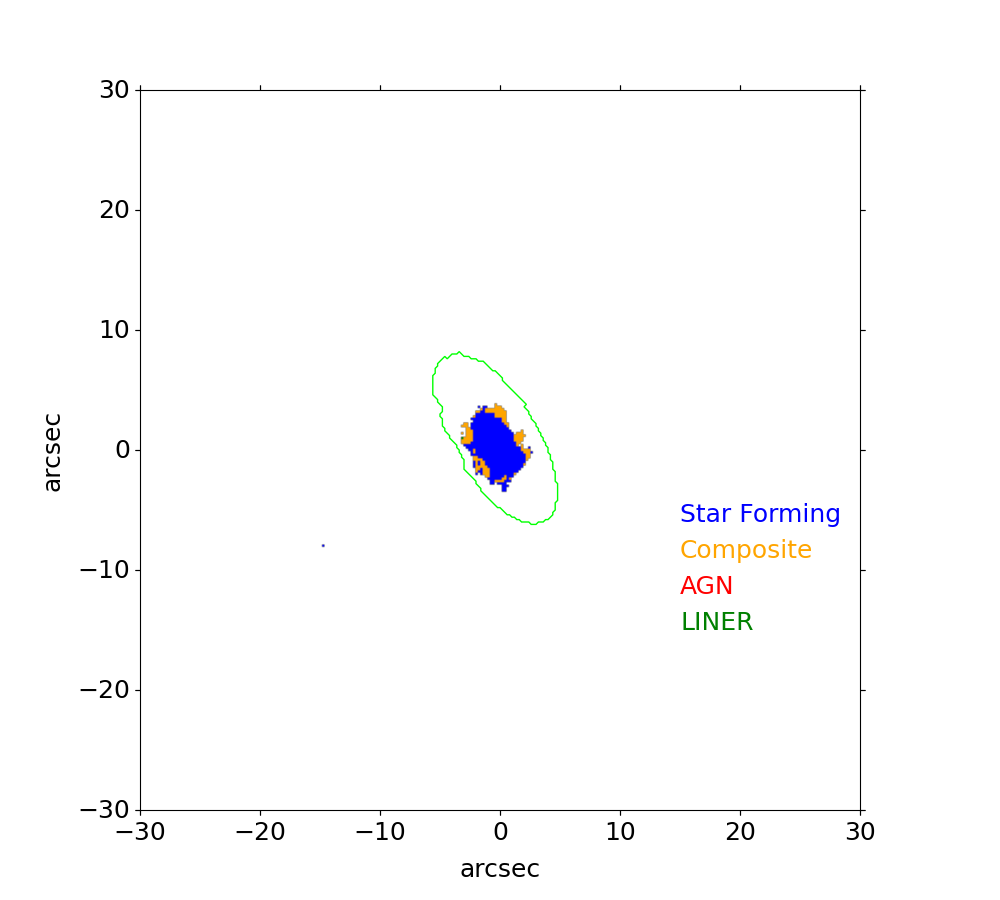}\par
 \end{multicols}
 \end{figure*}
 \begin{figure*}
 \begin{multicols}{3}
    \includegraphics[width=6.0cm]{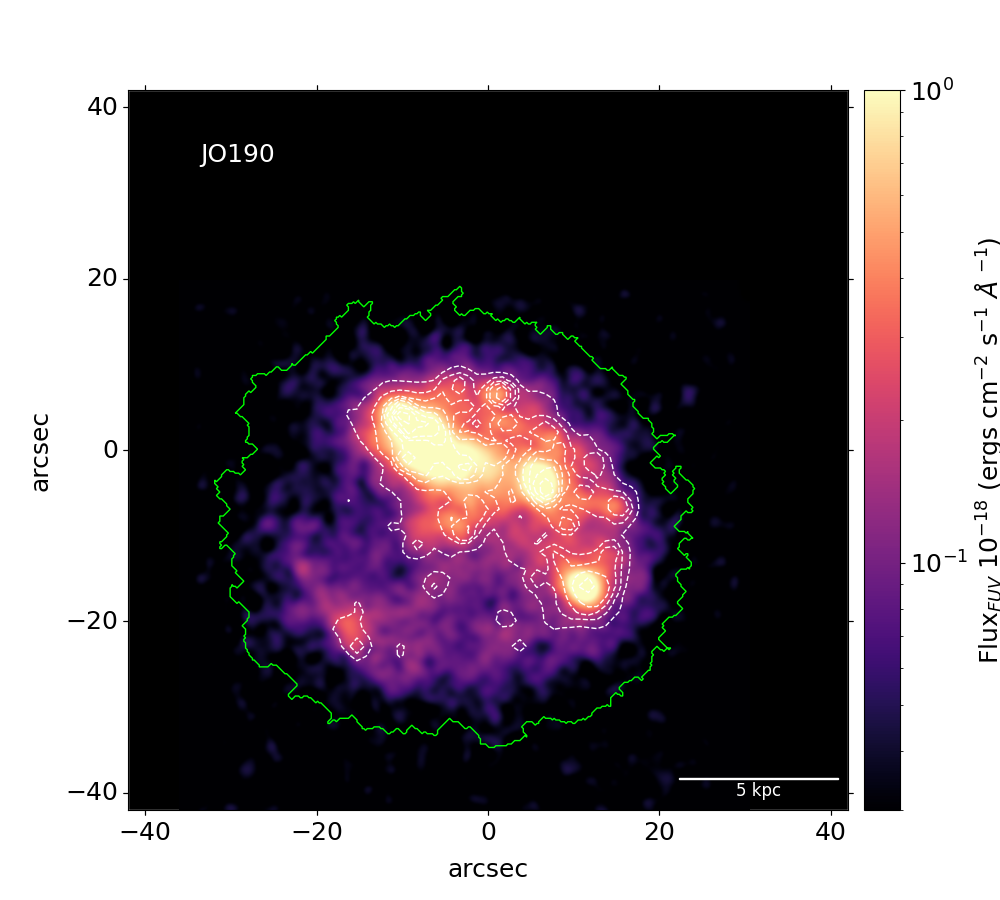}\par 
    \includegraphics[width=6.0cm]{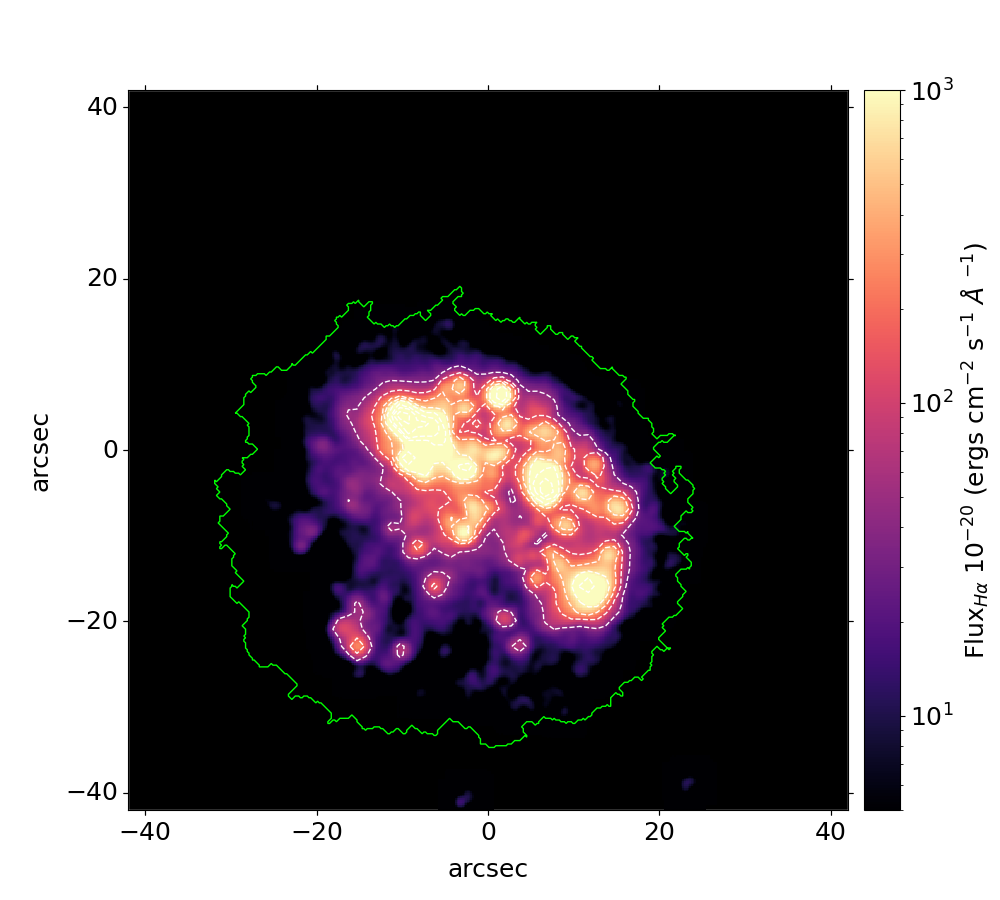}\par 
    \includegraphics[width=6.0cm]{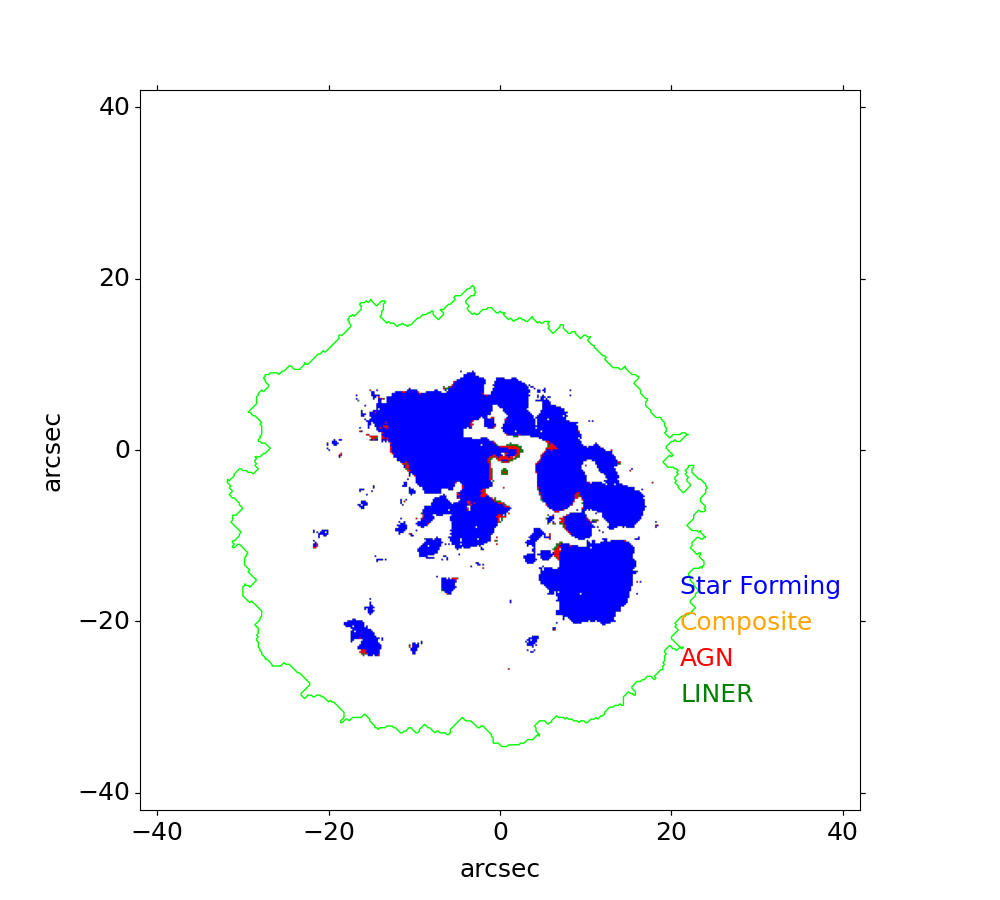}\par
 \end{multicols}
  \caption{Far-ultraviolet, H$\alpha$ flux image, and emission line diagnostic map of galaxies. The $\mathrm{H}{\alpha}$ flux contours in white colour are overlaid over FUV and H$\alpha$ images. The green line defines the galaxy's main body. Regions covered due to emission from LINER, composite (AGN+SF) and star
formation are assigned with different colours in the emission line map.}
 \label{fig:fuvhalphabptmapA1}
\end{figure*}

\section{Morphological analysis of each galaxy}

We present a morphological analysis of each galaxy, based on our visual inspection of FUV, H$\alpha$ flux, and emission line region maps.\\

\subsubsection{JO69}

The galaxy JO69 belongs to the Abell 2399 galaxy cluster and has a JType=1. The galaxy has a bar \citep{Sanchez-Garcia_2023}. The galaxy is seen face-on at an inclination of 43.7$\pm$4.9$\degree$, with an unwinding pattern for the spiral arms, likely formed due to ram-pressure stripping. The galaxy is covered by one MUSE pointing. FUV and H$\alpha$ emission are present throughout the stellar disc. The FUV and H$\alpha$ emission generally match spatially on the disc, but in the tails, the FUV flux is very weak. There is a good correspondence between FUV emitting regions and star formation in the emission line diagnostic map.

\subsubsection{JO70}

The galaxy JO70 belongs to the Abell 2399 galaxy cluster assigned with JType=1. The galaxy is a barred face-on spiral \citep{Sanchez-Garcia_2023}, with an inclination of 39.7$\pm$6.1$\degree$ \citep{Franchetto_2020} and an unwinding pattern in its spiral arms, likely formed due to ram-pressure stripping. The galaxy is covered in one MUSE pointing 
There is FUV and H$\alpha$ emission throughout the stellar disc. The FUV and H$\alpha$ emission in general spatially match on the disc and the tails of the galaxy. The FUV emitting regions shows good correspondence with regions of star formation in the emission line diagnostic map.

\subsubsection{JO93}

The galaxy JO93 belongs to the Abell 2593 galaxy cluster (JType=1 in Poggianti et al. 2024). The galaxy is seen face-on, with an inclination of 25.3$\pm$7.1$\degree$ \citep{Franchetto_2020}, and its spiral arms have a slightly unwinding pattern, likely formed due to ram-pressure stripping. The galaxy was observed with two MUSE pointings that covered the disc and stripped regions. There are regions in the stellar disc where neither FUV nor H$\alpha$ emission is detected. In general, the FUV and H$\alpha$ emission spatially match on the disc, but there are regions on the galaxy's tails with no FUV flux. There is good correspondence between the FUV emitting region and the region covered by star formation in the emission line diagnostic map.

\subsubsection{JO134}

The galaxy JO134 (JType=1), which is located in the field of the Abell 3530 galaxy cluster, is most likely part of a galaxy group. The galaxy shows asymmetry in its stellar kinematics, indicating a likely tidal interaction, and it is also experiencing ram-pressure stripping \citep{Vulcani_2021}. This makes it the only galaxy in the current study to be simultaneously affected by multiple mechanisms. The galaxy is seen face-on and covered in one MUSE pointing. FUV and H$\alpha$ emission are present throughout the stellar disc. In general, the FUV and H$\alpha$ emission spatially match on the disc, but on the tails, the FUV flux is very weak in regions with H$\alpha$ flux. There is good correspondence between FUV-emitting regions and star-forming regions on the emission line diagnostic map for this galaxy. There is a region outside the galaxy at the bottom that appears to be saturated in H$\alpha$ imaging. This region is detected as single knot in FUV but appears as multiple knots in H$\alpha$. The spectrum demonstrates that it is an A-type foreground star, and we didn’t include this region in our analysis.

\subsubsection{JO200}

The galaxy JO200 belongs to the Abell 85 galaxy cluster and has a JType=1 in Poggianti et al. (2025). The galaxy has a bar \citep{Sanchez-Garcia_2023}. The galaxy is viewed at an inclination of 45.7$\pm$3.4$\degree$ \citep{Franchetto_2020}, displaying a face-on orientation, with its spiral arms likely unwinding due to ram-pressure stripping. The galaxy was observed in two MUSE pointings that covered the disc and stripped regions. FUV and H$\alpha$ emission are present throughout the stellar disc. The FUV and H$\alpha$ emission in general spatially match on the disc and the tails of the galaxy. There is good correspondence between the FUV emitting region and the region covered by star formation in the emission line diagnostic map.

\subsubsection{JW56}

The galaxy JW56, which belongs to the Abell 1736 galaxy cluster, is assigned a JType of 2 in Poggianti et al. (2025). The galaxy disc is observed at an inclination of 65.2$\pm$1.2$\degree$ \citep{Franchetto_2020}, with stripping occurring in the west direction. No stripping feature is detected in FUV imaging outside the MUSE pointing. There are regions in the stellar disc (inside green contour) where no FUV or H$\alpha$ emission is detected. The FUV and H$\alpha$ emission generally match spatially on the disc, but in the galaxy's tails, there is little FUV emission, whereas H$\alpha$ emission is present. There is good correspondence between the FUV emitting region and the region covered by star formation in the emission line diagnostic map on the disc. The tail region displays emission regions related to star formation, but with no corresponding flux detection in FUV imaging irrespective of a deep integration time ($\sim$ 19 ks).

\subsubsection{JO85}

The galaxy JO85 belongs to the Abell 2589 galaxy cluster and shows extreme ram-pressure stripping features with JType=2. The galaxy appears lopsided, features a stellar bar, and has an AGN at its centre, and it is currently undergoing nearly edge-on stripping \citep{Gullieuszik_2020,Peluso_2022,Sanchez-Garcia_2023}. The galaxy is seen face-on at an inclination of 25.1$\pm$3.7$\degree$ \citep{Franchetto_2020}, with stripping happening in the north-east. The spiral arms of the galaxy follow an unwinding pattern, likely formed due to ram-pressure stripping. The galaxy was observed in two MUSE pointings that covered the disc and tail of the galaxy. Several features are detected in FUV imaging (see Fig \ref{fig:rgb}) outside the MUSE pointing (north direction of top pointing), which may be associated with the galaxy's stripped feature. The stellar disc of the galaxy is shown with a green contour covering regions with significant FUV and H$\alpha$ flux. The FUV and H$\alpha$ emission in general spatially match on the disc and tails of the galaxy. In general, we find that emission coincides with H$\alpha$ and FUV imaging in both the tail and disc of the galaxy. There is good correspondence between the FUV emitting region and the region covered by star formation in the emission line diagnostic map. The disc's centre and the unwinding spiral arms' sides feature regions with composite emission. An enhanced emission detected at the south-west side of the galaxy disc is seen in FUV and H$\alpha$ imaging. This is likely due to the enhanced star formation in the region that gets compressed where the galaxy hits the ICM. 

\subsubsection{JO147}

The galaxy JO147 (JType=2) belongs to the Abell 3558 galaxy cluster. The galaxy hosts an AGN \citep{Peluso_2022}. The star formation rate (SFR) from the tail of the galaxy is $\sim$ 4\% of the total \citep{Poggianti_2019b}. The galaxy is seen edge-on at an inclination of 81.6$\pm$2.3$\degree$ \citep{Franchetto_2020} with stripping happening in the north-west direction. The galaxy is covered in one MUSE pointing. Three knots are visible in the H$\alpha$ image outside the galaxy disc, one of which is also detected in FUV and is likely a star-forming dwarf satellite galaxy. The FUV and H$\alpha$ emission generally match spatially on the disc. The stripped tails are prominently detected in H$\alpha$, with no corresponding emission in FUV imaging data. This is further supported by the difference in the number of segments detected in the H$\alpha$ and FUV images of the galaxy. We detected 33 segments in the H$\alpha$ image of the tail, which are reduced to five segments in the tail region near the disc in FUV imaging. The emission line diagnostic map reveals only one clump on the tail, which is a star-forming region not detected in FUV imaging. The disc shows shows composite emission regions on the stripping side. We note that the H$\alpha$ emission is very diffuse along the tails. JO147 is a strong example of a ram-pressure stripped galaxy, characterised by significant ionisation along its stripped tails and very low star formation efficiency, as revealed by FUV imaging.

\subsubsection{JO149}

The galaxy JO149 (JType =2) belongs to the Abell 3558 galaxy cluster. The galaxy has the lowest stellar mass (logM$_\star$/M$_\odot$=8.87) out of all the galaxies studied here. The galaxy is seen face-on and is covered in one MUSE pointing. There is FUV and H$\alpha$ emission throughout the stellar disc. The FUV and H$\alpha$ emission generally spatially match on the disc, but on the tails, the FUV flux is very weak. On the emission line diagnostic map for this galaxy, there is a good match between star-forming regions on the disc and FUV emitting regions. The FUV emitting regions on the tail show little correspondence with the emission line diagnostic map for this galaxy, except for two segments.

\subsubsection{JO160}

The galaxy JO160 belongs to the Abell 3558 galaxy cluster and is showing extreme ram-pressure stripping with JType=2. The galaxy has a bar \citep{Sanchez-Garcia_2023}. The stripped tail is found to contribute less than 1\% of the galaxy's total SFR \citep{Poggianti_2019b}. The galaxy is observed almost face-on, with an inclination of 59.4$\pm$1.5$\degree$ \citep{Franchetto_2020}, and stripping is occurring to the west of the disc, forming a cometary-shaped tail. There are diffuse features associated with the tail detected in FUV imaging that appear outside the MUSE pointing (see Fig \ref{fig:rgb}). There are regions in the stellar disc (shown with a green contour) that do not have FUV, nor H$\alpha$, emission. In general, the FUV and H$\alpha$ emission spatially match on the disc galaxy. We note that the stripped tail, where there is an unresolved H$\alpha$ flux emitting region, shows little or no FUV emission. There is good correspondence between the FUV emitting region and the region covered by star formation in the emission line diagnostic map (mainly on the disc).

\subsubsection{JO171}

The galaxy JO171, part of the Abell 3667 galaxy cluster, is classified as JType=2 in Poggianti et al. (2025), which is typical of extreme jellyfish galaxies undergoing ram-pressure stripping. The galaxy hosts an AGN and resembles Hoag's galaxy in terms of morphology, while the kinematic and stellar population analysis of the MUSE data reveals the stripping of gas from a ring galaxy that is falling into the cluster for the first time \citep{Moretti_2018b,Peluso_2022}. The tail accounts for $\sim$ 20 \% of the total SFR of the galaxy \citep{Poggianti_2019b}. The galaxy disc is seen face-on at an inclination of 18.4$\degree$ \citep{Franchetto_2020} with stripping likely happening in the north direction. No stripping feature is detected in FUV imaging outside the MUSE pointing. The central part contains the red spheroid component, which is dominated by an evolved stellar population. There is significant flux coming from features outside the galaxy in FUV imaging due to ongoing star formation in the stripped tails. As seen in Fig \ref{fig:rgb}, a UV cavity is present at the galaxy's centre, resulting from the ring-like structure, and there is a significant flux reduction on the southern side of the disc, caused by the progression of star formation quenching in the disc due to ram-pressure stripping. The main body of the galaxy, as determined by isophotal analysis, is shown in green, and the H$\alpha$ contour is shown in white. We overlay the H$\alpha$ contour over both FUV and H$\alpha$ images for easy comparison. The FUV and H$\alpha$ flux-emitting regions are displaced with respect to the stellar disc, shown in green. In general, the FUV and H$\alpha$ emission spatially match on the disc and tails of the galaxy. In certain regions of the galaxy's disc, UV emission is detected, but H$\alpha$ is not, while in the tails, H$\alpha$ is present, but FUV emission is absent. There is a good correspondence between the FUV emitting region and the region of star formation in the emission line diagnostic map. A few spaxels exhibit composite emission, both on the disc and tail of the galaxy. We note that unresolved H$\alpha$ flux emitting regions is having little or no FUV emission. The southern region of the disc is devoid of both UV and H$\alpha$ emission which indicates that quenching due to ram-pressure stripping started on the disc at least $\sim$ 100-200 Myr ago. There is a region close to the disc with only FUV flux (detected as two segments in segmentation map created from FUV image) with no H$\alpha$ which suggests the presence of stellar populations with ages $>$ 10 Myr. As the galaxy moves within the cluster, gas is stripped, which may cause star formation quenching to progress on the disc.

\subsubsection{JO36}

The galaxy JO36, belonging to the Abell 160 galaxy cluster, was assigned a JType=3 in Poggianti et al. (2025), typical for galaxies with truncated discs. Morphologically classified as an Sc spiral, the galaxy is seen almost edge-on at an inclination of 80.7$\pm$2.6$\degree$, hosting an AGN and a truncated ionised gas disc \citep{Fritz_2017,Peluso_2022}. \citet{Fritz_2017} report that the central 20 kpc region of the galaxy hosts star formation, while the external parts of the disc are dominated by an older population (less than 500 Myr). The galaxy does not have an extended tail and may be at an advanced stage of ram-pressure stripping, with little gas left, resulting in the truncation of ionised gas regions on the disc. Based on LOFAR 144 MHz and uGMRT 675 MHz observations, the galaxy is found to undergo an interaction with the radio plume of another galaxy 200-500 Myr ago that resulted in an enhanced SFR \citep{Ignesti_2023}. The FUV and H$\alpha$ emission is confined to the central regions of the stellar disc (shown with green contour). The FUV emission from the disc is truncated, following a similar trend to that in H$\alpha$. We note that the FUV emission is coming from a narrow region compared to H$\alpha$, which is more extended on the disc of the galaxy. The emission line diagnostic map shows regions with composite emission in these region with extended H$\alpha$ emission. The FUV emission on the tail is showing a very weak correspondence with the H$\alpha$ emission. There is one star-forming region in the tail emission line map that has matching emission in FUV.

\subsubsection{JO23}

The galaxy JO23 belongs to the Abell 151 galaxy cluster and is assigned a JType=3 suggesting the presence of a truncated disc. The galaxy is seen almost edge-on at an inclination of 72.6$\pm$1.6$\degree$ and is covered in one MUSE pointing. There is FUV and H$\alpha$ emission confined to the central regions of the stellar disc (shown with green contour). The FUV emission on the disc is more truncated with respect to H$\alpha$ emission. The emission line diagnostic map shows regions with composite emission matching with the region having extended H$\alpha$ emission. The H$\alpha$ image shows diffuse emission along the tail, but there is no corresponding feature in the FUV image. In the emission line map, the galaxy's disc displays star formation regions confined to the centre, with no detection along the tail. 

\subsubsection{JO190}

The galaxy JO190 belongs to the Abell 3880 galaxy cluster. The galaxy was selected for observation because it was classified as an extreme case of ram-pressure stripping, with a JClass of 5, based on visual identification from B-band imaging. According to Poggianti et al. (2025), the galaxy is assigned a JType of -9, indicating that it is not experiencing ram-pressure stripping, and its underlying nature remains unclear. The galaxy features a bar, as reported by \citep{Sanchez-Garcia_2023}. The galaxy is viewed face-on, with features visible on its spiral arms in the south-west direction. The galaxy is covered by a single MUSE pointing. FUV and H$\alpha$ emission are present throughout the stellar disc. In general, the FUV and H$\alpha$ emission spatially match on the disc. There is good correspondence between FUV emitting regions and star-forming regions on the emission line diagnostic map for this galaxy. We present the FUV and H$\alpha$ imaging details of the galaxy here but do not include this galaxy in analysis.  

\end{appendix}

\end{document}